\documentclass[11pt]{article}
\usepackage[colorlinks=true,linkcolor=black,citecolor=black,hypertexnames=false]{hyperref}
\usepackage{genres}
\bibpunct[, ]{(}{)}{;}{a}{}{,}

\usepackage{amssymb,amsfonts,amsmath,amsthm,amscd,dsfont,mathrsfs,bm,mathtools}
\usepackage{graphicx,float,psfrag,epsfig,amssymb}
\usepackage{nameref}
\usepackage{makecell}
\usepackage[font={small}]{caption}
\usepackage{subcaption}
\usepackage{pdflscape}
\usepackage{afterpage}
\usepackage{verbatimbox}
\usepackage{gensymb}
\usepackage{placeins}
\usepackage{array}
\usepackage{graphicx, comment}
\usepackage[utf8]{inputenc}
\usepackage{multirow}
\usepackage{xr}
\usepackage{tabularx}
\usepackage{nicefrac}
\usepackage{cleveref}
\usepackage{booktabs}
\usepackage{siunitx}
\usepackage{algorithm}
\usepackage[noend]{algorithmic}

\def\mycitet#1{\citep{#1}}

\usepackage[top=1in,bottom=1in,left=1in,right=1in]{geometry}

\usepackage{color}

\makeatletter
\def\enumhook{}

\def\itemhook{}

\def\descripthook{}
\def\enumerate{%
  \ifnum \@enumdepth >\thr@@\@toodeep\else
    \advance\@enumdepth\@ne
    \edef\@enumctr{enum\romannumeral\the\@enumdepth}%
      \expandafter
      \list
        \csname label\@enumctr\endcsname
        {\usecounter\@enumctr\def\makelabel##1{\hss\llap{##1}}%
          \enumhook \csname enumhook\romannumeral\the\@enumdepth\endcsname}%
  \fi}
\def\itemize{%
  \ifnum \@itemdepth >\thr@@\@toodeep\else
    \advance\@itemdepth\@ne
    \edef\@itemitem{labelitem\romannumeral\the\@itemdepth}%
    \expandafter
    \list
      \csname\@itemitem\endcsname
      {\def\makelabel##1{\hss\llap{##1}}%
        \itemhook \csname itemhook\romannumeral\the\@itemdepth\endcsname}%
  \fi}

\renewcommand{\enumhook}{
  \setlength{\topsep}{2pt}%
  \setlength{\itemsep}{0pt}
}
\renewcommand{\descripthook}{
  \setlength{\topsep}{2pt}%
  \setlength{\itemsep}{2pt}
}
\renewcommand{\itemhook}{
  \setlength{\topsep}{2pt}%
  \setlength{\itemsep}{0pt}
}

\allowdisplaybreaks

\makeatletter
\renewcommand\section{\@startsection {section}{1}{\z@}%
                                   {-3.5ex \@plus -1ex \@minus -.2ex}%
                                   {2.3ex \@plus.2ex}%
                                   {\fontfamily{phv}\selectfont\large\Large\bfseries}}
\renewcommand\subsection{\@startsection {subsection}{1}{\z@}%
                                   {-3.5ex \@plus -1ex \@minus -.2ex}%
                                   {1ex \@plus.2ex}%
                                    {\fontfamily{phv}\selectfont\large}}
\makeatother

\DeclareCaptionLabelFormat{subrefformat}{\sf{(#2)}}
\DeclareMathAlphabet{\mathpzc}{OT1}{pzc}{m}{it}

\newcommand{\sref}[1]{\ref{#1}}
\newcommand{\tref}[1]{Table~\ref{#1}}
\newcommand{\fref}[1]{Figure~\ref{#1}}

\newcommand{\thmref}[1]{Theorem~\ref{#1}}

\newtheorem{propo}{Proposition}[section]

\newtheorem{thm}{Theorem}

\newtheorem{remark}[propo]{Remark}

\newcommand\independent{\protect\mathpalette{\protect\independenT}{\perp}}
\def\independenT#1#2{\mathrel{\rlap{$#1#2$}\mkern2mu{#1#2}}}

\def\cA{{\cal A}}
\def\cD{{\cal D}}

\def\cC{{\cal C}}

\def\cE{{\cal E}}

\def\cT{{\cal T}}

\def\Bino{{\sf Binomial}}

\def\reals{{\mathbb R}}

\def\prob{{\mathbb P}}
\def\E{{\mathbb E}}

\def\cB{{\cal B}}

\def\L0{{L_0}}

\def\de{{\rm d}}
\def\<{\langle}
\def\>{\rangle}
\def\diag{{\rm diag}}

\def\F{{\sf F}}

\def\F{{\sf F}}

\def\sT{{\sf T}}

\def\cov{{\rm Cov}}

\def\v*{v_0}
\def\T*{T_0}

\def\u*{u_0}
\def\F*{F_0}

\definecolor{olivegreen}{rgb}{0,0.6,0.4}

\def\cL{\mathcal{L}}

\def\diag{{\rm diag}}

\def\cL{\mathcal{L}}

\def\hmu{\hat{\mu}}
\def\heta{\hat{\eta}}
\def\cD{\mathcal{D}}
\def\cE{\mathcal{E}}
\def\u{\mathbf{1}}

\def\logit{{\rm logit}}
\def\bq{\mathbf{q}_\ell^{\rm cen}}
\def\tbq{\mathbf{\tilde{q}}_\ell^{\rm cen}}

\DeclareMathOperator*{\argmin}{arg\,min}

\def\bfa{\mathbf{a}}
\def\bfb{\mathbf{b}}

\def\bfu{\mathbf{u}}
\def\bfy{\mathbf{y}}
\def\bfz{\mathbf{z}}

\def\bfdelta{\mathbf{\delta}}
\def\bfzero{\mathbf{0}}
\newcommand{\defeq}{\vcentcolon=}
\newcommand{\eqdef}{=\vcentcolon}

\newcommand{\ajcomment}[1]{}

\makeatletter
\newcommand{\labitem}[2]{%
\def\@itemlabel{\text{#1}}
\item
\def\@currentlabel{#1}\label{#2}}
\makeatother

\addtocontents{toc}{\protect\setcounter{tocdepth}{2}}

\def\localizationAlgorithm{{\sf GAP}}
\def\associationTestingAlgorithm{{\sf SCGAP}}

\def\GCAT{{\sf GCAT}}
\def\SpaceMix{{\sf SpaceMix}}
\def\SPA{{\sf SPA}}
\def\SCAT{{\sf SCAT}}
\def\OriGen{{\sf OriGen}}
\def\HO{{\sf Human Origins}}
\def\POPRES{{\sf POPRES}}
\def\GLOBETROTTER{{\sf GLOBETROTTER}}
\def\NFBC{{\sf NFBC}}

\title{Novel probabilistic models of spatial genetic ancestry with applications to stratification correction in genome-wide association studies}
\author{
Anand Bhaskar\footnote{These authors contributed equally to this work and are ordered alphabetically.} \footnote{Correspondence should be addressed to A.B. (abhaskar@stanford.edu) or A.J. (ajavanma@marshall.usc.edu).} \thanks{Department of Genetics, Stanford University, Stanford, CA 94305} \thanks{Howard Hughes Medical Institute, Stanford University, Stanford, CA 94305} \and 
Adel Javanmard$^{*\dagger}$\thanks{Marshall School of Business, University of Southern California, Los Angeles, CA 90089} \and
Thomas A. Courtade\thanks{Department of Electrical Engineering and Computer Sciences, University of California, Berkeley, CA 94720} \and 
David Tse$^{\|}$\thanks{Department of Electrical Engineering, Stanford University, Stanford, CA 94305}
}

\begin{document}

\begin{titlepage}
\clearpage
\maketitle
\thispagestyle{empty}

\section*{Abstract}
Genetic variation in human populations is influenced by geographic ancestry due to spatial locality in historical mating and migration patterns. Spatial population structure in genetic datasets has been traditionally analyzed using either model-free algorithms, such as principal components analysis (PCA) and multidimensional scaling, or using explicit spatial probabilistic models of allele frequency evolution. We develop a general probabilistic model and an associated inference algorithm that unify the model-based and data-driven approaches to visualizing and inferring population structure. Our algorithm {\sf Geographic Ancestry Positioning} (\localizationAlgorithm{}) relates local genetic distances between samples to their spatial distances, and can be used for visually discerning population structure as well as accurately inferring the spatial origin of individuals on a two-dimensional continuum. On both simulated and several real datasets from diverse human populations, \localizationAlgorithm{} exhibits substantially lower error in reconstructing spatial ancestry coordinates compared to PCA.

Our spatial inference algorithm can also be effectively applied to the problem of population stratification in genome-wide association studies (GWAS), where hidden population structure can create fictitious associations when population ancestry is correlated with both the genotype and the trait. We develop an association test that uses the ancestry coordinates inferred by \localizationAlgorithm{} to accurately account for ancestry-induced correlations in GWAS. Based on simulations and analysis of a dataset of 10 metabolic traits measured in a Northern Finland cohort, which is known to exhibit significant population structure, we find that our method has superior power to current approaches.

\vspace{5mm}
\noindent \textit{Software:} Our software implementation is available at \url{https://github.com/anand-bhaskar/gap}.

\end{titlepage}

\section{Introduction}
Modern human genomic datasets routinely contain samples from geographically diverse populations \citep{nelson:2008,1000genomes:2010}, and
analyses of these datasets has shown that the patterns of genetic variation across human populations encodes substantial information about their geographic ancestry \citep{cavalli:1994,ramachandran:2005,novembre:2008}.
Inferring such spatial population structure from genetic data is of fundamental importance to many problems in population genetics --- identifying genomic regions under selective pressure \citep{lewontin:1973,coop:2009,yang:2012}, correcting for population structure in genome-wide association studies \citep{price:2006}, and shedding light on ancient human history \citep{jakobsson:2008}, to name a few. 

A fundamental technique for studying spatial demography is the visualization and inference of population structure through low-dimensional representations of genomic data. Methods like principal components analysis (PCA) \citep{lao:2008,novembre:2008} and multidimensional scaling (MDS) \citep{jakobsson:2008} were among the first approaches that demonstrated that genotype data could be used to accurately recapitulate geographic ancestry. Moreover, their performance and interpretation has been backed by theoretical work \citep{patterson:2006,paschou:2007,mcvean:2009,novembre:2008b}. 
There is also a wide spectrum of spatial genetic models and methods \citep{wasser:2004,yang:2012,ranola:2014,baran:2015,bradburd:2016} which have been developed for inferring geographic ancestry coordinates. The \SPA{} model \cite{yang:2012} uses a logistic function over space to parameterize the allele frequency at each SNP, while methods like \SCAT{} \citep{wasser:2004} and \SpaceMix{} \citep{bradburd:2016} consider allele frequency covariance functions which decay exponentially with geographic distance. The \OriGen{} algorithm of \cite{ranola:2014}, while not positing a specific functional form for the allele frequency function, performs an optimization which encourages smoothness in allele frequency over space. 
In all these spatial models, the inference of ancestry coordinates is performed using maximum likelihood or expectation-maximization algorithms that are tailored to the details of the model.

In this work, we marry the previously mentioned model-free and model-based approaches to geographic ancestry localization by developing a flexible spatial stochastic process model that subsumes previously developed parametric allele frequency models such as \SPA{}, \SCAT{} and \SpaceMix{} as special cases. Furthermore, we develop a data-driven spatial reconstruction algorithm {\sf Geographic Ancestry Positioning} (\localizationAlgorithm{}), that exploits the structural properties of our stochastic process while being agnostic to its minutiae. Our localization algorithm is inspired by principles from manifold learning, and can be viewed as a generalization of PCA. The idea behind our approach is to infer the local spatial distances between sampled individuals using their genotypes, and to then create a global spatial embedding that is faithful to the local geometry information. Our probabilistic process and associated inference algorithm bridge the long threads of work in data-driven and model-based ancestry localization from genotypic data. Through extensive simulations, we demonstrate that \localizationAlgorithm{} often performs substantially better than PCA at both visually discerning spatially structured populations (\fref{fig:logisticExpDecayCov_n2000_p50k}) as well as inferring the spatial coordinates of individuals (\tref{tab:logisticExpDecayCov_MDS_PCA_beta1_main_text}). We also prove theoretically that, under our probabilistic model, \localizationAlgorithm{} performs at least as well as PCA in reconstructing the spatial coordinates of genetic samples. We apply \localizationAlgorithm{} to three public genotype datasets from the \HO{} \citep{lazaridis:2014}, \GLOBETROTTER{} \citep{hellenthal:2014}, and \POPRES{} \citep{nelson:2008} projects. Compared to PCA, \localizationAlgorithm{} exhibits 31\% lower error in spatial reconstruction of the subpopulations in the \HO{} dataset, 10\% lower error on the \GLOBETROTTER{} dataset, and 56\% lower error on the \POPRES{} dataset.\footnote{If we use only a subset of SNPs with minor allele frequency $\geq 10\%$, \localizationAlgorithm{} and PCA perform similarly. See Supplementary Information \S1.7 for details.}

Population structure also has serious implications for genome-wide association studies (GWAS). In the GWAS setting, one is interested in finding loci that are causal for the trait, while being resilient to false associations arising from hidden population structure and environmental confounders. Spurious associations can arise due to ancestry-induced correlations between causal and non-causal loci, or when ancestry is correlated with both the genotype and the trait \citep{campbell:2005}. 
PCA \citep{price:2006} and linear mixed models (LMM) \citep{kang:2010} are two popular classes of methods for correcting ancestry confounding in human genetics studies.
Both of these methods test genetic associations in prospective models describing the distribution of the trait conditional on the genotype. On the other hand, retrospective models describing the distribution of genotypes conditional on the trait are more natural in the setting of case-control studies and have been shown to be equivalent to prospective models under suitable assumptions \citep{prentice:1979, song:2015}. Based on this, Song \emph{et al.} \citep{song:2015} developed a testing procedure, \GCAT{}, that controls for ancestral confounding by using a latent factor model \citep{hao:2016} to estimate the allele frequencies at each SNP across the sample.
We propose an alternative allele frequency estimation procedure and association test, {\sf Stratification Correction via \localizationAlgorithm{}} (\associationTestingAlgorithm{}), that can effectively correct for ancestry confounding by using the spatial coordinates inferred by \localizationAlgorithm{}. \associationTestingAlgorithm{} employs an allele frequency smoothing procedure using the inferred coordinates from \localizationAlgorithm{} in order to estimate the allele frequency at each SNP across the sample. Our association testing procedure uses these estimates of the allele frequency to test each SNP in an inverse regression of the genotype against the trait, conditional on the estimated allele frequency. 
Through simulations, we show that our allele frequency estimation procedure when used with the ancestry coordinates from our localization algorithm \localizationAlgorithm{} has almost as high power as if the true ancestry coordinates were known, and has considerably higher power than if the ancestry coordinates were inferred using PCA. 
We applied \associationTestingAlgorithm{} to a birth cohort from Northern Finland containing several quantitative metabolic traits and observe that it compares favorably to state-of-the-art computationally intensive approaches such as LMMs. For instance, \associationTestingAlgorithm{} and \GCAT{} are the only methods to identify a SNP (rs2814982) associated with height in this dataset.

\section{Methods}
\subsection{Model}
Suppose that we are given genotype or sequence data from $n$ individuals at $p$ SNPs. 
We will use $X$ to denote the $n \times p$ genotype matrix, where entry $x_{i\ell} \in \{0,1,2\}$ is the number of alleles at SNP $\ell$ in individual $i$. We let $\bfz_i \in \reals^2$ denote the unknown ancestral origin of individual $i$. In our spatial probabilistic model, the allele frequencies for different SNPs are assumed to be drawn from independent stochastic processes defined over the two-dimensional geographical space. Specifically, letting $q_\ell(\bfz)$ be the allele frequency of SNP $\ell$ at location $\bfz$, we let $\mu_{\ell} \defeq \E[q_\ell(\bfz)]$ denote the mean allele frequency of SNP $\ell$ in the population. The covariance $\cov(q_\ell(\bfz),q_\ell(\bfz'))$ in allele frequencies between pairs of locations $\bfz$ and $\bfz'$ is captured by a covariance decay function $\eta$ as follows,
\begin{align}
\cov(q_\ell(\bfz), q_\ell(\bfz')) = \E[(q_\ell(\bfz) - \mu_\ell)(q_\ell(\bfz') - \mu_\ell)] &\eqdef \eta(\bfz - \bfz'). \label{eq:cov_process}
\end{align}
This assumption allows us to model the phenomenon of isolation by distance, where the covariance in allele frequencies decays with geographic separation, while also allowing for different rates of covariance decay in different spatial directions. Such observations of anisotropic isolation by distance have been reported by previous studies of African, Asian, and European populations \citep{jay:2013}. Moreover, several previous spatial genetic models \citep{wasser:2004,bradburd:2016,yang:2012} can be recast as specific parametrizations of the autocovariance function $\eta$ of our model (Supplementary Information \S1.4). However, in contrast to these models, we do not impose any explicit parametric form on the autocovariance function $\eta$. 
Figure S2 (Supplementary Information) shows example allele frequency surfaces from two such previously proposed spatial processes \citep{wasser:2004,yang:2012} that are captured by our model formulation.

\subsection{Algorithm}
Our localization algorithm \localizationAlgorithm{} takes a data-driven approach while exploiting the structure of the autocovariance function in \eqref{eq:cov_process}. The idea behind our algorithm is to define a genetic squared-distance $d^2$ between each pair of sampled individuals as follows,
\begin{align}
d^2_{ij} = \eta(\bfzero) - \eta(\bfz_i-\bfz_j).\label{eq:genetic_squared_distance}
\end{align}
We can exploit the structure of our model in \eqref{eq:cov_process} to relate the genetic squared-distances $d^2_{ij}$ between genetically similar pairs of individuals to their spatial squared-distances $\|\bfz_i-\bfz_j\|^2$. More precisely, we show that,
\begin{align}
	d^2_{ij} \approx \|J(\bfz_i-\bfz_j)\|^2\,, \textrm{for } i,j \textrm{ where } d_{ij} \textrm{ is small enough}. \label{eq:genetic_spatial_squared_distance_relation}
\end{align}
In \eqref{eq:genetic_spatial_squared_distance_relation}, $J$ is a $2 \times 2$ invertible matrix that is determined by the underlying stochastic process. We use \eqref{eq:genetic_spatial_squared_distance_relation} only for those pairs of individuals $i$ and $j$ where $d_{ij}$ is smaller than some threshold parameter $\tau$. 

Our localization algorithm consists of three main steps, which we describe below, leaving some of the involved details to \S1.2 of the Supplementary Information: \\
(1) Using the genotype matrix $X$, we construct provably consistent estimators $\hat{\eta}_0$ and $\hat{\eta}_{ij}$ for $\eta(\bfzero)$ and $\eta(\bfz_i - \bfz_j)$ respectively. These estimators are given in Theorem 1 in the Supplementary Information. \\
(2) We estimate the genetic squared-distances $\hat{d}^2_{ij}$ according to \eqref{eq:genetic_squared_distance} using the estimates for $\eta(\bfzero)$ and $\eta(\bfz_i - \bfz_j)$ computed in the previous step. Applying relation \eqref{eq:genetic_spatial_squared_distance_relation}, local genetic distances are good proxies for the spatial distances. We therefore keep estimates $\hat{d}_{ij}$ only for those pairs of individuals where $\hat{d}_{ij} \leq \tau$. \\
(3) We find a \emph{global} embedding of individuals in the geographic space from their estimated \emph{local} pairwise distances.  To this end, we borrow tools  from the area of manifold learning. In this work, we have used the ISOMAP algorithm \citep{Tenenbaum:ISOMAP} for this step. However, other algorithms developed for manifold learning can be applied in this step too, some of which are discussed in the Supplementary Information. 

The spatial reconstruction accuracy of our procedure will depend on the threshold $\tau$ that is chosen in step (2) above. The optimal choice of the threshold $\tau$ will depend on the dataset and the validity of the second-ordinary stationarity assumption of our model. One can pick $\tau$ in a similar manner to how parameter tuning is done in machine learning. Specifically, we will use a small subset (20\%) of the samples as a training set with known spatial coordinates, and perform cross-validation over a grid of $\tau$ (see Supplementary Information \S1.2 and \S1.3).

\section{Results}
\subsection{Simulations}
We considered two simulation scenarios to model isotropic and direction-dependent allele frequency covariance decay. For both simulation scenarios, we simulated $n = 2{,}000$ individuals at $p = 50{,}000$ SNPs. The true geographic origin $\bfz_i$ of individual $i$ was simulated by sampling each coordinate according to a Beta($\beta, \beta$) distribution from the unit square. This distribution lets us smoothly interpolate between dense sampling of individuals in the interior of the space to dense sampling at the boundaries (Figures 1(a) and 1(d)), with the setting $\beta = 1$ representing uniform sampling. The spatial allele frequencies at each SNP were generated by applying the logistic function to sample paths from a spatial Gaussian process. Assuming Hardy-Weinberg equilibrium, the genotypes of each individual $i$ were drawn according to a binomial distribution from the allele frequencies at their geographic origin $\bfz_i$. 

\begin{itemize}
 \item {\bf Isotropic covariance decay}: The allele frequency $q_{\ell}(\bfz)$ of SNP $\ell$ at location $\bfz$ is given by $q_{\ell}(\bfz) = 1 / ( 1 + \exp(G_{\ell}(\bfz)))$, where $G_{\ell}(\bfz)$ is a sample path of a two dimensional stationary Gaussian process with mean zero and covariance kernel $K(\bfz, \bfz') = \exp(- (\alpha_1 \|\bfz - \bfz'\|)^{\alpha_2})/\alpha_0$. Such models have been previously considered by the \SCAT{} \citep{wasser:2004} and \SpaceMix{} \citep{bradburd:2016} methods. In order for $K(\bfz, \bfz')$ to be a valid covariance kernel, $0 \leq \alpha_2 \leq 2$.
 \item{\bf Directional covariance decay}: The allele frequency $q_{\ell}(\bfz)$ of SNP $\ell$ at location $\bfz$ is given by $q_{\ell}(\bfz) = 1 / ( 1 + \exp(G_{\ell}(\bfz)) )$, where $G_{\ell}(\bfz)$ is a sample path of a two dimensional stationary Gaussian process with mean zero and covariance kernel $K(\bfz, \bfz') = \exp(- (\alpha_1 |\< \bfu, \bfz - \bfz' \>|)^{\alpha_2})/\alpha_0$, and $\bfu$ is a unit-length direction vector in $\reals^2$. The resulting allele frequencies are equal along directions perpendicular to $\bfu$, and this model can thus be viewed as a generalization of the \SPA{} model \citep{yang:2012} (Supplementary Information \S1.4).
  In the simulations, we drew 100 different direction vectors $\bfu_k$ from a von Mises distribution (a circular analogue of the Normal distribution), and simulated 500 SNPs using each of these direction vectors. 
\end{itemize}

Figure S2 shows example allele frequency surfaces from these two covariance decay models. For each parameter combination in the above models, we simulated 10 random datasets, and used PCA and our algorithm \localizationAlgorithm{} to infer the spatial coordinates $\bfz_i$. PCA infers the locations up to an orthogonal transformation, while \localizationAlgorithm{} infers these locations up to an invertible linear transformation which is related to the curvature of the allele frequency variance $\eta(\bfzero)$. We use the true geographic locations of a random subset of 20\% of the simulated individuals to rescale the coordinates inferred by PCA and \localizationAlgorithm{}. As a measure of spatial reconstruction accuracy, we use the root mean squared error (RMSE) between the inferred locations $\hat{\bfz}_i$ and the true locations $\bfz_i$ measured as $\sqrt{1/n \sum_{i=1}^n \| \bfz_i - \hat{\bfz}_i \|^2}$.

\begin{figure*}[t]
\begin{center}
\begin{subfigure}[]{0.30\linewidth}
\includegraphics[width=\linewidth]{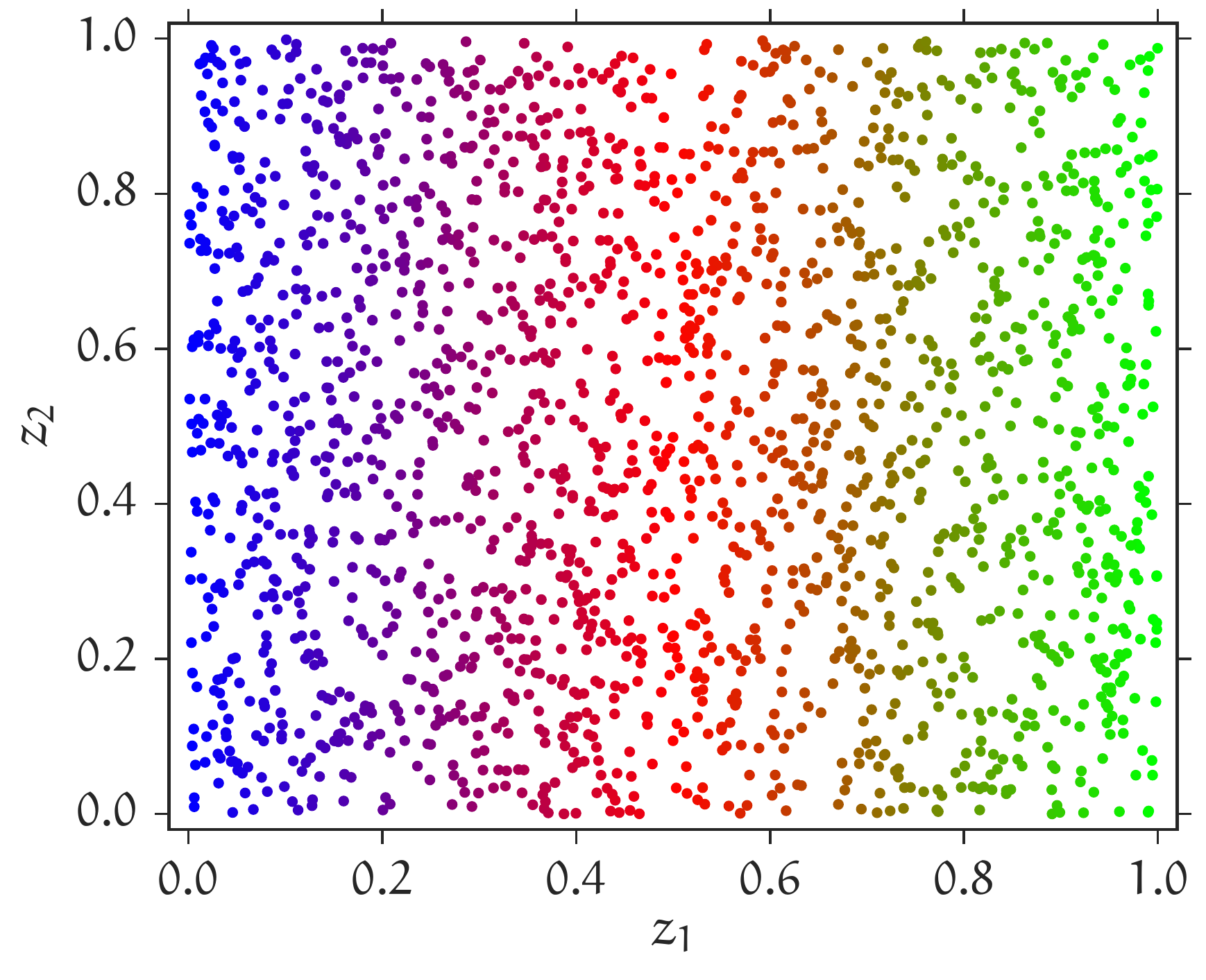}
\caption{}
\label{fig:logisticExpDecayCov_n2000_p50k_beta1_loc}
\end{subfigure}
\begin{subfigure}[]{0.30\linewidth}
\includegraphics[width=\linewidth]{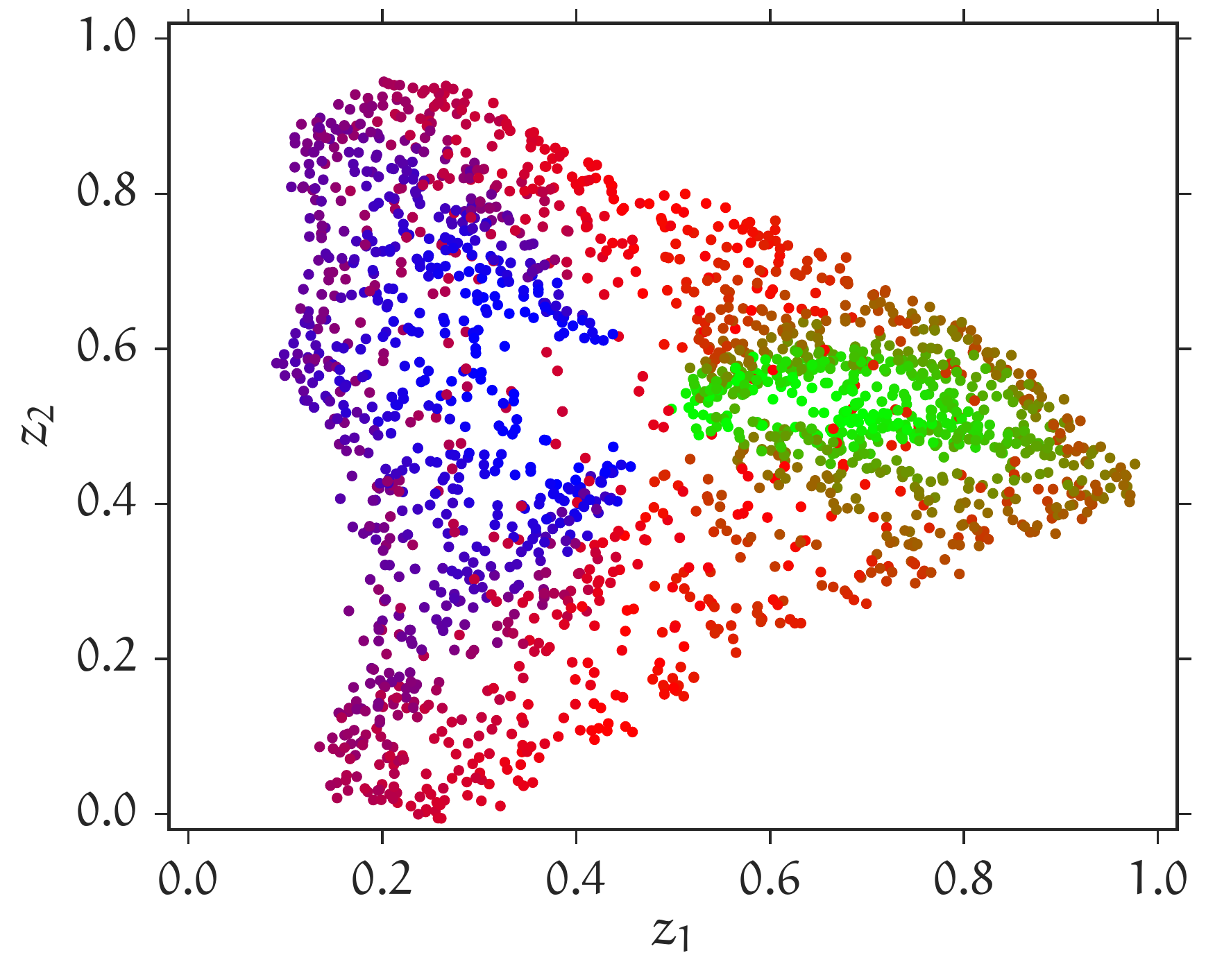}
\caption{}
\label{fig:logisticExpDecayCov_n2000_p50k_beta1_pca}
\end{subfigure}
\begin{subfigure}[]{0.30\linewidth}
\includegraphics[width=\linewidth]{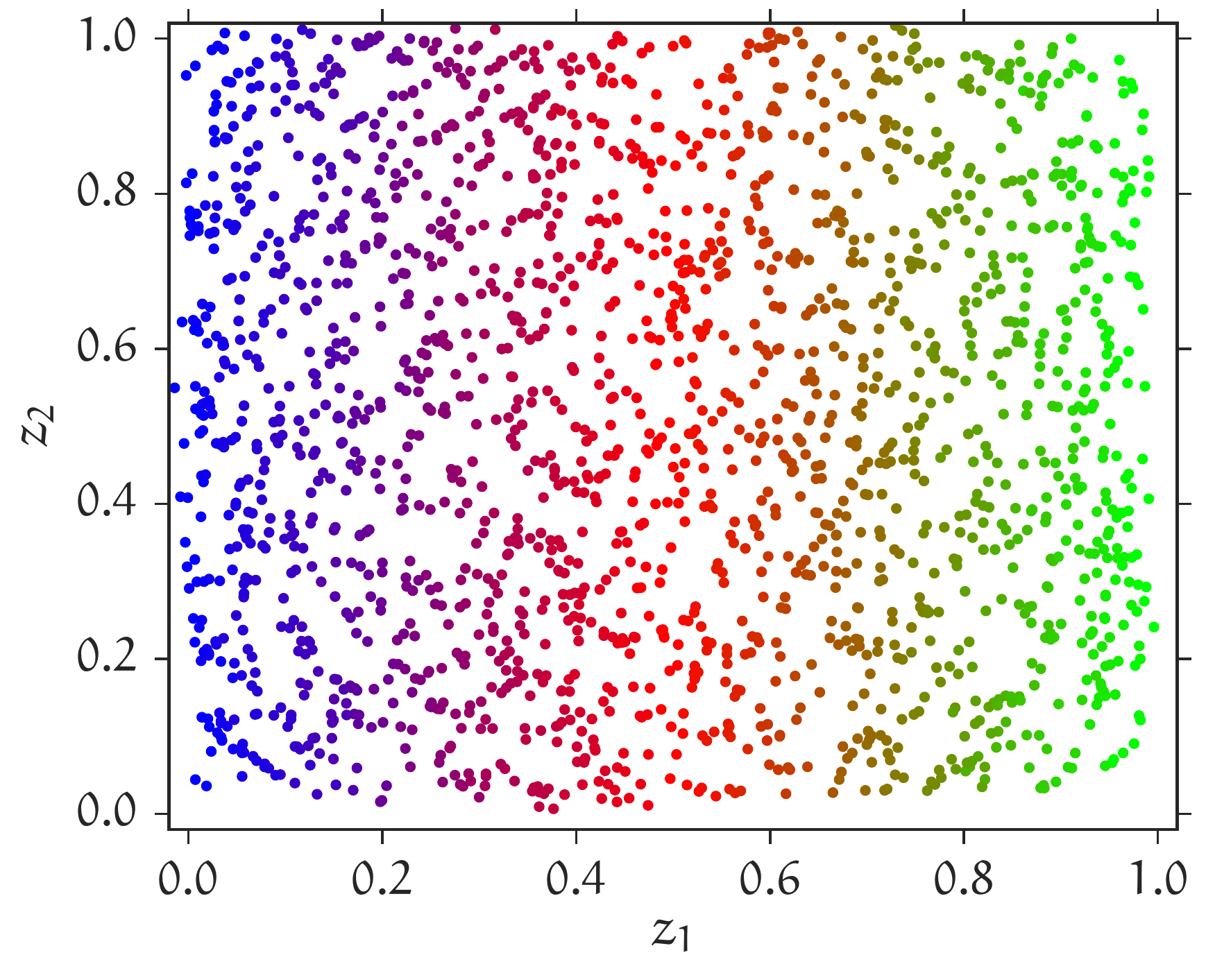}
\caption{}
\label{fig:logisticExpDecayCov_n2000_p50k_beta1_mds}
\end{subfigure}

\vspace{5mm}
\begin{subfigure}[]{0.30\linewidth}
\includegraphics[width=\linewidth]{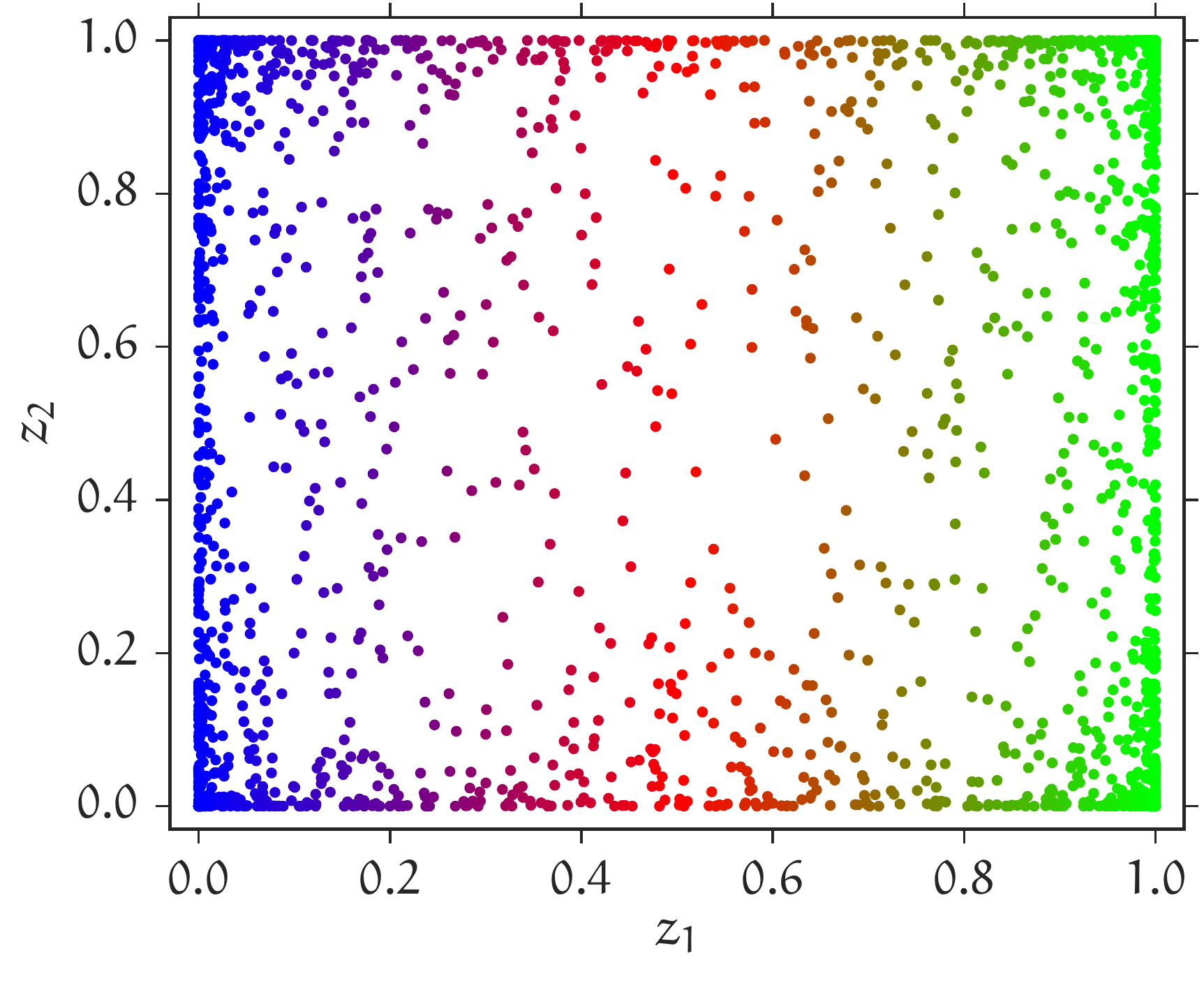}
\caption{}
\label{fig:logisticExpDecayCov_n2000_p50k_beta0.25_loc}
\end{subfigure}
\begin{subfigure}[]{0.30\linewidth}
\includegraphics[width=\linewidth]{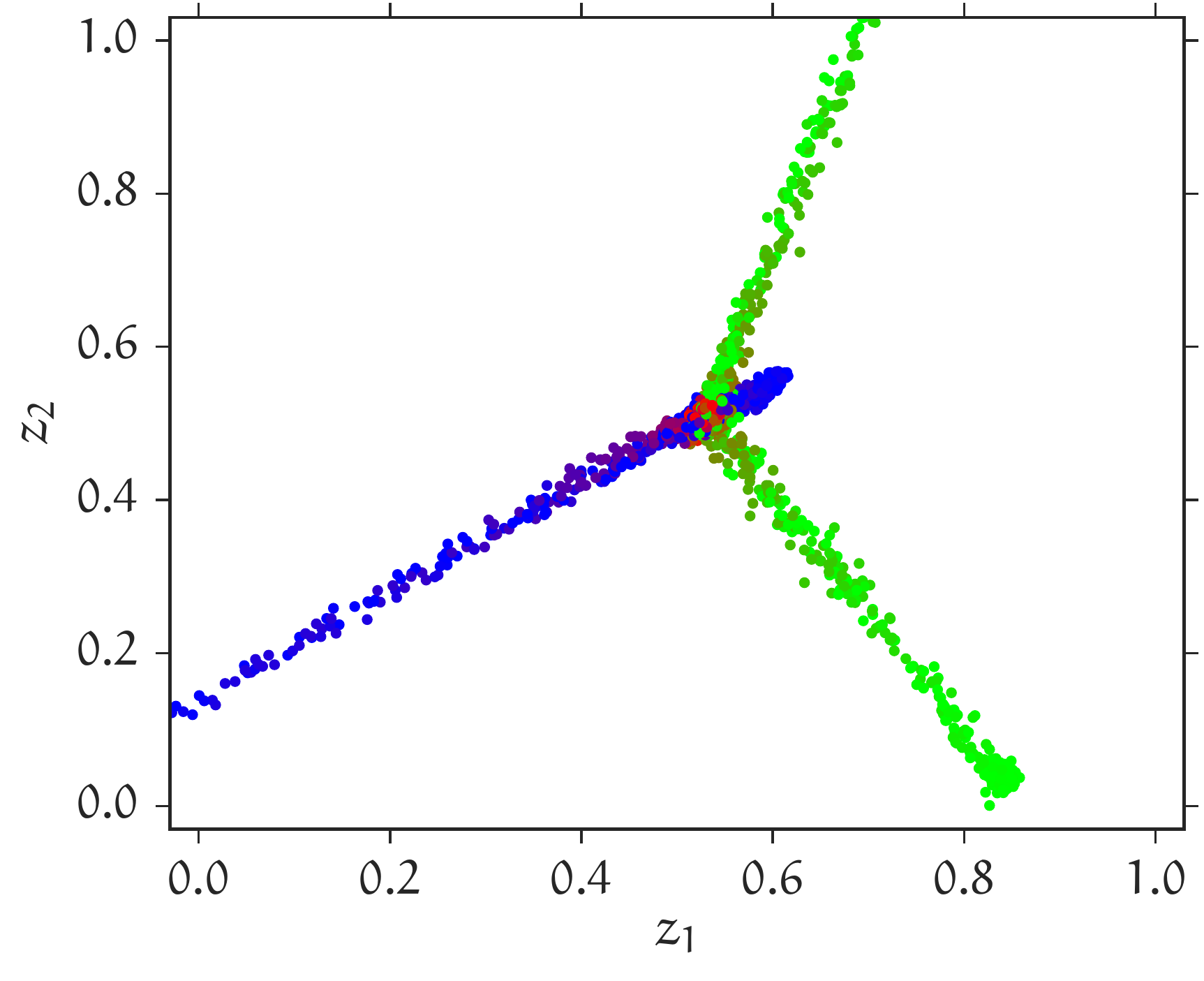}
\caption{}
\label{fig:logisticExpDecayCov_n2000_p50k_beta0.25_pca}
\end{subfigure}
\begin{subfigure}[]{0.30\linewidth}
\includegraphics[width=\linewidth]{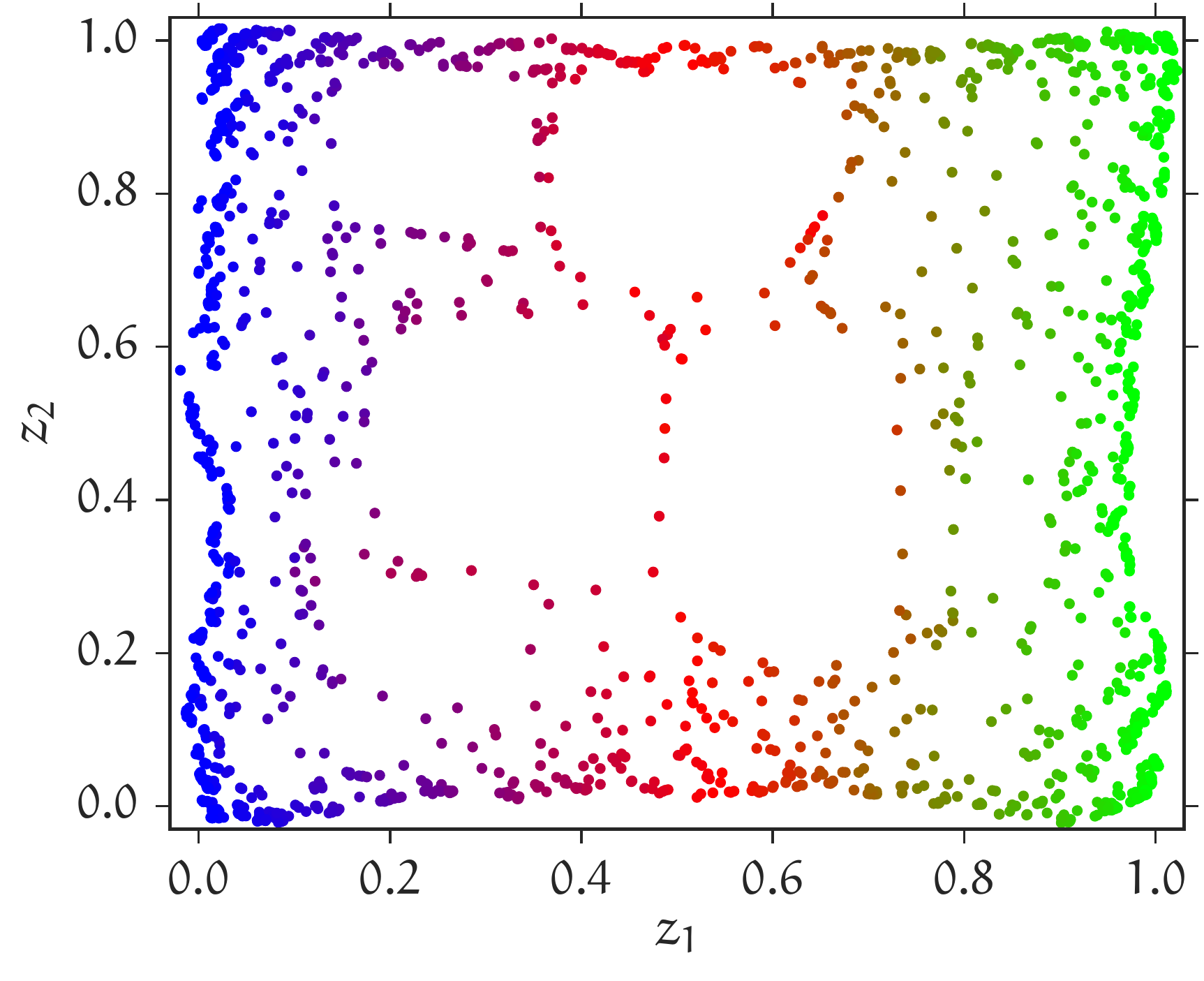}
\caption{}
\label{fig:logisticExpDecayCov_n2000_p50k_beta0.25_mds}
\end{subfigure}
\end{center}
\caption{Simulated datasets with the PCA and \localizationAlgorithm{} reconstructions. The genotype data were simulated using the isotropic covariance decay model, where the $z_1$ and $z_2$ coordinates of each individual were sampled independently and {\sf \subref{fig:logisticExpDecayCov_n2000_p50k_beta1_loc}} uniformly, {\sf \subref{fig:logisticExpDecayCov_n2000_p50k_beta0.25_loc}} according to a Beta(0.25, 0.25) distribution from the unit square. The covariance decay parameters for the simulation are $\alpha_0 = \alpha_2 = 1$ and $\alpha_1=16$.
{\sf \subref{fig:logisticExpDecayCov_n2000_p50k_beta1_loc}} \& {\sf \subref{fig:logisticExpDecayCov_n2000_p50k_beta0.25_loc}} True locations of sampled individuals; {\sf \subref{fig:logisticExpDecayCov_n2000_p50k_beta1_pca}} \& {\sf \subref{fig:logisticExpDecayCov_n2000_p50k_beta0.25_pca}} Reconstructed locations using PCA (RMSE 0.2554 and 0.4390, respectively), {\sf \subref{fig:logisticExpDecayCov_n2000_p50k_beta1_mds}} \& {\sf \subref{fig:logisticExpDecayCov_n2000_p50k_beta0.25_mds}} Reconstructed locations using \localizationAlgorithm{} (RMSE 0.0245 and 0.0293, respectively).
}
\label{fig:logisticExpDecayCov_n2000_p50k}
\end{figure*}

For most parameter combinations in both the covariance decay models, the RMSE of \localizationAlgorithm{} is substantially lower than that of PCA (Tables \ref{tab:logisticExpDecayCov_MDS_PCA_beta1_main_text} and S6). In fact, we prove that under our spatial probabilistic model, \localizationAlgorithm{} performs at least as well as PCA in the asymptotic regime where the sample size $n$ goes to infinity (Supplementary Information \S1.4). For some simulation parameter combinations, the RMSE of PCA is slightly better than the RMSE of \localizationAlgorithm{} by a few percent (Tables \ref{tab:logisticExpDecayCov_MDS_PCA_beta1_main_text} \& S6), which is due to the sample size being finite.
\fref{fig:logisticExpDecayCov_n2000_p50k} illustrates the conceptual difference between \localizationAlgorithm{} and PCA. PCA tries to embed individuals into a two dimensional space which preserves the pairwise genetic distance between all pairs of individuals as estimated from their genotype data. On the other hand, \localizationAlgorithm{} takes a more local approach by using the genotype data from only genetically similar pairs of individuals to estimate their spatial distance. This leads to a qualitatively better low-dimensional embedding.
Simulation results for other parameter settings for these isotropic and directional covariance decay models bear out this intuition (Tables S1--S10).

\begin{table}
\caption{\textbf{Isotropic covariance decay model}}
\label{tab:logisticExpDecayCov_MDS_PCA_beta1_main_text}
\begin{center}
\begin{tabular}{ccccc}
\toprule
\multirow{2}{*}{$\alpha_2$}  & \multirow{2}{*}{$\alpha_1$}  & \multirow{2}{*}{$\displaystyle\frac{\text{RMSE \localizationAlgorithm{}}}{\text{RMSE PCA}}$}	&	\multirow{2}{*}{RMSE PCA} & \multirow{2}{*}{RMSE \localizationAlgorithm{}}	\\[0.4cm]
\midrule
\multirow{5}{*}{0.5}	&	1	&	1.010	&	0.0879	&	0.0888 	\\
			&	2	&	\bf{0.978}	&	0.1030	&	0.1008 	\\
			&	4	&	\bf{0.699}	&	0.1001	&	0.0700 	\\
			&	8	&	\bf{0.414}	&	0.1151	&	0.0477 	\\
			&	16	&	\bf{0.307}	&	0.1372	&	0.0421 	\\
\midrule
\multirow{5}{*}{1}	&	1	&	1.018	&	0.0716	&	0.0729 	\\
			&	2	&	\bf{0.880}	&	0.0929	&	0.0818 	\\
			&	4	&	\bf{0.359}	&	0.1285	&	0.0461 	\\
			&	8	&	\bf{0.094}	&	0.1733	&	0.0163 	\\
			&	16	&	\bf{0.096}	&	0.2554	&	0.0245 	\\
\midrule
\multirow{5}{*}{1.5}	&	1	&	1.028	&	0.0555	&	0.0570 	\\
			&	2	&	\bf{0.857}	&	0.0983	&	0.0843 	\\
			&	4	&	\bf{0.212}	&	0.1647	&	0.0349 	\\
			&	8	&	\bf{0.100}	&	0.2842	&	0.0285 	\\
			&	16	&	\bf{0.100}	&	0.3114	&	0.0311 	\\
\bottomrule
\end{tabular}
\end{center}
Comparison of the localization accuracy of \localizationAlgorithm{} and PCA in the isotropic model simulation setup described in {\sf Simulations}, with parameters $\beta = \alpha_0 = 1$.
\end{table}

\subsection{Analysis of diverse human population datasets}
We applied \localizationAlgorithm{} to three public genotype datasets that have been previously analyzed in studies of population structure --- (a) the \HO{} dataset containing 198 diverse populations that has been used to analyze ancient admixture \citep{lazaridis:2014}, (b) the \GLOBETROTTER{} dataset of 95 populations \citep{hellenthal:2014}, and (c) the Population Reference Sample (POPRES) dataset \citep{nelson:2008} (Supplementary Information \S1.6).

\smallskip

\noindent $\bullet$ {\HO{}: The publicly available release \citep{lazaridis:2014} contains 1{,}945 individuals genotyped at 600{,}841 SNPs. We used a subset of 863 individuals from 91 diverse populations from North Africa and Western Eurasia in order to have fairly uniform sampling over the relevant geographic region. We considered autosomal SNPs which were filtered using {\sf plink} for deviation from Hardy-Weinberg equilibrium, and also excluded SNPs in linkage disequilibrium by pruning SNP pairs which had a pairwise genotypic linkage disequilibrium $r^2$ of greater than 10\% within sliding windows of 50 SNPs (with a 5-SNP increment between windows). This left us with a set of $127,922$ SNPs. PCA applied to this dataset produced a visually poor separation of the populations in Eastern Europe and Western Asia (\fref{fig:HO}(c)). However, population structure is better discerned using our localization algorithm \localizationAlgorithm{}, which also shows a strong correlation between the true sampling locations and the inferred population locations (\fref{fig:HO}(b)). 
This pulling together of individuals from geographically disparate regions by PCA is consistent with our observations in simulated data (\fref{fig:logisticExpDecayCov_n2000_p50k}), where we see that the genetic correlation between distant samples is not as informative about spatial ancestry as that between spatially proximate samples. On the other hand, our approach of using local genetic distances alleviates this issue and better preserves the separation between geographically dispersed populations.

\begin{figure*}
\begin{center}
\begin{subfigure}[]{0.45\linewidth}
\includegraphics[width=\linewidth]{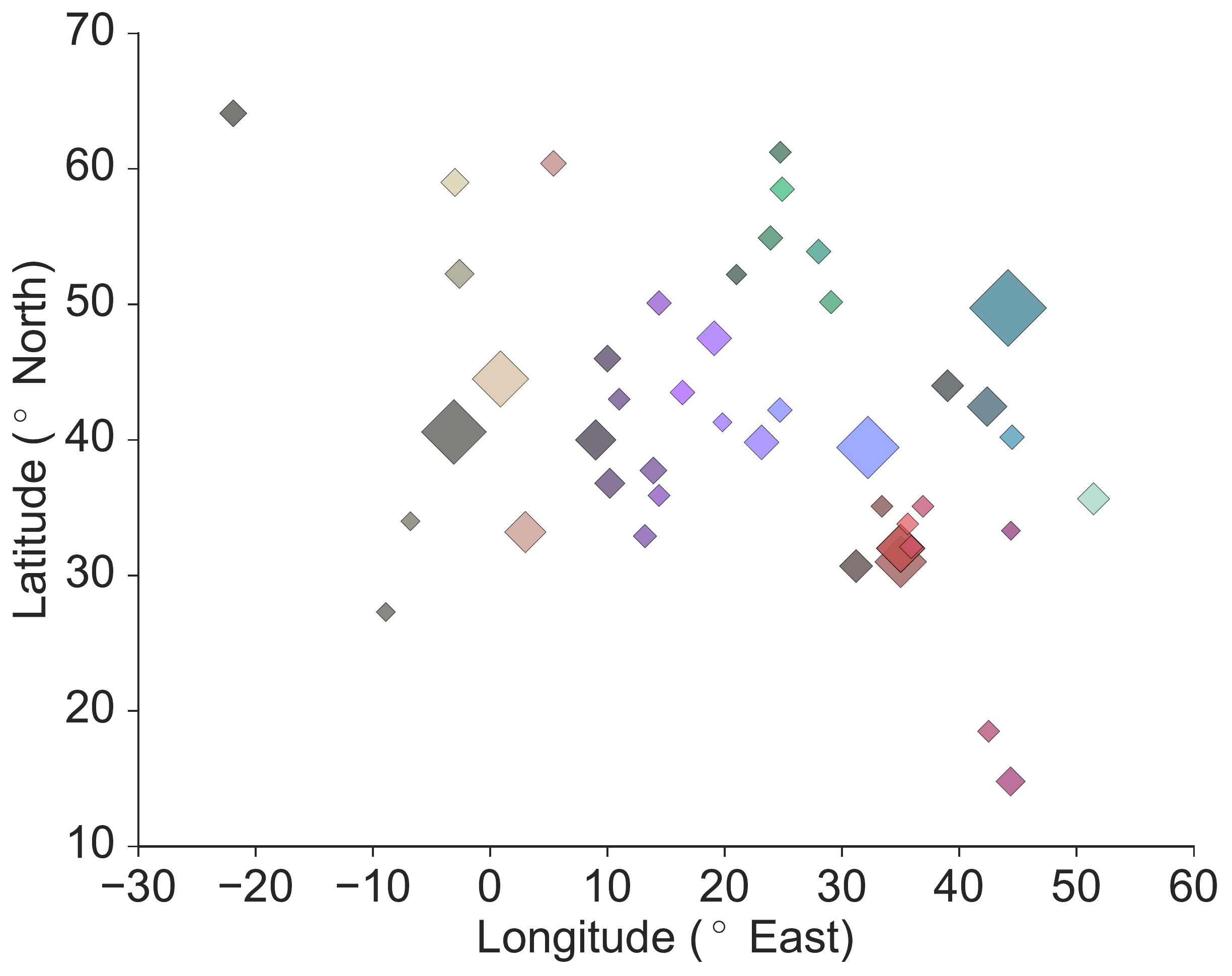}
\caption{}
\label{fig:HO_true_locations}
\end{subfigure}
\begin{subfigure}[]{0.45\linewidth}
\includegraphics[width=\linewidth]{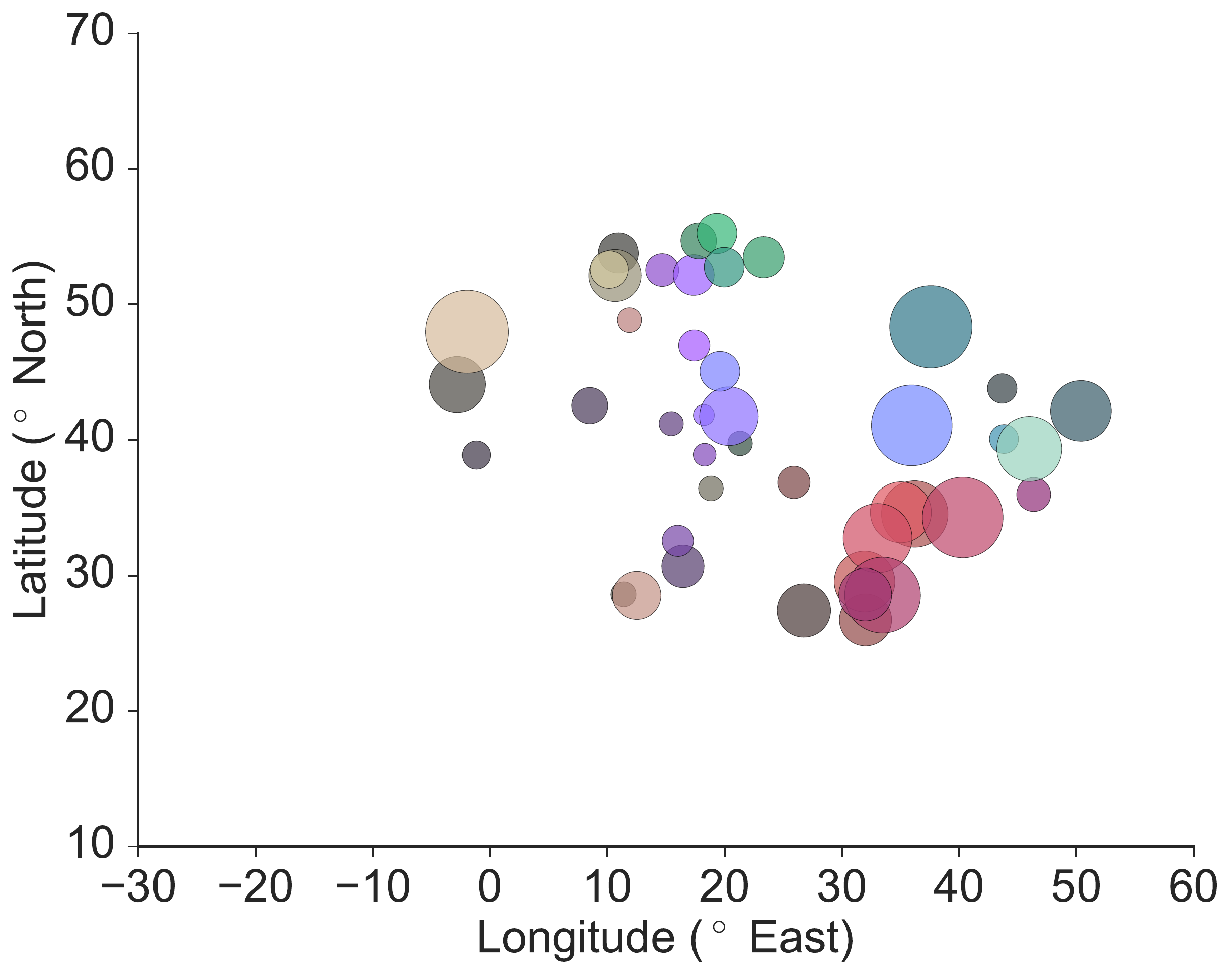}
\caption{}
\label{fig:HO_GAP}
\end{subfigure}

\begin{subfigure}[]{0.45\linewidth}
\includegraphics[width=\linewidth]{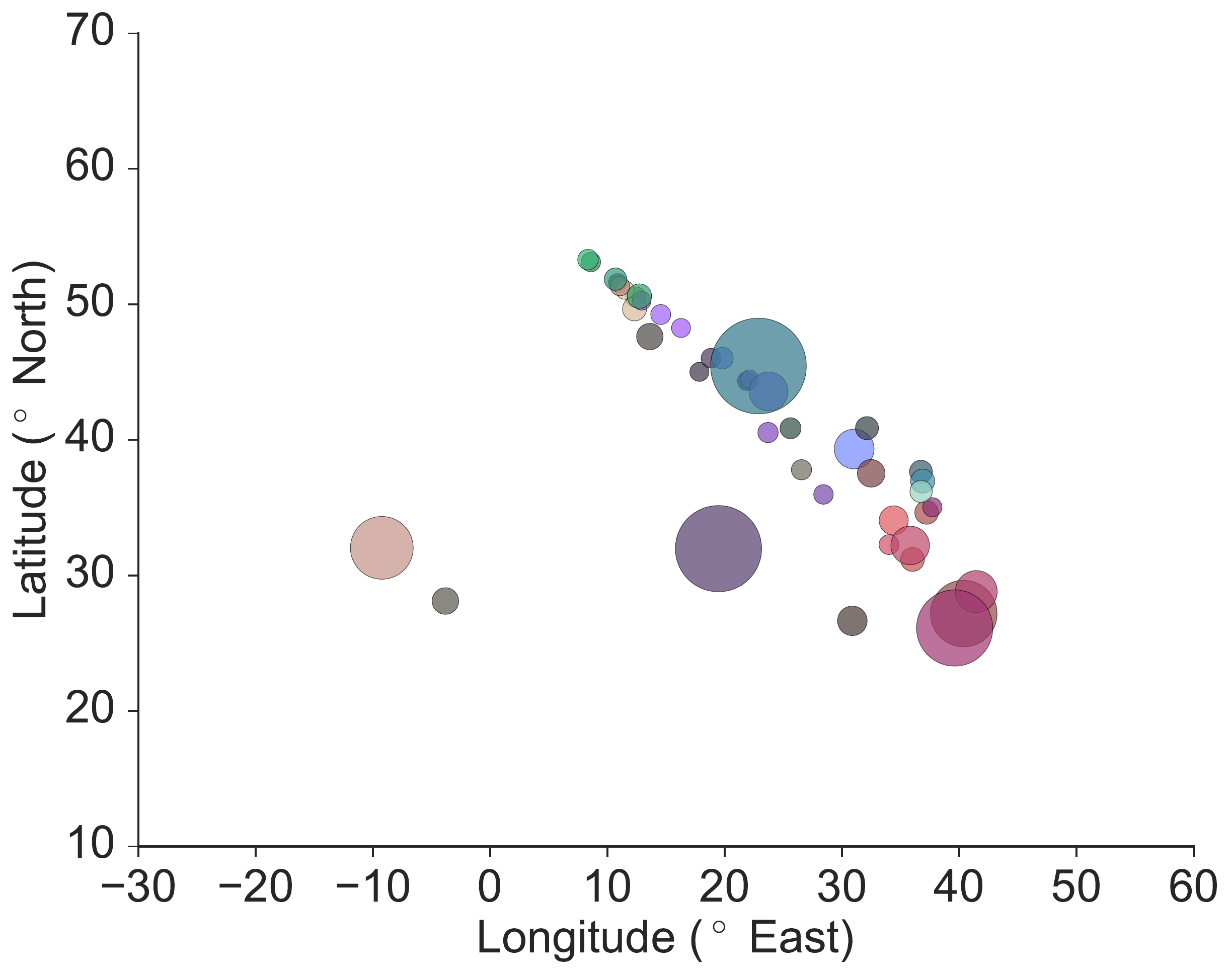}
\caption{}
\label{fig:HO_PCA}
\end{subfigure}
\begin{subfigure}[]{0.45\linewidth}
\includegraphics[width=\linewidth, trim=2in 3in 2in 4in, clip]{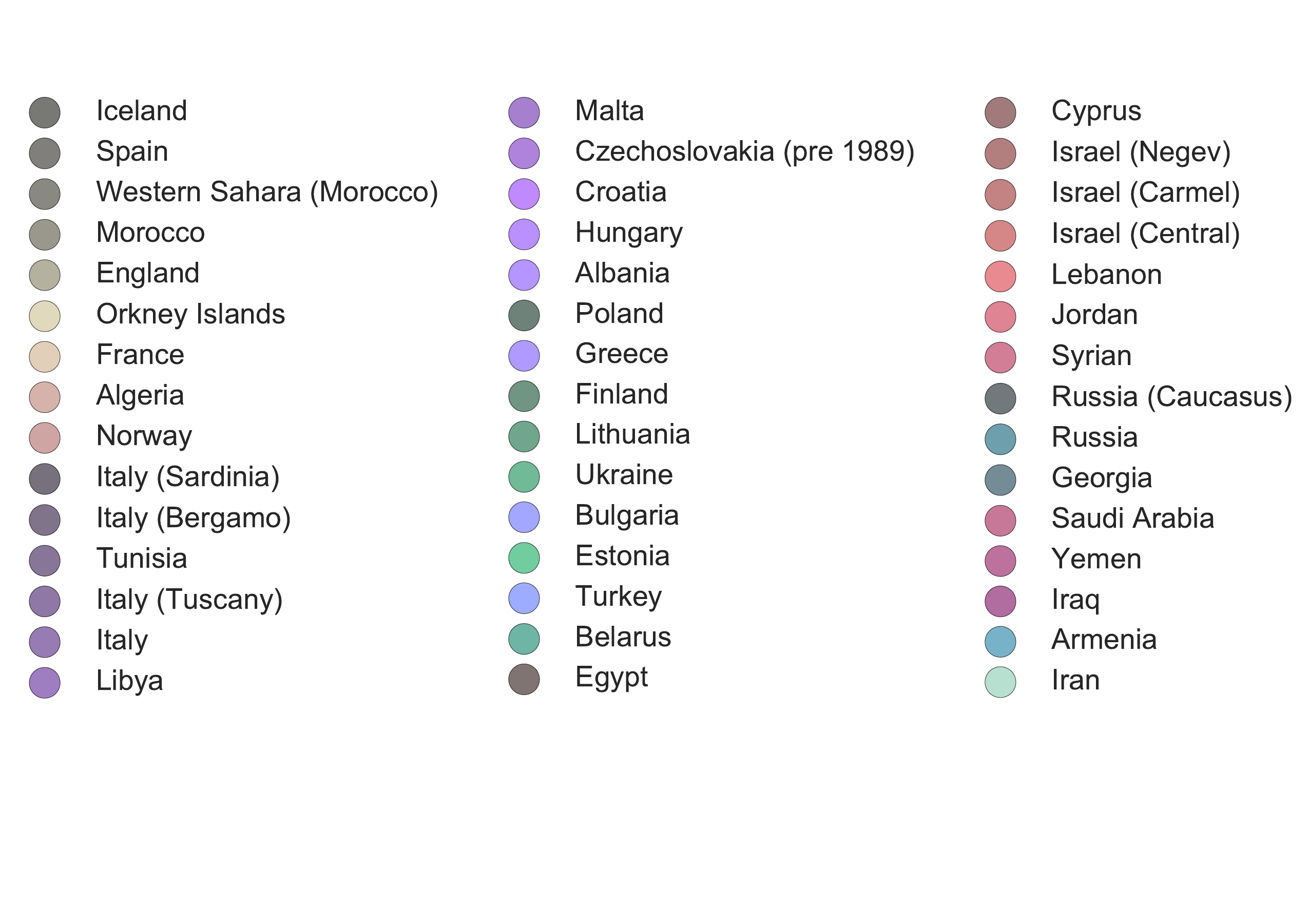}
\caption{}
\label{fig:HO_legends}
\end{subfigure}
\end{center}

\caption{PCA and \localizationAlgorithm{} visualization of the North African and West Eurasian samples in the \HO{} dataset. Each data point corresponds to the sampling location of a population. \subref{fig:HO_true_locations} True sampling locations, 
\subref{fig:HO_GAP} \localizationAlgorithm{} reconstructed locations, 
\subref{fig:HO_PCA} PCA reconstructed locations, 
and \subref{fig:HO_legends} population legends. The areas of the circles are proportional to the estimated variance in the reconstructed locations of the samples in each subpopulation, while the areas of the diamonds are proportional to the number of sampled individuals from the population.
}
\label{fig:HO}
\end{figure*}

\smallskip
\noindent $\bullet$ \GLOBETROTTER{}: This dataset contains 486{,}669 SNPs from 1{,}530 individuals from 95 diverse human populations. We considered the subset of individuals sampled from Europe, the Middle East, North and East Africa, and Western, Central and South Asia in order to have a fairly uniform sampling across geography. 
We filtered SNPs in linkage disequilibrium and violating Hardy-Weinberg equilibrium in the same manner as for the \HO{} dataset, resulting in a final dataset of 71{,}736 SNPs from 1{,}046 individuals from 59 subpopulations.
Applying PCA and our localization algorithm \localizationAlgorithm{} to this dataset (Figure S3), we noticed a similar pattern as in the \HO{} dataset, where PCA pulls together individuals from Southern Europe, North Africa, and the Middle East into a cluster, much more so than \localizationAlgorithm{}.

\subsection{Spatial reconstruction accuracy}
The spatial coordinates assigned by \localizationAlgorithm{} are effective at visually discerning population structure (Figures \ref{fig:HO}, S3 \& S4), even where PCA has difficulty distinguishing them. We also evaluated the performance of \localizationAlgorithm{} in assigning spatial coordinates to new samples given access to the sampling locations of a subset of individuals.
We used a random subset of 20\% of the subpopulations in each dataset to transform the PC coordinates from PCA and the coordinates inferred by \localizationAlgorithm{} into latitude-longitude coordinates (Supplementary Information \S1.6). Since there is substantial variability in the reconstruction error depending on the subset that is used for rescaling coordinates, we used 100 random training data subsets and computed the reconstruction RMSE on each of them. We find that \localizationAlgorithm{} exhibits 31\% lower error for the \HO{} dataset than PCA, and 10\% lower error for the \GLOBETROTTER{} dataset (Table S11).

\subsection{Application to correcting ancestry confounding in GWAS}
Consider the following prospective model for a quantitative phenotype $y$,
\begin{align}
  y_i = \alpha + \sum_{\ell=1}^{p} \beta_\ell x_{i\ell} + \lambda_i + \varepsilon_i, \label{eq:forward_model_quantitative}
\end{align}
where $\alpha$ is an intercept term, $\beta_\ell$ is the effect size of SNP $\ell$, and $\lambda_i$ and $\varepsilon_i$ are the environmental and noise contributions respectively. 
The linear model in \eqref{eq:forward_model_quantitative} can also be adapted to binary phenotypes using the following generalized linear model,
\begin{align}
  y_i \sim \Bino(2, \logit^{-1}(\alpha + \sum_{\ell=1}^{p} \beta_\ell x_{i\ell} + \lambda_i)). \label{eq:forward_model_binary} %
\end{align}
Population structure can induce correlations between the genotypes at different SNPs $\ell$ and $\ell'$, and also between the genotypes and environmental contribution $\lambda$. Unaccounted structure can thus lead to spurious genotype-phenotype associations \citep{campbell:2005}.

PCA-correction \citep{price:2006} and linear mixed models (LMM) \citep{kang:2010} are popular approaches for dealing with ancestral confounding which use the above prospective models for testing if $\beta_{\ell} = 0$. Song \emph{et al.} \citep{song:2015} showed that testing $\beta_{\ell} = 0$ in \eqref{eq:forward_model_quantitative} or \eqref{eq:forward_model_binary} is equivalent to testing $R_\ell = 1$ in the following inverse regression model, 
\begin{align}
\label{eq:inverse_reqression}
\begin{split}
& x_{i\ell} \mid y_i,\bfz_i \sim \Bino(2,\theta_{i\ell}) \\
& \theta_{i\ell} = \frac{\kappa_\ell R_\ell^{y_i}q_\ell(\bfz_i)}{1-q_\ell(\bfz_i) + \kappa_\ell R_\ell^{y_i} q_{\ell}(\bfz_i)}.
\end{split}
\end{align}
In \eqref{eq:inverse_reqression}, $R_\ell$ is the genetic risk factor of the alternate allele at SNP $\ell$ and $\kappa_\ell$ is an intercept term that absorbs the effects of the other SNPs. The retrospective model in \eqref{eq:inverse_reqression} accounts for population structure by testing the distribution of the genotype conditional on the ancestry-dependent allele frequency function $q_\ell(\bfz)$. Our association testing procedure, {\sf Stratification Correction via \localizationAlgorithm{}} (\associationTestingAlgorithm{}), also operates in the retrospective model of \eqref{eq:inverse_reqression}. We first estimate the ancestry coordinates $\hat{\bfz}_i$ for each individual $i$ in the sample using our localization algorithm \localizationAlgorithm{}. We then estimate $\kappa_\ell$ under the null hypothesis (while setting $R_{\ell} = 1$), and estimate $R_{\ell}$ and $\kappa_{\ell}$ under the alternate hypothesis for each SNP as follows: \\
(1) Estimate the spatial allele frequency function $q_{\ell}(\bfz)$ for each SNP $\ell$ by assuming that the allele frequencies vary smoothly over space. To this end, we use the squared exponential kernel,
\begin{align}
\hat{q}_{\ell}(\hat{\bfz}_i) \propto \sum_{j=1}^n \frac{x_{j\ell}}{2} \exp\left(-\frac{1}{2} \|H^{-1}(\hat{\bfz}_i-\hat{\bfz}_j)\|^2\right) \,. \label{eq:allele_frequency_smoothing}
\end{align}
The two-dimensional kernel bandwidth matrix $H$ in \eqref{eq:allele_frequency_smoothing} is chosen using Scott's rule \citep{scott:1979}. \\
(2) Estimate the genetic risk factor $R_\ell$ and intercept term $\kappa_\ell$ using Newton's method (Supplementary Information \S2).

In principle, one could perform the above two steps using the coordinates $\hat{\bfz}_i$ inferred from some other algorithm, and we thus also compare performance using the (unknown) true ancestry coordinates $\bfz_i$ and using the coordinates inferred by PCA. We refer to the work of \citet{song:2015} for a comparison of the retrospective model association test common to their method \GCAT{} and our work, with the prospective model association tests performed by the PCA-correction and LMM approaches.

\subsubsection{Simulations}
We simulated genotype data for $n=2{,}000$ individuals at $p=50{,}000$ SNPs using the isotropic and directional allele frequency covariance decay models described earlier. We generated phenotype data using a linear model with genotypic, ancestry-dependent, and random environmental effects, with 20\%, 10\%, and 70\% contributions respectively to the phenotypic variance, and with 10 SNPs randomly chosen to have non-zero effects drawn from a standard Normal distribution.
In our simulations, we inferred the two-dimensional ancestral coordinates using \localizationAlgorithm{} and PCA, and also compared them against an oracle which has access to the true ancestry coordinates. Since our hypothesis test conditional on estimated allele frequencies operates in the same retrospective model as the \GCAT{} method \mycitet{song:2015}, we also compare our results using \GCAT{} with 2 and 6 latent factors. 
For most parameter combinations, \associationTestingAlgorithm{} has higher power than PCA or \GCAT{} (Figures \ref{fig:ROCCurveExamples} \& \S5--S10 and Tables S12--S13), and has similar power as the oracle procedure that uses the true ancestry coordinates in our association test.

\begin{figure}
\begin{center}
\includegraphics[width=0.6\linewidth]{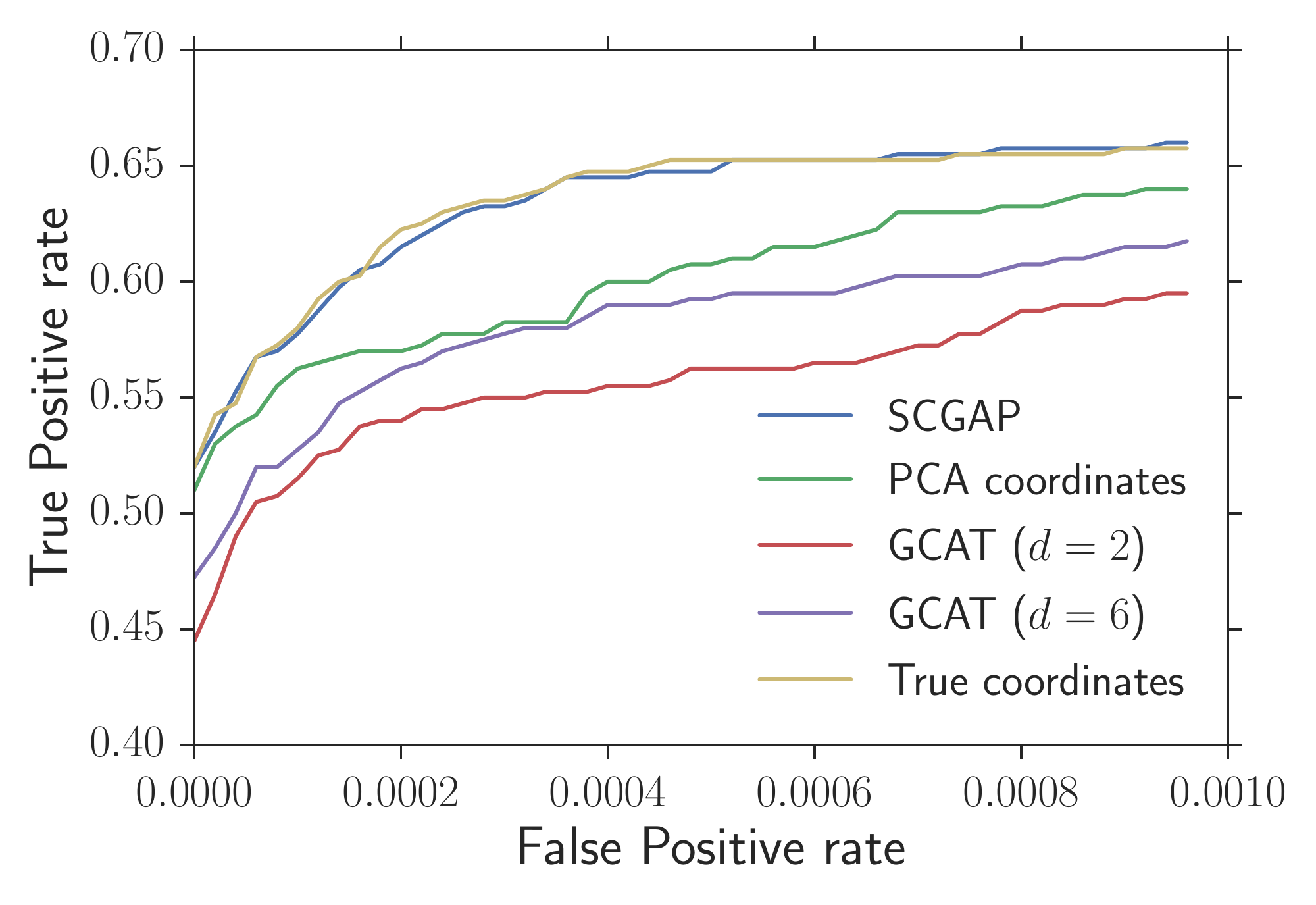}
\end{center}
\vspace{-5mm}
\caption{ROC curves for our stratification correction procedure with ancestral coordinates inferred using \localizationAlgorithm{} (\associationTestingAlgorithm{}), PCA, or using the true coordinates. We also compared our results with the \GCAT{} method, which uses a latent factor model (indexed by $d$) to estimate the allele frequencies for each individual at each locus.
Genotypes were drawn according to the isotropic covariance decay model with $\alpha_0 = \alpha_2 = 1$ and $\alpha_1 = 16$ (same simulation parameters as \fref{fig:logisticExpDecayCov_n2000_p50k_beta1_loc}).
}
\label{fig:ROCCurveExamples}
\end{figure}

\subsubsection{Analysis of Northern Finland Birth Cohorts dataset}
We analyzed a dataset of 10 quantitative metabolic traits from 364{,}590 SNPs of 5{,}402 individuals of a birth cohort from Northern Finland (\NFBC{}) \citep{sabatti:2009}. 
We filtered individuals and SNPs using the same criteria described by \cite{song:2015}, and were left with 335{,}143 SNPs and 5{,}246 individuals. We added features for known confounders such as sex, oral contraceptive use, pregnancy status, and fasting status according to the same procedure described in the first analysis of this dataset by \cite{sabatti:2009}, and performed a Box-Cox transform on the median 95\% of trait values to make the distribution of traits as close to a normal distribution as possible\footnote{We also performed the association test on the untransformed values for the C-reactive protein and Triglyceride level traits, since these traits appear exponentially distributed and the equivalence of the retrospective and prospective models that we rely on also holds for exponentially distributed traits \citep{song:2015}.}. After applying genomic control \citep{devlin:1999} to correct for inflation of the log-likelihood ratios from our association test (Figure S11), we identified $17$ significant loci (Table S14, S15), $16$ of which were also reported by \GCAT{}, at a significance threshold of $p<7.2 \times 10^{-8}$ that has been used in previous works on this data. Other association tests that operate in the prospective model identify between 11 and 14 loci \citep{kang:2010}. Out of these 17 loci identified by \associationTestingAlgorithm{}, 15 have been identified in independent association studies on different samples (Table S16).

\section{Discussion}
In this paper, we developed a novel spatial probabilistic model of allele frequency evolution that avoids imposing any explicit parametric form for the dependence of allele frequencies on geographic location. The flexibility of our model allows us to generalize several popular parametric models of allele frequency evolution. Based on our model, we develop an ancestry localization algorithm \localizationAlgorithm{} that generalizes parameter-free dimensionality reduction approaches such as PCA, and that we prove performs at least as well as PCA for large sample sizes. 
Our algorithm, which can be viewed as a form of manifold learning, also adds to the rich literature on theoretical population genetic models \citep{mcvean:2009,patterson:2006,paschou:2007} that can motivate the application of PCA for detecting population structure from genotype data. Our algorithm is also very efficient: for any candidate distance threshold used for estimating local spatial distances from local genetic distances, our algorithm has computational complexity $O(n^2 p)$ (assuming $p \gg n$), which is the cost of computing the inner product matrix from a genotype matrix with $n$ individuals and $p$ SNPs. This is also the same computational complexity required by dimension-reduction methods such as PCA.

Our spatial probabilistic model can also be extended to incorporate other demographic and evolutionary forces, and we leave these for future work.
One can potentially infer the spatial origin of the ancestors of admixed samples by applying our method to the output of local-ancestry inference algorithms \citep{price:2009}. Admixture might also be directly incorporated into our spatial probabilistic model by jointly inferring admixture proportions along with the spatial covariance function, similar in approach to the \SpaceMix{} \citep{bradburd:2016} and \SPA{} \citep{yang:2012} models.

Our probabilistic model assumes that the decay in allele frequency covariance around a given spatial location does not depend on the location itself. This is an idealized model of isolation by distance where there are no sharp geographic or genetic barriers to random mating between spatially proximate populations. We leave the extension of our model to handle such spatial heterogeneity for future work.

We also developed an association testing procedure, \associationTestingAlgorithm{}, for genotype-trait association that uses the ancestral coordinates inferred by \localizationAlgorithm{} and an exponential kernel to estimate smooth allele frequency functions for every SNP. Our association test is based on a retrospective model that tests the distribution of the genotype conditional on the phenotype and the estimated allele frequency function \citep{song:2015}. We find that using the ancestry coordinates inferred by \localizationAlgorithm{} in our association test performs almost as well as knowing the true spatial ancestry coordinates. Moreover, for simulated datasets, our association test exhibits slightly better performance than the \GCAT{} test that uses a different allele frequency estimation procedure \citep{hao:2016}. On the \NFBC{} dataset, our method recovers the same set of associations as \GCAT{}. However, our procedure can control for ancestry confounding using just two ancestry components, as opposed to \GCAT{} which was used with six latent factors and an intercept, in both the simulations and for the \NFBC{} dataset. Our maximum likelihood estimation for the hypothesis test at each SNP is also very efficient, employing the quadratically converging Newton method to estimate the intercept and genetic risk factors at each SNP.

\section*{Acknowledgements}
We are grateful to the Simons Institute at UC Berkeley, where part of this work was completed during the Information Theory program. We thank George Busby for sharing the \GLOBETROTTER{} dataset. A.J. was partially supported by a CSoI fellowship during the course of this work (NSF Grant CCF-0939370). A.B. was supported in part by NIH grant HG008140 to Jonathan K Pritchard and a Stanford CEHG fellowship. \\
\textit{Data:} The \POPRES{} and \NFBC{} datasets were obtained from dbGaP (Study Accession phs000145.v4.p2 and phs000276.v2.p1, respectively). The \HO{} dataset was obtained from the Reich lab\footnote{\url{https://genetics.med.harvard.edu/reich/Reich_Lab/Datasets_files/EuropeFullyPublic.tar.gz}}, and the \GLOBETROTTER{} dataset was made available by George Busby.

\newpage
\setcounter{section}{0}
\renewcommand{\thesection}{\S\the\value{section}}
\setcounter{figure}{0}
\renewcommand{\thefigure}{S\the\value{figure}}
\setcounter{table}{0}
\renewcommand{\thetable}{S\the\value{table}}

\begin{center}
{\LARGE Supplementary Information: \\ Novel probabilistic models of spatial genetic ancestry with applications to stratification correction in genome-wide\\ association studies \par}
\end{center}

\vspace{5mm}

\section{Probabilistic model and localization algorithm}\label{sec:loc_model}

Suppose that we are given genotypes from $n$ individuals at $p$ SNPs distributed across the geographic region under study.
We denote by $x_{i\ell}\in\{0,1,2\}$ the observed number of alleles at SNP $\ell$ in individual $i$ for $\ell=1,2,\dotsc, p$ and $i=1,2,\dotsc, n$.
Further, let $X$ be the $n\times p$ genotype matrix, where the $(i,\ell)$ entry is $x_{i\ell}$. 

In order to capture the spatial structure of the genotype matrix, we let $\bfz_i$ be the geographical location of individual $i$ and for each SNP $\ell$, we view allele frequency $q_\ell$ as a function of location $\bfz_i$, i.e., $q_\ell(\bfz_i)$. Note that $\bfz_i$ are unobserved ancestry coordinates with the implicit assumption that random mating and localized migration has been occuring between proximate locations. %

We define a general flexible probabilistic model of allele frequencies that generalize several previously developed parametric models of spatial genetic variation such as \SPA{} \citep{yang:2012}, \SCAT{} \citep{wasser:2004} and \SpaceMix{} \citep{bradburd:2016}. In our model, we consider an arbitrary stochastic process over the geographical region under consideration. The allele frequencies $\{q_\ell\}_{\ell=1}^p$ for different SNPs are independent sample paths drawn from this stochastic process. 

Throughout, we use the shorthand $q_{i\ell} \equiv q_\ell(\bfz_i)$ to represent the allele frequency for SNP $\ell$ 
conditional on $\bfz_i$. Assuming Hardy-Weinberg equilibrium, genotypes are generated by binomial sampling as $$x_{i\ell} \mid q_{i\ell} \sim \Bino(2,q_{i\ell})\,.$$ 

Before presenting our localization algorithm, we provide a brief overview of the PCA method for recovering geographic ancestry. We explain the rationale behind PCA from a perspective that motivates our algorithm and clarifies its superiority over PCA.

\subsection{Why PCA?}
We denote the genotypes for individual $i$ by $x^i = (x_{i,1}, x_{i,2}, \dotsc, x_{i,p})$. This can be viewed as a representation of individuals in the $p$-dimensional space. In this way, the localization task seeks for an embedding of individuals from the $p$-dimensional space into the two-dimensional geographical region.

PCA with two principal components gives the best rank two approximation of the genotype matrix $X$ in the following sense.\footnote{In general, the top $k$ principal components give the best rank $k$ approximation in a similar sense.} Define
\begin{align}\label{eq:PCA1}
X_* = \underset{Y\in \reals^{n\times 2}}{\arg\min}\,\|XX^\sT - YY^\sT\|_F\,,
\end{align}
where for a matrix $A =(a_{i,\ell})$, $\|A\|_F = (\sum_{i,\ell}a_{i,\ell}^2)^{1/2}$ indicates the Frobenius norm.  The solution to \eqref{eq:PCA1} is given by the top singular vectors of $X$. Specifically, let $X= U\Sigma V^T$ be the singular value decomposition of $X$ where $\Sigma = \diag(\sigma_1, \sigma_2,\dotsc,\sigma_n)$ with $\sigma_1\ge \sigma_2 \ge \dotsc\sigma_n$. The solutions to \eqref{eq:PCA1} are given by $X_* = U_2 \Sigma_2 Q_2$, where $U_2$ denotes the first two left singular vectors, $\Sigma_2 = \diag(\sigma_1,\sigma_2)$, and $Q_2$ is an arbitrary $2 \times 2$ orthogonal matrix (i.e. $Q_2 Q_2^T = I$).

We next recall the following identity that for a given set of points relates their pairwise inner products to their pairwise distances.
 Consider the centering matrix $L = I_{n\times n}- \u \u^\sT$, where $I_{n\times n}$ is the identity matrix of size $n$ and 
 $\u = (1/\sqrt{n},\dotsc,1/\sqrt{n})$. For a vector $v$, $Lv$ centers the entries of $v$ by subtracting the mean of the entries of $v$. Further, we denote by $D$ the squared distance matrix $D_{ij} = \|x^i-x^j\|^2$. Using this notation, we have the following,
\begin{align*}
LXX^\sT L = -\frac{1}{2} LDL\,.
\end{align*}
A common preprocessing step for PCA is to center each column in the genotype matrix. This centered genotype matrix is precisely $LX$.
It is then straightforward to see that PCA provides a lower dimensional representation of points $\{x^i\}_{i=1}^n$ such that
their squared distance matrix $D_*$ solves the following optimization problem,
\begin{align}\label{eq:PCA2}
D_* = \underset{\tilde{D}\in \mathcal{D}}{\arg\min}\,\|LDL - L\tilde{D}L\|_F\,.
\end{align}
Here $\mathcal{D}$ is the set of squared distance matrices for all possible two-dimensional embeddings of the points $\{x^i\}_{i=1}^n$. In other words, PCA seeks a low-dimensional representation of the $n$ points that best approximates \emph{all} pairwise distances.

However, an important question that is unanswered is the following:
\begin{quote}
\emph{How are the spatial distances between individuals reflected in their genotype information?}
\end{quote}

The PCA approach merely assumes that for any two individuals, their genotype distance is a good approximation of their spatial distance and hence it returns the embedding of individuals on the map that best preserves all pairwise genotype distances. However, a more profound answer to the above question requires a model that relates genetic distances to spatial distances. The PCA approach to ancestry localization lacks such a model. In the following, we use our proposed probabilistic model for the allele frequencies to answer the above question. 

According to our model, the spatial allele frequencies at each SNP $\ell$ come from some spatial stochastic process. 
Our model posits that the underlying spatial processes are second-order stationary, in the sense that for each SNP $\ell$, $\E(q_\ell(\bfz)) = \mu_\ell$ for all locations $\bfz$, and the allele frequency covariance functions ${\rm Cov}(q_\ell(\bfz),q_\ell(\bfz'))$ depend solely on $\bfz-\bfz'$ as follows,
\begin{eqnarray}
{\rm Cov}(q_\ell(\bfz),q_\ell(\bfz')) = \E[(q_\ell(\bfz)-\mu_\ell)(q_\ell(\bfz') - \mu_\ell)] := \eta(\bfz-\bfz')\,.
\end{eqnarray}
Note that while the processes for different SNPs can have different means $\mu_\ell$, they share the same covariance function $\eta(\cdot)$. 
The implicit structure imposed by the second-order stationarity are used by our localization algorithm \localizationAlgorithm{}, which consists of three main steps:
\begin{enumerate}
\item[(1)] Construct consistent estimators of $\eta(\cdot)$ and $\mu(\cdot)$ using the genotype information from $p$ SNPs ($p \gg n$). 
\item[(2)] Use $\eta(\cdot)$ and $\mu(\cdot)$ functions to approximate \emph{local} spatial distances between the individuals. 
\item[(3)] Find a \emph{global} embedding of individuals on the geographical map that respects the estimated \emph{local} distances.
\end{enumerate}

\subsection{\localizationAlgorithm{} algorithm}\label{sec:localization_algorithm}
In the following, we discuss the details of each step.

\bigskip
\noindent{\bf Step $(1)$: Estimating mean and autocorrelation.}
Estimates for $\eta(\cdot)$ and $\mu(\cdot)$ functions are given by Theorem \ref{thm:mu-eta} below.
\begin{thm}\label{thm:mu-eta}
Consider the proposed probabilistic model for the allele frequency functions and define the following quantities:
\begin{eqnarray}
\hmu_\ell  &=& \frac{1}{2n} \sum_{i=1}^n x_{i\ell}\,,\label{eq:mu-est}\\
\heta_{i,j} &=& \frac{1}{p}\sum_{\ell=1}^p \left(\frac{x_{i\ell}}{2}-\hmu_\ell \right) \left(\frac{x_{j\ell}}{2}-\hmu_\ell \right)\,,\label{eq:eta-est}\\
\heta_0 &=& \frac{1}{p} \sum_{\ell=1}^p\left(\frac{1}{2n}\sum_{i=1}^n(x_{i\ell}^2-x_{i\ell}) -\hmu_\ell^2 \right)\,.\label{eq:eta0-est} 
\end{eqnarray}
Let ${\bf 1} = (1/\sqrt{n}, \dotsc, 1/\sqrt{n})^\sT$ and let $K\in \reals^{n \times n}$ with $K_{ij} = \eta(\bfz_i - \bfz_j)$.
Further, set $\kappa:= {\bf 1}^\sT K {\bf 1}$.
Then with probability at least  $1-(n+1)^{-2}$, the following statements are true:
\begin{eqnarray}
|\heta_{i,j} - \eta(\bfz_i-\bfz_j)| &\le& 5\sqrt{\frac{2\log(n+1)}{p}} + 16 \sqrt{\frac{\kappa}{n}} + \frac{8}{n}\,,\quad \quad \forall 1\le i\neq j\le n\,,\label{eq:eta-app}\\
|\heta_0 - \eta(0)| &\le& 5\sqrt{\frac{2\log (n+1)}{p}} + 16 \sqrt{\frac{\kappa}{n}} + \frac{8}{n}\,.\label{eq:eta0-app}
\end{eqnarray}
\end{thm}
\begin{remark}
The estimates in \eqref{eq:eta-est} and \eqref{eq:eta0-est} are consistent, i.e., as the number of individuals $n$ increases indefinitely, the resulting sequence of estimates converges in probability to the quantities of interest, provided that $\log(n+1)/p \to 0$ and $\kappa/n\to 0$. (Note that $\kappa$ is bounded by the spectral radius of $K$.)
\end{remark}

We provide the proof of \thmref{thm:mu-eta} in \sref{sec:proofs}.

\bigskip
\noindent{\bf Step $(2)$: Estimating local spatial distances.}
The next step consists in showing how the local spatial distances can be inferred from functions $\eta(\cdot)$ and $\mu(\cdot)$.
To do so, we write the Taylor expansion of $\eta(\cdot)$ around the origin,
\begin{align}
\eta(\bfz_i-\bfz_j) - \eta(\bfzero) =\nabla\eta(\bfzero)^\sT (\bfz_i-\bfz_j) + \frac{1}{2} (\bfz_i-\bfz_j)^\sT \nabla^2\eta(\bfzero)(\bfz_i-\bfz_j) + O(d_{ij}^3)\,, \label{eq:eta_taylor_expansion}
\end{align}
where $d_{ij} = \|\bfz_i - \bfz_j\|$ represents the spatial distance between individuals $i$ and $j$. Further, $\nabla\eta$ and $\nabla^2\eta$
respectively denote the gradient and the Hessian of the $\eta(\cdot)$ function. Recall that the autocorrelation function of a stationary 
distribution achieves its maximum at zero and therefore $\nabla \eta(\bfzero) = \bfzero$. Further, $\nabla^2\eta(\bfzero)$ is negative semidefinite. We let $J$ be a square root of $(-1/2)\nabla^2\eta(\bfzero)$.
For `local distances' where $d_{ij}$ is small enough
we can neglect the higher order term $O(d_{ij}^3)$ in \eqref{eq:eta_taylor_expansion} and therefore, 
\begin{align}
\eta(\bfzero) - \eta(\bfz_i-\bfz_j) \approx \|J(\bfz_i-\bfz_j)\|^2\,.
\end{align} 
Using our estimates from the previous step we obtain
\begin{align}
\heta_0 - \heta_{i,j} \approx \|J(\bfz_i-\bfz_j)\|^2\,. \label{eq:eta_z_relation}
\end{align}
We correct for the transformation $J$ using some anchor individuals whose locations are known apriori.\footnote{PCA also reconstructs locations only up to an orthogonal transformation. In particular, if $X_*$ is a solution to the optimization problem \eqref{eq:PCA1}, then $X_* Q_2$ for any $2 \times 2$ orthogonal matrix $Q_2$ is also a solution.}
Hence, we obtain consistent estimates for local pairwise distances.

It is worth noting that the above argument fails if $d_{ij}$ is large because the higher order term $O(d_{ij}^3)$ cannot be neglected in our estimation procedure. We thus employ a threshold value $\tau$ and only use the estimated distances in \eqref{eq:eta_z_relation} for individuals $i$ and $j$ for which $\hat{d}_{ij} = (\heta_0-\heta_{i,j})^{1/2} \le \tau$.  When some of the estimated local distances $\heta_0-\heta_{i,j}$ are negative, we shift all of the estimates by the smallest constant which makes them non-negative.
We discuss the procedure for choosing this threshold $\tau$ in \sref{sec:choosing_tau}.
\bigskip

\noindent{\bf Step $(3)$: Global embedding.}
The final step is finding a `global' embedding of individuals from their estimated local pairwise distances.
There has been a great deal of research on this task as it appears in various applications such as network localization~\citep{shang2003localization,SensorNetworks1}
and reconstruction of protein conformations from NMR measurements. It is also directly related to dimensionality reduction of high dimensional data under the topic of manifold learning. Several interesting algorithms have been proposed in the literature for this task. Probably the most well-known is the ISOMAP algorithm \citep{Tenenbaum:ISOMAP}. It first estimates the missing pairwise distances by computing the shortest path between all pairs of nodes, via local distances. It then applies multidimensional scaling (MDS) to infer the locations from the pairwise distances. Some other methods for this task are Locally linear embedding (LLE) \citep{saul:2003}, Laplacian eigenmap \citep{Belkin02laplacianeigenmaps}, Hessian eigenmap \citep{DonohoGrimes}, and Locally rigid embedding \citep{Singer:positioning}.
Another group of algorithms formulate the localization task as a non-convex optimization problem and then consider different convex relaxations to solve it.  A famous example of this type is the relaxation to semidefinite programming (SDP) \citep{Biswas:SDP,Alfakih:SDP,Weinberger06anintroduction,javanmard2013localization}.

One can use any of the above proposed methods for this step. In the remainder of this paper, we use the ISOMAP algorithm to infer the locations from the estimated local distances. For the reader's convenience, we summarize the steps of ISOMAP below.

Let $\hat{d}_{ij} = (\heta_0 - \heta_{i,j})^{1/2}$ be estimated pairwise distances. Construct a graph $G$ with $n$ nodes such that
$i$ and $j$ are connected by an edge of weight $\hat{d}_{ij}$ if they are within the local distance threshold, i.e., $\hat{d}_{ij} \le \tau$. The steps of ISOMAP follow:
\begin{enumerate}
\item[(1)] Compute pairwise shortest paths in the (weighted) graph $G$.
\item[(2)] Let $D_{\tau}$ be the matrix of squared shortest paths distances in $G$.
\item[(3)] Let $(u_1,u_2)$ and $(\sigma_1,\sigma_2)$ be the top two eigenvectors and eigenvalues of $(-1/2)LD_{\tau}L$, where $L = I_{n\times n} - \u\u^\sT$ is the centering matrix.
\item[(4)] Return the estimated locations $\bfz_i = (\sqrt{\sigma_1} u_{1,i}, \sqrt{\sigma_2} u_{2,i})$, for $i=1,2,\dotsc, n$.
\end{enumerate}

\begin{remark}
It is straightforward to verify that output of \localizationAlgorithm{} is unaltered if we relabel the alleles at any SNP. In other words,
for any SNP $\ell$, if we replace the genotypes 0 and 2 for all individuals at that SNP, \localizationAlgorithm{} returns the same locations.
\end{remark}

\subsection{Choosing the local distance threshold $\tau$}\label{sec:choosing_tau}
We describe two strategies for choosing the distance threshold $\tau$ that we use in order to estimate spatial distances from genetic distances in \eqref{eq:eta_z_relation}.
When we have the true sampling locations for the individuals in the dataset, we can use a subset of these known locations as training data for choosing $\tau$. In particular, in the simulations for \localizationAlgorithm{} in \sref{sec:localization_simulations} and in the simulations for the association testing procedure \associationTestingAlgorithm{} in \sref{sec:association_test_simulations}, we used the known locations of a random subset of 20\% of the samples as training data, and chose the value of $\tau$ that minimized the spatial reconstruction RMSE on the training set. For evaluation on the real datasets where we have the sampling coordinates for each subpopulation, we used a leave-one-out cross-validation procedure on the training set to choose the value of $\tau$. For the simulation scenarios, we optimized $\tau$ using the RMSE on the training set instead of performing leave-one-out cross-validation for computational reasons.
\fref{fig:rmse_vs_tau} shows an example of the dependence of the spatial reconstruction RMSE (of the entire dataset) on the choice of $\tau$ in the isotropic covariance decay model. In general, different covariance decay models $\eta$ will exhibit different dependence of the spatial reconstruction RMSE on the threshold $\tau$, and one can tune this parameter using the kind of training and cross-validation procedures that are commonly employed in machine learning.

The second-order stationarity assumption of our model, i.e. $\cov(q_{\ell}(\bfz), q_{\ell}(\bfz')) = \eta(\bfz - \bfz')$, implicitly assumes that the covariance decay function $\eta$ is the same across space. However, we expect different covariance decay functions in different geographic regions \citep{ramachandran:2011,jay:2013} due to geographic barriers, historical migrations, and other factors that introduce spatial heterogeneity. The distance threshold $\tau$ is used to determine the regime in which the second-order Taylor expansion of the $\eta$ function given in \eqref{eq:eta_taylor_expansion} can be considered to be valid. As a result, when the estimated covariances $\hat{\eta}_{ij}$ and distances $\hat{d}_{ij}$ change, say due to refocusing an earlier analysis on a subset of the samples, it would make sense to retune the local distance threshold $\tau$ via the above cross-validation procedure using samples from the relevant geographic region.

\begin{figure}
\includegraphics[width=\linewidth, trim=0 0 0 0]{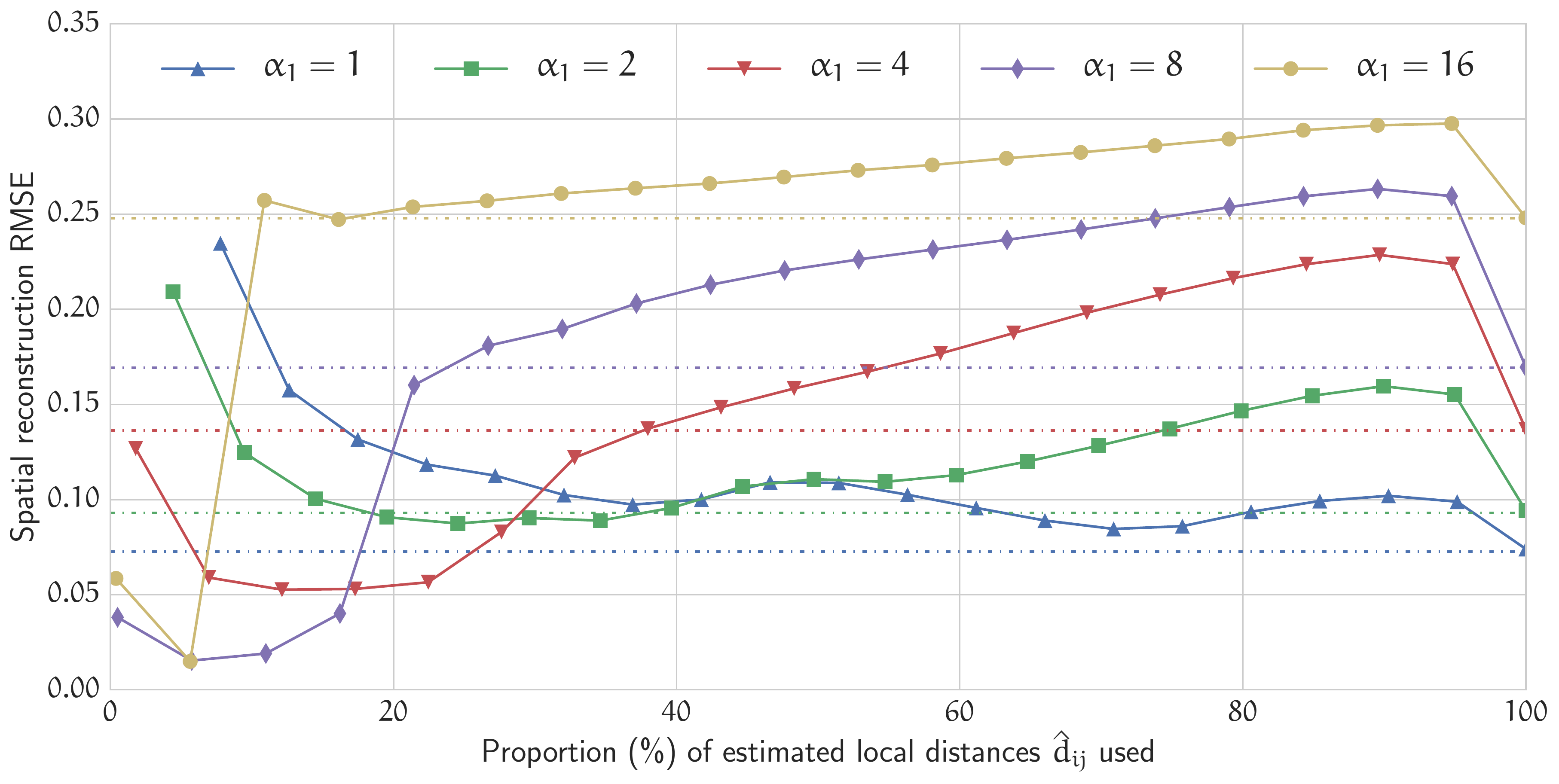}
\caption{{\bf Sensitivity of the spatial reconstruction RMSE to the local distance threshold $\tau$ for the isotropic covariance decay model.} The simulations were performed with $n=2{,}000$ individuals sampled from the unit square $[0,1]^2$, with $p=50{,}000$ SNPs according to the isotropic covariance decay model with parameter combinations $\alpha_0 = \alpha_2 = 1$ and different choices of $\alpha_1$. Solid lines indicate the spatial reconstruction RMSE of \localizationAlgorithm{} as a function of the distance threshold $\tau$, while the horizontal dashed lines indicate the reconstruction RMSE of PCA. In order to put the different ranges for $\tau$ for each parameter setting of $\alpha_1$ on the same scale, the $x$-axis is measured using the percentage of the estimated $\binom{n}{2}$ genetic distances $\hat{d}_{ij} \leq \tau$ which are used by \localizationAlgorithm{}.}
\label{fig:rmse_vs_tau}
\end{figure}

When we do not have any individuals with known locations, as is the case with the Northern Finland Birth Cohort GWAS dataset analyzed in \sref{sec:nfbc}, we use the following procedure for picking $\tau$. 
For any given threshold $\tau$, $D_{\tau}$ is the squared shortest paths matrix produced in the second step of the ISOMAP algorithm. Let $\tilde{D}_{\tau}$ denote the pairwise squared distance matrix of the two-dimensional embedding produced in the fourth step of the ISOMAP algorithm. We choose $\tau$ to maximize the value of $||L\tilde{D}_{\tau}L||_*/||LD_{\tau}L||_*$, where $\|A\|_*$ is the nuclear norm of the matrix $A$ and is given by the sum of the singular values of $A$.

\subsection{Relation to previous spatial models}\label{sec:relation_to_previous_works}
Here, we show that several previously proposed spatial genetic models and ancestry localization algorithms can be viewed as a special case of our probabilistic model and algorithm.

\begin{enumerate}
\item \SpaceMix{} \citep{bradburd:2016}: This model posits that the distribution of alleles among individuals 
comes from a spatial process such that the covariance function $F$ between normalized allele frequencies for individuals $i$ and $j$ 
has an exponential decay with respect to their spatial distance:
\begin{align}
F(\bfz,\bfz') = \frac{1}{\alpha_0} \exp(-(\alpha_1 \|\bfz-\bfz'\|)^{\alpha_2})\,.
\end{align}
This is clearly a special case of our probabilistic model since $F(\bfz,\bfz')$ is a function of $\bfz - \bfz'$.

\item \SCAT{} \citep{wasser:2004}: For the case of two alleles at each locus (similar to the setting considered in the present paper), this model is based on writing the allele frequencies as
\begin{align}
q_\ell(\bfz) = \frac{1}{1+\exp(\theta_{\ell}(\bfz))}\,,
\end{align}
where the $\theta_\ell(\cdot)$ values for different SNPs $\ell$ are assumed to be independent Gaussian processes. For each $\ell$, $\theta_\ell$ is a Gaussian spatial process with $\E(\theta_\ell(\bfz)) = \mu_\ell$ and covariance kernel $K_{\theta_\ell}(\bfz,\bfz') = (1/\alpha_0)\exp(-(\alpha_1\|\bfz -\bfz'\|)^{\alpha_2})$. Note that process $\theta_\ell$ is translation invariant. More specifically, for any collection of locations $\{\bfz_i\}_{i=1}^n$, the distribution of $(\theta_\ell(\bfz_1+\bfdelta),\cdots, \theta_\ell(\bfz_n+\bfdelta))$
is invariant to $\bfdelta$. This property is preserved after applying any one-to-one deterministic function, and in particular, the logistic function. Therefore, the process $q_\ell(\bfz)$ is also translation invariant. As a result, the covariance of allele frequencies $\cov(q_{\ell}(\bfz),q_{\ell}(\bfz'))$ only depends on $\bfz-\bfz'$ and can be written as $\cov(q_{\ell}(\bfz),q_{\ell}(\bfz')) = \eta(\bfz-\bfz')$ for some function $\eta$. This is clearly 
a special case of our probabilistic model.   

\item \SPA{} \citep{yang:2012}: In the \SPA{} model, allele frequencies are given by a logistic function
\begin{align}
q_\ell(\bfz) = \frac{1}{1+\exp(-\<\bfa_\ell, \bfz\> -b_\ell)}\,,
\end{align}
where $\bfa_\ell$ and $b_\ell$ are coefficients for SNP $\ell$. Under such a model, the allele frequencies at each SNP $\ell$ are constant along lines perpendicular to the vector $\bfa_\ell$.
The directional covariance decay model introduced in the Simulation results section of the main text also possesses this property. In the directional covariance decay model, the allele frequency at SNP $\ell$ and location $\bfz$ is given by $q_\ell(\bfz) = 1/(1+\exp(G_{\ell}(\bfz)))$ where $G_\ell(\cdot)$ is a sample path from a Gaussian spatial process with mean 0 and covariance kernel 
$K(\bfz,\bfz') = (1/\alpha_0)\exp(-(\alpha_1|\<\bfu,\bfz -\bfz'\>|)^{\alpha_2})$. 
For any two locations $\bfz, \bfz'$ such that $\bfz-\bfz' \perp \bfu$, we have $K(\bfz,\bfz') = 1/\alpha_0$. Further, $K(\bfz,\bfz) = K(\bfz',\bfz') = 1/\alpha_0$. In words, $G_\ell(\bfz)$ and $G_\ell(\bfz')$ have equal variance and are perfectly correlated, therefore $G_\ell(\bfz) = G_\ell(\bfz')$ almost surely. This argument shows that the lines perpendicular to the direction vector $\bfu$ are level sets for the allele frequency. This is also apparent from \fref{fig:logisticExpDecayCov_n2000_p50k} in the main text.

\item PCA \citep{price:2006,novembre:2008}: We next show that under our probabilistic model for allele frequencies, \localizationAlgorithm{}
asymptotically \emph{always} dominates PCA. Specifically, if we choose the local distance threshold $\tau$ to be large enough, then \localizationAlgorithm{} and PCA return the same outputs in the asymptotic regime $n\to \infty$, and hence PCA can be 
viewed as a special case of \localizationAlgorithm{}. In Tables \ref{tab:logisticExpDecayCov_MDS_PCA_beta0.25}--\ref{tab:logisticDirectionalExpDecayCov_MDS_PCA_beta4}, the ratio of the RMSE of \localizationAlgorithm{} to the RMSE of PCA exceeds 1 by a very small amount for some parameter combinations, which is due to the effect of finite sample size. However, this effect of the finite sample size is already very small for $n\ge 2{,}000$. 
To corroborate our claim, recall that PCA estimates locations using the two top eigenvectors of $LXX^\sT L$, where $L$ is the centering matrix $L = I - \u\u^\sT$, with $\u = (1/\sqrt{n}, \dotsc, 1/\sqrt{n})^\sT$ the unit norm vector with equal entries. Often, the columns of the centered genotype matrix are normalized to have unit variance before applying PCA (This is also done in our simulations.) In the asymptotic regime $n \to \infty$, the normalization factors for all columns concentrate at $\eta(\bfzero)$ and as such PCA uses the (scaled) two top eigenvectors of $(1/\eta(\bfzero)) LXX^\sT L$. On the other hand,
in Step (1) of \localizationAlgorithm{}, the estimates $\hat{\eta}_{i,j}$ can be written as the $(i,j)$ entry of $LXX^\sT L$. Let $\widehat{D} = (\hat{d}_{ij}^2)$ where $\hat{d}_{ij} = (\hat{\eta}_0 -\hat{\eta}_{i,j})^{1/2}$
are the estimated local spatial distances in Step (2). We thus have the matrix representation $\widehat{D} = \hat{\eta}_0 \u\u^\sT -LXX^\sT L$.
 If $\tau$ is chosen to be larger than the range of pairwise distances $\hat{d}_{ij}$, all of them will be treated as local distances and no thresholding occurs. Therefore, in Step (3) the constructed graph $G$ is a complete graph and the squared shortest path distances $\widehat{D}$ are indeed the squared local distances $D$. The ISOMAP employed in the last step reduces to PCA applied to 
 $$-\frac{1}{2}L\widehat{D}L = -\frac{1}{2} L(\hat{\eta}_0\u\u^\sT - LXX^\sT L) L = \frac{1}{2}LXX^\sT L\,,$$   
 where the last equality holds because $L\u = \bfzero$ and $L^2 = L$.
 It is now clear that \localizationAlgorithm{} and PCA are the same procedure in this case (up to a scaling factor which is corrected for using some individuals with known locations).
\end{enumerate}

\begin{figure}
\begin{subfigure}[]{\textwidth}
\centering
\caption{}\label{fig:logisticExpDecayCovAlleleFreqFnExample}
\includegraphics[width=\linewidth, trim=0 0 0 0]{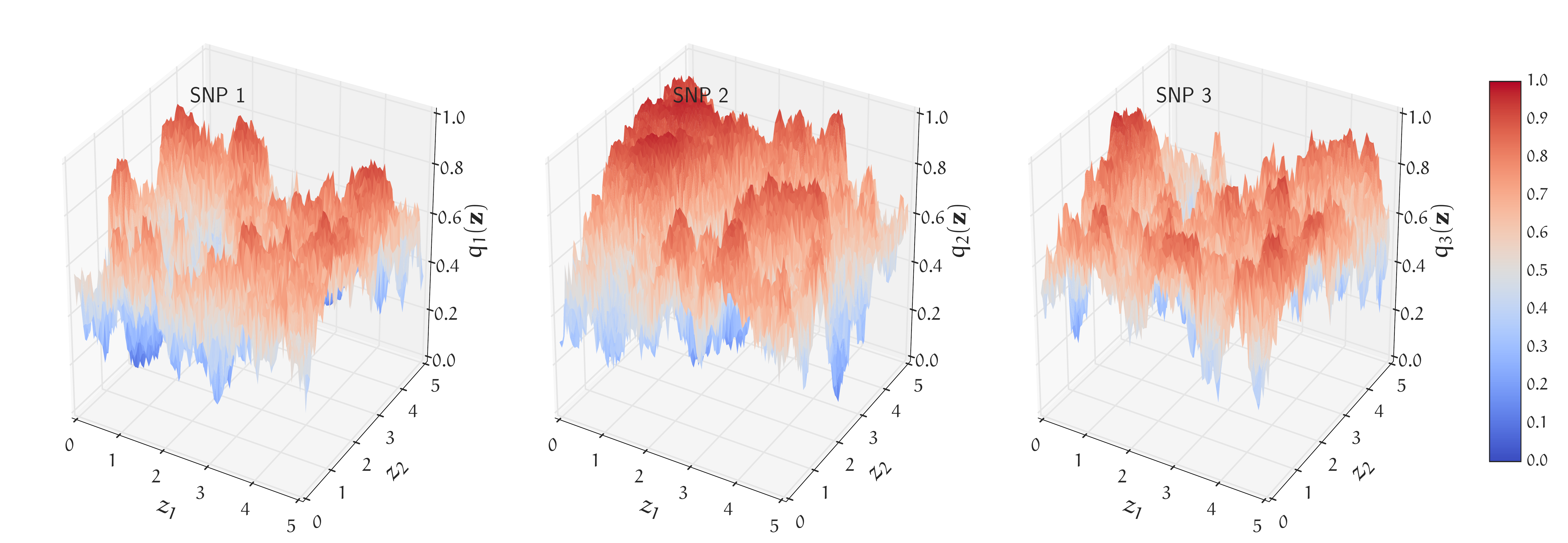}
\vspace{4mm}
\end{subfigure}
\begin{subfigure}[]{\textwidth}
\centering
\caption{}\label{fig:logisticDirectionalExpDecayCovAlleleFreqFnExample}
\includegraphics[width=\linewidth, trim=0 0 0 0]{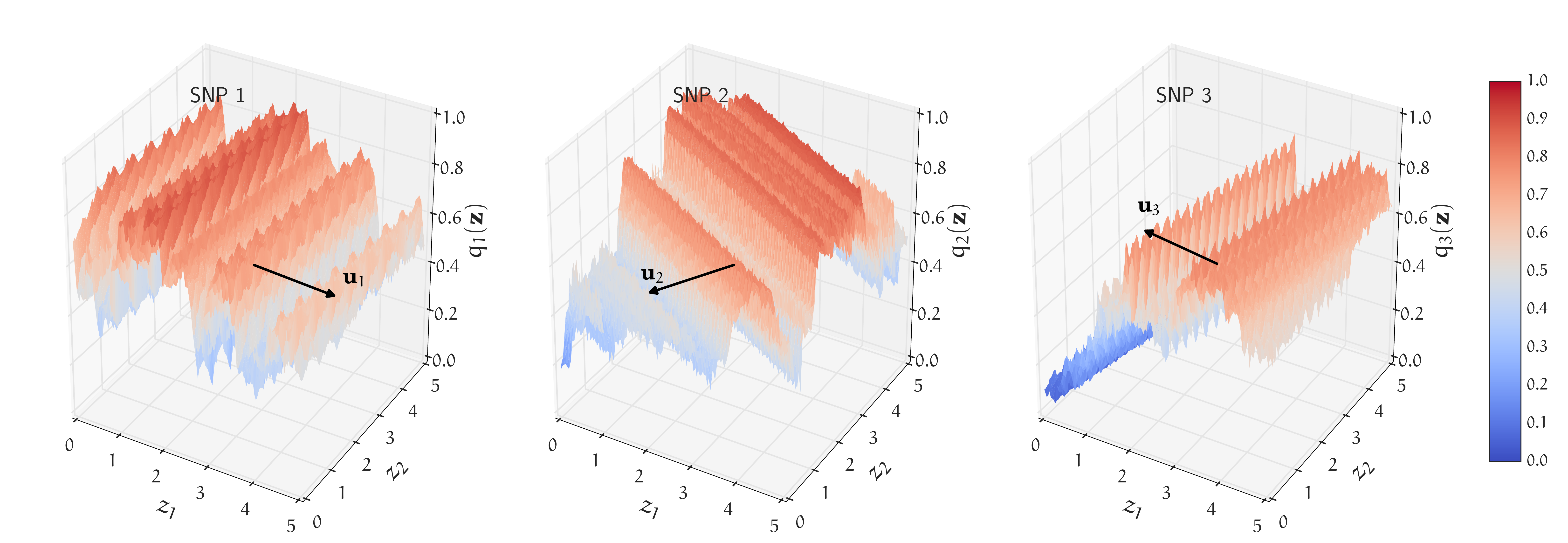}
\end{subfigure}
\caption{{\bf Sample spatial allele frequencies from our probabilistic model.} Each figure corresponds to an allele frequency covariance function $\eta$ in the underlying collection of spatial stochastic processes over the two-dimensional space represented by a $5 \times 5$ grid. \\
\subref{fig:logisticExpDecayCovAlleleFreqFnExample} {\bf Isotropic covariance decay.} The allele frequency surface is generated by a logistic function applied to a Gaussian process. In particular, the allele frequency $q_{\ell}(\bfz)$ of SNP $\ell$ at location $\bfz$ is given by $q_{\ell}(\bfz) = 1 / ( 1 + \exp(G_{\ell}(\bfz)) )$, where $G_{\ell}(\cdot)$ is a sample path from a stationary Gaussian process with mean 0 and covariance kernel $K(\bfz, \bfz') = \exp(- (\alpha_1 \|\bfz - \bfz'\|)^{\alpha_2})/\alpha_0$. In this example, $\alpha_0 = \alpha_1 = \alpha_2 = 1$. Such families of allele frequency functions are also used in previous probabilistic models such as \SCAT{} \citep{wasser:2004} and \SpaceMix{} \citep{bradburd:2016}. \\
\subref{fig:logisticDirectionalExpDecayCovAlleleFreqFnExample} {\bf Directional covariance decay.} The allele frequency $q_{\ell}(\bfz)$ of SNP $\ell$ at location $\bfz$ is given by $q_{\ell}(\bfz) = 1 / ( 1 + \exp(G_{\ell}(\bfz)) )$, where $G_{\ell}(\cdot)$ is a sample path from a stationary Gaussian process with mean 0 and covariance kernel $K(\bfz, \bfz') = \exp(- (\alpha_1 |\< \bfu, \bfz - \bfz' \>|)^{\alpha_2})/\alpha_0$. This form for the Gaussian process kernel leads to level sets of equal allele frequency in directions perpendicular to $\bfu$. In this example, $\alpha_0 = \alpha_1 = \alpha_2 = 1$ and the directions $\bfu$, shown as black arrows, were randomly chosen for each SNP. Such allele frequency functions can be viewed as a generalization of the logistic allele frequency surfaces considered in the \SPA{} model \citep{yang:2012}.
}
\label{fig:alleleFrequencyExamples}
\end{figure}

\subsection{Ancestry localization simulations}\label{sec:localization_simulations}
As described in the main text, we considered two sets of simulation scenarios to model isotropic and direction-dependent decay rates for the allele frequency covariance. For both simulation scenarios, we simulated $n = 2{,}000$ individuals at $p = 50{,}000$ SNPs. 
The true geographic origin $\bfz_i$ of individual $i$ was simulated by sampling each coordinate according to a Beta($\beta, \beta$) distribution from the unit square. This distribution lets us smoothly interpolate between dense sampling of individuals in the interior of the space to dense sampling at the boundaries (Figures \ref{fig:logisticExpDecayCov_n2000_p50k_beta1_loc} and \ref{fig:logisticExpDecayCov_n2000_p50k_beta0.25_loc} in the main text), with $\beta = 1$ representing uniform sampling. We considered $\beta \in \{ 0.25, 0.5, 1, 2, 4 \}$.
The spatial allele frequencies at each SNP were generated by applying the logistic function to sample paths from a spatial Gaussian process. The genotypes of each individual $i$ were then drawn according to a binomial distribution from the allele frequencies at their geographic origin $\bfz_i$.
\begin{itemize}
  \item {\bf Isotropic covariance decay}: The allele frequency $q_{\ell}(\bfz_i)$ of SNP $\ell$ at location $\bfz_i$ is given by $q_{\ell}(\bfz_i) = 1 / ( 1 + \exp(Z_{\ell,i}) )$, where $Z_{\ell, \cdot}$ is an $n$-dimensional Normal random variable with mean $\bfzero$ and covariance $\cov(Z_{\ell,i}, Z_{\ell,j}) = \exp(- (\alpha_1 \|\bfz_i - \bfz_j\|)^{\alpha_2})/\alpha_0$. Such covariance decay models have been previously used by \citet{wasser:2004} and \citet{bradburd:2016}. \fref{fig:logisticExpDecayCovAlleleFreqFnExample} in the main text shows example allele frequency surfaces drawn from this model.
  \item {\bf Directional covariance decay}: Given a unit norm direction vector $\bfu_k \in \reals^2$, the allele frequency $q_{\ell}(\bfz_i)$ of SNP $\ell$ at location $\bfz_i$ is given by $q_{\ell}(\bfz_i) = 1 / ( 1 + \exp(Z_{\ell,i}) )$, where $Z_{\ell, \cdot}$ is an $n$-dimensional Normal random variable with mean $\bfzero$ and covariance $\cov(Z_{\ell,i}, Z_{\ell,j}) = \exp(- (\alpha_1 |\< \bfu_k, \bfz_i - \bfz_j \>|)^{\alpha_2})/\alpha_0$.  \fref{fig:logisticDirectionalExpDecayCovAlleleFreqFnExample} in the main text shows example allele frequency surfaces of this form.
  Such models can be viewed as a generalization of the SPA model of \cite{yang:2012} (see \sref{sec:relation_to_previous_works}). In the simulations, we sampled 100 different direction vectors $\bfu_k$ from a von Mises distribution, which is a circular analogue of the Normal distribution. For each such direction vector $\bfu_k$, we simulated 500 SNPs, which will have level sets of equal allele frequency in directions perpendicular to $\bfu_k$.
\end{itemize}

For each parameter combination in the above simulation scenarios, we simulated 10 random datasets, and used PCA and our algorithm \localizationAlgorithm{} to infer the spatial coordinates $\bfz_i$. PCA can estimate the coordinates up to an orthogonal transformation, while \localizationAlgorithm{} estimates coordinates up to the invertible linear transformation $J$ in \eqref{eq:eta_z_relation}.
We use the true geographic locations of a random subset of 20\% of the simulated individuals to rescale the coordinates inferred by PCA and \localizationAlgorithm{}. As a measure of inference accuracy, we use the root mean squared error (RMSE) between the inferred locations $\hat{\bfz}_i$ and the true locations $\bfz_i$ as follows,
\begin{align}
  \textrm{RMSE} = \sqrt{\frac{1}{n} \sum_{i=1}^n \| \bfz_i - \hat{\bfz}_i \|^2} \,.
\end{align}
In order to choose the threshold parameter $\tau$ that is used when estimating the local spatial distances from the genetic distances, we picked a uniform grid (of 20 points) over the quantiles of the $\binom{n}{2}$ estimated pairwise genetic distances $\hat{d}_{ij}$. We picked the value of $\tau$ which minimized the reconstruction RMSE over the aforementioned random subset containing 20\% of the samples whose locations were assumed known. 
For most parameter combinations in both the simulation models, and for the range of sampling distribution parameters $\beta $, the RMSE of \localizationAlgorithm{} is substantially lower than that of PCA (Tables \ref{tab:logisticExpDecayCov_MDS_PCA_beta1}--\ref{tab:logisticDirectionalExpDecayCov_MDS_PCA_beta4}).

\vspace{1cm}
\begin{center}
\begin{tabular}{ccccccc}
\toprule
\multicolumn{7}{c}{$\beta = 1$} \\
\multirow{2}{*}{$\alpha_2$}  & \multirow{2}{*}{$\alpha_1$}  & \multirow{2}{*}{$\displaystyle\frac{\text{RMSE \localizationAlgorithm{}}}{\text{RMSE PCA}}$}	&	\multirow{2}{*}{RMSE PCA} & \multirow{2}{*}{RMSE \localizationAlgorithm{}}	& \multirow{2}{*}{$\tau$}  & \multirow{2}{1in}{Proportion of distances used}	\\[0.5cm]
\midrule
\multirow{5}{*}{0.5}	&	1	&	1.010	&	0.0879	&	0.0888	&	0.2507	&	100.0\% 	\\
			&	2	&	\bf{0.978}	&	0.1030	&	0.1008	&	0.2305	&	17.2\% 	\\
			&	4	&	\bf{0.699}	&	0.1001	&	0.0700	&	0.2349	&	11.4\% 	\\
			&	8	&	\bf{0.414}	&	0.1151	&	0.0477	&	0.2416	&	6.2\% 	\\
			&	16	&	\bf{0.307}	&	0.1372	&	0.0421	&	0.2612	&	5.8\% 	\\
\midrule
\multirow{5}{*}{1}	&	1	&	1.018	&	0.0716	&	0.0729	&	0.2700	&	100.0\% 	\\
			&	2	&	\bf{0.880}	&	0.0929	&	0.0818	&	0.2599	&	24.4\% 	\\
			&	4	&	\bf{0.359}	&	0.1285	&	0.0461	&	0.2584	&	11.7\% 	\\
			&	8	&	\bf{0.094}	&	0.1733	&	0.0163	&	0.2587	&	5.7\% 	\\
			&	16	&	\bf{0.096}	&	0.2554	&	0.0245	&	0.2816	&	5.7\% 	\\
\midrule
\multirow{5}{*}{1.5}	&	1	&	1.028	&	0.0555	&	0.0570	&	0.2806	&	100.0\% 	\\
			&	2	&	\bf{0.857}	&	0.0983	&	0.0843	&	0.2800	&	33.5\% 	\\
			&	4	&	\bf{0.212}	&	0.1647	&	0.0349	&	0.2610	&	12.2\% 	\\
			&	8	&	\bf{0.100}	&	0.2842	&	0.0285	&	0.2595	&	5.7\% 	\\
			&	16	&	\bf{0.100}	&	0.3114	&	0.0311	&	0.1963	&	0.5\% 	\\
\bottomrule
\end{tabular}
\captionof{table}{{\bf Isotropic covariance decay model}: Comparison of the localization accuracy of \localizationAlgorithm{} and PCA for simulated datasets with $n=2{,}000$ samples and $p=50{,}000$ SNPs. 
The geographic locations $\bfz_i$ of the individuals are simulated by sampling each coordinate according to a Beta($\beta, \beta$) distribution from the unit square. In this table, $\beta = 1$, which is equivalent to sampling the individuals uniformly from the unit square. The allele frequency for individual $i$ at locus $\ell$ is given by $q_{i\ell} = 1 / (1 + \exp(Z_{i\ell}))$, where $Z_{\cdot,\ell}$ is an $n$-dimensional multivariate Gaussian random variable with mean 0 and covariance between the $i$-th and $j$-th entries given by $\exp(- (\alpha_1 \|\bfz_i - \bfz_j\|)^{\alpha_2})/\alpha_0$. In these simulations, $\alpha_0 = 1$. 
The columns for PCA and \localizationAlgorithm{} show the root mean squared error (RMSE) in the reconstruction accuracy for PCA and MDS, respectively. The column $\tau$ indicates the threshold on the genetic distance that was used when applying \localizationAlgorithm{}. This threshold $\tau$ was optimized using the known ancestral locations of a random subset of 20\% of the simulated points. The last column of the table indicates the percentage of entries in the pairwise genetic distance matrix less than the threshold value $\tau$.}
\label{tab:logisticExpDecayCov_MDS_PCA_beta1}
\end{center}

\begin{center}
\begin{tabular}{ccccccc}
\toprule
\multicolumn{7}{c}{$\beta = 0.5$} \\
\multirow{2}{*}{$\alpha_2$}  & \multirow{2}{*}{$\alpha_1$}  & \multirow{2}{*}{$\displaystyle\frac{\text{RMSE \localizationAlgorithm{}}}{\text{RMSE PCA}}$}	&	\multirow{2}{*}{RMSE PCA} & \multirow{2}{*}{RMSE \localizationAlgorithm{}}	& \multirow{2}{*}{$\tau$}  & \multirow{2}{1in}{Proportion of distances used}	\\[0.5cm]
\midrule
\multirow{5}{*}{0.5}	&	1	&	\bf{0.925}	&	0.0605	&	0.0560	&	0.2223	&	23.4\% 	\\
			&	2	&	\bf{0.774}	&	0.0650	&	0.0503	&	0.2371	&	23.1\% 	\\
			&	4	&	\bf{0.681}	&	0.0697	&	0.0475	&	0.2326	&	6.9\% 	\\
			&	8	&	\bf{0.448}	&	0.0718	&	0.0321	&	0.2609	&	12.1\% 	\\
			&	16	&	\bf{0.444}	&	0.0826	&	0.0367	&	0.2633	&	6.2\% 	\\
\midrule
\multirow{5}{*}{1}	&	1	&	\bf{0.966}	&	0.0623	&	0.0602	&	0.2309	&	37.1\% 	\\
			&	2	&	\bf{0.593}	&	0.0770	&	0.0457	&	0.2524	&	24.4\% 	\\
			&	4	&	\bf{0.205}	&	0.1090	&	0.0224	&	0.2293	&	6.8\% 	\\
			&	8	&	\bf{0.200}	&	0.1174	&	0.0234	&	0.2544	&	5.9\% 	\\
			&	16	&	\bf{0.072}	&	0.3645	&	0.0263	&	0.2290	&	1.2\% 	\\
\midrule
\multirow{5}{*}{1.5}	&	1	&	1.027	&	0.0567	&	0.0583	&	0.2817	&	100.0\% 	\\
			&	2	&	\bf{0.466}	&	0.0921	&	0.0429	&	0.2797	&	36.9\% 	\\
			&	4	&	\bf{0.195}	&	0.1336	&	0.0261	&	0.2451	&	11.6\% 	\\
			&	8	&	\bf{0.071}	&	0.3815	&	0.0269	&	0.2619	&	5.9\% 	\\
			&	16	&	\bf{0.102}	&	0.4187	&	0.0425	&	0.2144	&	0.9\% 	\\
\bottomrule
\end{tabular}
\captionof{table}{{\bf Isotropic covariance decay model}: The geographic locations $\bfz_i$ of the individuals are simulated by sampling each coordinate according to a Beta($\beta, \beta$) distribution from the unit square, with $\beta = 0.5$. The other simulation settings are as described in \tref{tab:logisticExpDecayCov_MDS_PCA_beta1}.}
\label{tab:logisticExpDecayCov_MDS_PCA_beta0.5}
\end{center}
\vspace{4mm}

\begin{center}
\begin{tabular}{ccccccc}
\toprule
\multicolumn{7}{c}{$\beta = 0.25$} \\
\multirow{2}{*}{$\alpha_2$}  & \multirow{2}{*}{$\alpha_1$}  & \multirow{2}{*}{$\displaystyle\frac{\text{RMSE \localizationAlgorithm{}}}{\text{RMSE PCA}}$}	&	\multirow{2}{*}{RMSE PCA} & \multirow{2}{*}{RMSE \localizationAlgorithm{}}	& \multirow{2}{*}{$\tau$}  & \multirow{2}{1in}{Proportion of distances used}	\\[0.5cm]
\midrule
\multirow{5}{*}{0.5}	&	1	&	\bf{0.772}	&	0.0476	&	0.0367	&	0.2005	&	15.1\% 	\\
			&	2	&	\bf{0.590}	&	0.0524	&	0.0309	&	0.2212	&	15.2\% 	\\
			&	4	&	\bf{0.491}	&	0.0715	&	0.0351	&	0.2232	&	9.3\% 	\\
			&	8	&	\bf{0.252}	&	0.1190	&	0.0300	&	0.2445	&	9.1\% 	\\
			&	16	&	\bf{0.221}	&	0.1938	&	0.0428	&	0.2628	&	8.4\% 	\\
\midrule
\multirow{5}{*}{1}	&	1	&	1.010	&	0.0534	&	0.0540	&	0.2758	&	100.0\% 	\\
			&	2	&	\bf{0.545}	&	0.0532	&	0.0290	&	0.2169	&	15.2\% 	\\
			&	4	&	\bf{0.228}	&	0.1736	&	0.0396	&	0.2462	&	14.0\% 	\\
			&	8	&	\bf{0.132}	&	0.3293	&	0.0433	&	0.2616	&	8.5\% 	\\
			&	16	&	\bf{0.067}	&	0.4390	&	0.0293	&	0.2455	&	3.1\% 	\\
\midrule
\multirow{5}{*}{1.5}	&	1	&	\bf{0.991}	&	0.0464	&	0.0460	&	0.2315	&	40.3\% 	\\
			&	2	&	\bf{0.499}	&	0.0795	&	0.0397	&	0.2409	&	22.7\% 	\\
			&	4	&	\bf{0.097}	&	0.3640	&	0.0352	&	0.2522	&	12.8\% 	\\
			&	8	&	\bf{0.165}	&	0.3683	&	0.0609	&	0.2678	&	7.6\% 	\\
			&	16	&	\bf{0.089}	&	0.4571	&	0.0407	&	0.2711	&	4.3\% 	\\
\bottomrule
\end{tabular}
\captionof{table}{{\bf Isotropic covariance decay model}: The geographic locations $\bfz_i$ of the individuals are simulated by sampling each coordinate according to a Beta($\beta, \beta$) distribution from the unit square, with $\beta = 0.25$. The other simulation settings are as described in \tref{tab:logisticExpDecayCov_MDS_PCA_beta1}.}
\label{tab:logisticExpDecayCov_MDS_PCA_beta0.25}
\end{center}

\begin{center}
\begin{tabular}{ccccccc}
\toprule
\multicolumn{7}{c}{$\beta = 2$} \\
\multirow{2}{*}{$\alpha_2$}  & \multirow{2}{*}{$\alpha_1$}  & \multirow{2}{*}{$\displaystyle\frac{\text{RMSE \localizationAlgorithm{}}}{\text{RMSE PCA}}$}	&	\multirow{2}{*}{RMSE PCA} & \multirow{2}{*}{RMSE \localizationAlgorithm{}}	& \multirow{2}{*}{$\tau$}  & \multirow{2}{1in}{Proportion of distances used}	\\[0.5cm]
\midrule
\multirow{5}{*}{0.5}	&	1	&	1.009	&	0.0891	&	0.0899	&	0.2508	&	100.0\% 	\\
			&	2	&	1.008	&	0.0940	&	0.0947	&	0.2617	&	100.0\% 	\\
			&	4	&	\bf{0.910}	&	0.1064	&	0.0968	&	0.2319	&	11.4\% 	\\
			&	8	&	\bf{0.657}	&	0.1241	&	0.0816	&	0.2358	&	5.8\% 	\\
			&	16	&	\bf{0.347}	&	0.1377	&	0.0478	&	0.2295	&	1.1\% 	\\
\midrule
\multirow{5}{*}{1}	&	1	&	1.020	&	0.0663	&	0.0677	&	0.2647	&	100.0\% 	\\
			&	2	&	1.014	&	0.0859	&	0.0871	&	0.3076	&	100.0\% 	\\
			&	4	&	\bf{0.695}	&	0.1182	&	0.0821	&	0.2683	&	17.5\% 	\\
			&	8	&	\bf{0.202}	&	0.1725	&	0.0349	&	0.2554	&	5.9\% 	\\
			&	16	&	\bf{0.085}	&	0.2438	&	0.0207	&	0.2401	&	1.5\% 	\\
\midrule
\multirow{5}{*}{1.5}	&	1	&	1.030	&	0.0483	&	0.0498	&	0.2662	&	100.0\% 	\\
			&	2	&	1.018	&	0.0757	&	0.0771	&	0.3356	&	100.0\% 	\\
			&	4	&	\bf{0.468}	&	0.1416	&	0.0663	&	0.2869	&	18.8\% 	\\
			&	8	&	\bf{0.092}	&	0.2205	&	0.0203	&	0.2513	&	5.7\% 	\\
			&	16	&	\bf{0.066}	&	0.2845	&	0.0187	&	0.2318	&	1.5\% 	\\
\bottomrule
\end{tabular}
\captionof{table}{{\bf Isotropic covariance decay model}: The geographic locations $\bfz_i$ of the individuals are simulated by sampling each coordinate according to a Beta($\beta, \beta$) distribution from the unit square, with $\beta = 2$. The other simulation settings are as described in \tref{tab:logisticExpDecayCov_MDS_PCA_beta1}.}
\label{tab:logisticExpDecayCov_MDS_PCA_beta2}
\end{center}
\vspace{4mm}

\begin{center}
\begin{tabular}{ccccccc}
\toprule
\multicolumn{7}{c}{$\beta = 4$} \\
\multirow{2}{*}{$\alpha_2$}  & \multirow{2}{*}{$\alpha_1$}  & \multirow{2}{*}{$\displaystyle\frac{\text{RMSE \localizationAlgorithm{}}}{\text{RMSE PCA}}$}	&	\multirow{2}{*}{RMSE PCA} & \multirow{2}{*}{RMSE \localizationAlgorithm{}}	& \multirow{2}{*}{$\tau$}  & \multirow{2}{1in}{Proportion of distances used}	\\[0.5cm]
\midrule
\multirow{5}{*}{0.5}	&	1	&	1.011	&	0.0775	&	0.0783	&	0.2448	&	100.0\% 	\\
			&	2	&	1.009	&	0.0809	&	0.0816	&	0.2550	&	100.0\% 	\\
			&	4	&	1.006	&	0.0922	&	0.0927	&	0.2723	&	100.0\% 	\\
			&	8	&	\bf{0.815}	&	0.0994	&	0.0810	&	0.2340	&	5.8\% 	\\
			&	16	&	\bf{0.467}	&	0.1131	&	0.0528	&	0.2247	&	1.1\% 	\\
\midrule
\multirow{5}{*}{1}	&	1	&	1.020	&	0.0540	&	0.0551	&	0.2441	&	100.0\% 	\\
			&	2	&	1.017	&	0.0694	&	0.0706	&	0.2986	&	100.0\% 	\\
			&	4	&	\bf{0.958}	&	0.0940	&	0.0900	&	0.2759	&	18.8\% 	\\
			&	8	&	\bf{0.417}	&	0.1257	&	0.0524	&	0.2489	&	6.3\% 	\\
			&	16	&	\bf{0.105}	&	0.1798	&	0.0189	&	0.2561	&	3.7\% 	\\
\midrule
\multirow{5}{*}{1.5}	&	1	&	1.029	&	0.0369	&	0.0380	&	0.2419	&	100.0\% 	\\
			&	2	&	1.025	&	0.0538	&	0.0552	&	0.3239	&	100.0\% 	\\
			&	4	&	\bf{0.962}	&	0.0956	&	0.0920	&	0.2848	&	18.0\% 	\\
			&	8	&	\bf{0.225}	&	0.1576	&	0.0355	&	0.2742	&	10.5\% 	\\
			&	16	&	\bf{0.108}	&	0.2030	&	0.0219	&	0.2786	&	6.3\% 	\\
\bottomrule
\end{tabular}
\captionof{table}{{\bf Isotropic covariance decay model}: The geographic locations $\bfz_i$ of the individuals are simulated by sampling each coordinate according to a Beta($\beta, \beta$) distribution from the unit square, with $\beta = 4$. The other simulation settings are as described in \tref{tab:logisticExpDecayCov_MDS_PCA_beta1}.}
\label{tab:logisticExpDecayCov_MDS_PCA_beta4}
\end{center}

\begin{center}
\begin{tabular}{ccccccc}
\toprule
\multicolumn{7}{c}{$\beta = 1$} \\
\multirow{2}{*}{$\kappa$}  & \multirow{2}{*}{$\alpha_1$}  & \multirow{2}{*}{$\displaystyle\frac{\text{RMSE \localizationAlgorithm{}}}{\text{RMSE PCA}}$}	&	\multirow{2}{*}{RMSE PCA} & \multirow{2}{*}{RMSE \localizationAlgorithm{}}	& \multirow{2}{*}{$\tau$}  & \multirow{2}{1in}{Proportion of distances used}	\\[0.5cm]
\midrule
\multirow{5}{*}{0.1}	&	1	&	1.026	&	0.0680	&	0.0697	&	0.2383	&	100.0\% 	\\
			&	2	&	1.021	&	0.0808	&	0.0825	&	0.2779	&	100.0\% 	\\
			&	4	&	\bf{0.644}	&	0.1105	&	0.0711	&	0.2561	&	23.5\% 	\\
			&	8	&	\bf{0.345}	&	0.1392	&	0.0480	&	0.2397	&	6.0\% 	\\
			&	16	&	\bf{0.257}	&	0.1597	&	0.0410	&	0.2575	&	5.9\% 	\\
\midrule
\multirow{5}{*}{1}	&	1	&	1.024	&	0.0660	&	0.0676	&	0.2367	&	100.0\% 	\\
			&	2	&	1.032	&	0.0884	&	0.0912	&	0.2387	&	36.1\% 	\\
			&	4	&	\bf{0.655}	&	0.1062	&	0.0695	&	0.2617	&	23.3\% 	\\
			&	8	&	\bf{0.362}	&	0.1395	&	0.0505	&	0.2369	&	6.2\% 	\\
			&	16	&	\bf{0.167}	&	0.1552	&	0.0259	&	0.2621	&	5.8\% 	\\
\midrule
\multirow{5}{*}{10}	&	1	&	\bf{0.512}	&	0.2962	&	0.1518	&	0.1817	&	23.6\% 	\\
			&	2	&	\bf{0.518}	&	0.2956	&	0.1532	&	0.2135	&	19.3\% 	\\
			&	4	&	\bf{0.345}	&	0.2980	&	0.1028	&	0.2186	&	8.9\% 	\\
			&	8	&	\bf{0.411}	&	0.3067	&	0.1260	&	0.2252	&	6.4\% 	\\
			&	16	&	\bf{0.356}	&	0.3142	&	0.1118	&	0.1762	&	0.6\% 	\\
\bottomrule
\end{tabular}
\captionof{table}{{\bf Directional covariance decay model}: Comparison of the localization accuracy of \localizationAlgorithm{} and PCA for simulated datasets with $n=2{,}000$ samples and $p=50{,}000$ SNPs. 
The geographic locations $\bfz_i$ of the individuals are simulated by sampling each coordinate according to a Beta($\beta, \beta$) distribution from the unit square.
The geographic locations $\bfz_i$ of the individuals are simulated by sampling each coordinate according to a Beta($\beta, \beta$) distribution from the unit square, with $\beta = 1$.  In this table, $\beta = 1$, which is equivalent to sampling the individuals uniformly from the unit square.
100 different directions vectors $\bfu_k$ were chosen from a von Mises distribution with mean direction $(1, 0)^{T}$ (i.e. the $x$-axis) and dispersion parameter $\kappa$. For each such direction vector $\bfu_k$, 500 SNPs were simulated such that they have level sets of equal frequency along directions perpendicular to $\bfu_k$.
The allele frequency for individual $i$ at SNP $\ell$ for direction vector $\bfu_k$ is given by $q_{\ell}(\bfz_i) = 1 / (1 + \exp(Z_{\ell,i}))$, where $Z_{\ell,\cdot}$ is an $n$-dimensional multivariate Gaussian random variable with mean $\bfzero$ and covariance between the $i$-th and $j$-th entries given by $\exp(- (\alpha_1 |\<\bfu_k,\bfz_i - \bfz_j\>|)^{\alpha_2})/\alpha_0$. In these simulations, $\alpha_0 = \alpha_2 = 1$.
Bold values indicate those parameter combinations where the root mean squared error (RMSE) of \localizationAlgorithm{} is lower than that of PCA.}
\label{tab:logisticDirectionalExpDecayCov_MDS_PCA_beta1}
\end{center}
\clearpage

\begin{center}
\begin{tabular}{ccccccc}
\toprule
\multicolumn{7}{c}{$\beta = 0.5$} \\
\multirow{2}{*}{$\kappa$}  & \multirow{2}{*}{$\alpha_1$}  & \multirow{2}{*}{$\displaystyle\frac{\text{RMSE \localizationAlgorithm{}}}{\text{RMSE PCA}}$}	&	\multirow{2}{*}{RMSE PCA} & \multirow{2}{*}{RMSE \localizationAlgorithm{}}	& \multirow{2}{*}{$\tau$}  & \multirow{2}{1in}{Proportion of distances used}	\\[0.5cm]
\midrule
\multirow{5}{*}{0.1}	&	1	&	1.024	&	0.0648	&	0.0664	&	0.2366	&	100.0\% 	\\
			&	2	&	\bf{0.836}	&	0.0710	&	0.0593	&	0.2340	&	36.2\% 	\\
			&	4	&	\bf{0.549}	&	0.0984	&	0.0540	&	0.2304	&	13.6\% 	\\
			&	8	&	\bf{0.340}	&	0.0896	&	0.0305	&	0.2320	&	6.3\% 	\\
			&	16	&	\bf{0.102}	&	0.2766	&	0.0281	&	0.2603	&	6.3\% 	\\
\midrule
\multirow{5}{*}{1}	&	1	&	1.021	&	0.0663	&	0.0677	&	0.2357	&	100.0\% 	\\
			&	2	&	\bf{0.885}	&	0.0780	&	0.0690	&	0.2437	&	40.8\% 	\\
			&	4	&	\bf{0.479}	&	0.0803	&	0.0384	&	0.2430	&	18.6\% 	\\
			&	8	&	\bf{0.255}	&	0.0854	&	0.0218	&	0.2320	&	6.4\% 	\\
			&	16	&	\bf{0.109}	&	0.2747	&	0.0299	&	0.2558	&	6.0\% 	\\
\midrule
\multirow{5}{*}{10}	&	1	&	\bf{0.360}	&	0.3584	&	0.1291	&	0.1750	&	27.6\% 	\\
			&	2	&	\bf{0.272}	&	0.3571	&	0.0970	&	0.1927	&	17.4\% 	\\
			&	4	&	\bf{0.221}	&	0.3542	&	0.0783	&	0.1936	&	7.8\% 	\\
			&	8	&	\bf{0.278}	&	0.3579	&	0.0994	&	0.2049	&	6.3\% 	\\
			&	16	&	\bf{0.359}	&	0.3562	&	0.1279	&	0.1574	&	0.7\% 	\\
\bottomrule
\end{tabular}
\captionof{table}{{\bf Directional covariance decay model}: The geographic locations $\bfz_i$ of the individuals are simulated by sampling each coordinate according to a Beta($\beta, \beta$) distribution from the unit square, with $\beta = 0.5$. The other simulation settings are as described in \tref{tab:logisticDirectionalExpDecayCov_MDS_PCA_beta1}.}
\label{tab:logisticDirectionalExpDecayCov_MDS_PCA_beta0.5}
\end{center}
\vspace{4mm}

\begin{center}
\begin{tabular}{ccccccc}
\toprule
\multicolumn{7}{c}{$\beta = 0.25$} \\
\multirow{2}{*}{$\kappa$}  & \multirow{2}{*}{$\alpha_1$}  & \multirow{2}{*}{$\displaystyle\frac{\text{RMSE \localizationAlgorithm{}}}{\text{RMSE PCA}}$}	&	\multirow{2}{*}{RMSE PCA} & \multirow{2}{*}{RMSE \localizationAlgorithm{}}	& \multirow{2}{*}{$\tau$}  & \multirow{2}{1in}{Proportion of distances used}	\\[0.5cm]
\midrule
\multirow{5}{*}{0.1}	&	1	&	1.016	&	0.0433	&	0.0441	&	0.2384	&	100.0\% 	\\
			&	2	&	\bf{0.979}	&	0.0523	&	0.0512	&	0.1943	&	17.5\% 	\\
			&	4	&	\bf{0.559}	&	0.0511	&	0.0286	&	0.2194	&	14.4\% 	\\
			&	8	&	\bf{0.102}	&	0.2807	&	0.0286	&	0.2248	&	8.3\% 	\\
			&	16	&	\bf{0.115}	&	0.3723	&	0.0428	&	0.2618	&	8.7\% 	\\
\midrule
\multirow{5}{*}{1}	&	1	&	1.013	&	0.0425	&	0.0431	&	0.2401	&	100.0\% 	\\
			&	2	&	\bf{0.865}	&	0.0613	&	0.0531	&	0.1917	&	17.1\% 	\\
			&	4	&	\bf{0.553}	&	0.0740	&	0.0409	&	0.1964	&	9.9\% 	\\
			&	8	&	\bf{0.140}	&	0.2284	&	0.0321	&	0.2180	&	7.8\% 	\\
			&	16	&	\bf{0.117}	&	0.2678	&	0.0313	&	0.2596	&	8.3\% 	\\
\midrule
\multirow{5}{*}{10}	&	1	&	\bf{0.303}	&	0.4099	&	0.1243	&	0.1441	&	18.7\% 	\\
			&	2	&	\bf{0.193}	&	0.4172	&	0.0806	&	0.1650	&	12.7\% 	\\
			&	4	&	\bf{0.318}	&	0.4083	&	0.1300	&	0.1447	&	4.9\% 	\\
			&	8	&	\bf{0.192}	&	0.4163	&	0.0798	&	0.2159	&	8.0\% 	\\
			&	16	&	\bf{0.313}	&	0.3580	&	0.1120	&	0.1921	&	2.3\% 	\\
\bottomrule
\end{tabular}
\captionof{table}{{\bf Directional covariance decay model}: The geographic locations $\bfz_i$ of the individuals are simulated by sampling each coordinate according to a Beta($\beta, \beta$) distribution from the unit square, with $\beta = 0.25$. The other simulation settings are as described in \tref{tab:logisticDirectionalExpDecayCov_MDS_PCA_beta1}.}
\label{tab:logisticDirectionalExpDecayCov_MDS_PCA_beta0.25}
\end{center}

\begin{center}
\begin{tabular}{ccccccc}
\toprule
\multicolumn{7}{c}{$\beta = 2$} \\
\multirow{2}{*}{$\kappa$}  & \multirow{2}{*}{$\alpha_1$}  & \multirow{2}{*}{$\displaystyle\frac{\text{RMSE \localizationAlgorithm{}}}{\text{RMSE PCA}}$}	&	\multirow{2}{*}{RMSE PCA} & \multirow{2}{*}{RMSE \localizationAlgorithm{}}	& \multirow{2}{*}{$\tau$}  & \multirow{2}{1in}{Proportion of distances used}	\\[0.5cm]
\midrule
\multirow{5}{*}{0.1}	&	1	&	1.025	&	0.0627	&	0.0643	&	0.2286	&	100.0\% 	\\
			&	2	&	1.022	&	0.0780	&	0.0797	&	0.2732	&	100.0\% 	\\
			&	4	&	1.014	&	0.0969	&	0.0983	&	0.3022	&	100.0\% 	\\
			&	8	&	\bf{0.495}	&	0.1382	&	0.0684	&	0.2456	&	6.4\% 	\\
			&	16	&	\bf{0.222}	&	0.1574	&	0.0349	&	0.2571	&	6.0\% 	\\
\midrule
\multirow{5}{*}{1}	&	1	&	1.025	&	0.0676	&	0.0693	&	0.2256	&	100.0\% 	\\
			&	2	&	1.022	&	0.0750	&	0.0766	&	0.2735	&	100.0\% 	\\
			&	4	&	1.014	&	0.0968	&	0.0982	&	0.3118	&	100.0\% 	\\
			&	8	&	\bf{0.521}	&	0.1350	&	0.0703	&	0.2626	&	11.6\% 	\\
			&	16	&	\bf{0.166}	&	0.2412	&	0.0400	&	0.2519	&	5.9\% 	\\
\midrule
\multirow{5}{*}{10}	&	1	&	\bf{0.487}	&	0.2297	&	0.1119	&	0.1886	&	22.2\% 	\\
			&	2	&	\bf{0.533}	&	0.2298	&	0.1226	&	0.2354	&	24.8\% 	\\
			&	4	&	\bf{0.474}	&	0.2372	&	0.1124	&	0.2542	&	15.7\% 	\\
			&	8	&	\bf{0.372}	&	0.2453	&	0.0913	&	0.2376	&	7.3\% 	\\
			&	16	&	\bf{0.295}	&	0.2585	&	0.0763	&	0.2022	&	1.0\% 	\\
\bottomrule
\end{tabular}
\captionof{table}{{\bf Directional covariance decay model}: The geographic locations $\bfz_i$ of the individuals are simulated by sampling each coordinate according to a Beta($\beta, \beta$) distribution from the unit square, with $\beta = 2$. The other simulation settings are as described in \tref{tab:logisticDirectionalExpDecayCov_MDS_PCA_beta1}.}
\label{tab:logisticDirectionalExpDecayCov_MDS_PCA_beta2}
\end{center}
\vspace{4mm}

\begin{center}
\begin{tabular}{ccccccc}
\toprule
\multicolumn{7}{c}{$\beta = 4$} \\
\multirow{2}{*}{$\kappa$}  & \multirow{2}{*}{$\alpha_1$}  & \multirow{2}{*}{$\displaystyle\frac{\text{RMSE \localizationAlgorithm{}}}{\text{RMSE PCA}}$}	&	\multirow{2}{*}{RMSE PCA} & \multirow{2}{*}{RMSE \localizationAlgorithm{}}	& \multirow{2}{*}{$\tau$}  & \multirow{2}{1in}{Proportion of distances used}	\\[0.5cm]
\midrule
\multirow{5}{*}{0.1}	&	1	&	1.024	&	0.0506	&	0.0518	&	0.2161	&	100.0\% 	\\
			&	2	&	1.023	&	0.0625	&	0.0640	&	0.2642	&	100.0\% 	\\
			&	4	&	1.016	&	0.0765	&	0.0778	&	0.2991	&	100.0\% 	\\
			&	8	&	\bf{0.750}	&	0.1014	&	0.0761	&	0.2512	&	12.1\% 	\\
			&	16	&	\bf{0.363}	&	0.1285	&	0.0467	&	0.2577	&	6.9\% 	\\
\midrule
\multirow{5}{*}{1}	&	1	&	1.023	&	0.0532	&	0.0544	&	0.2188	&	100.0\% 	\\
			&	2	&	1.022	&	0.0603	&	0.0616	&	0.2646	&	100.0\% 	\\
			&	4	&	1.015	&	0.0776	&	0.0788	&	0.2983	&	100.0\% 	\\
			&	8	&	\bf{0.751}	&	0.1018	&	0.0764	&	0.2480	&	12.2\% 	\\
			&	16	&	\bf{0.254}	&	0.1879	&	0.0477	&	0.2551	&	7.2\% 	\\
\midrule
\multirow{5}{*}{10}	&	1	&	\bf{0.553}	&	0.1702	&	0.0941	&	0.1944	&	23.3\% 	\\
			&	2	&	\bf{0.534}	&	0.1743	&	0.0930	&	0.2408	&	25.8\% 	\\
			&	4	&	\bf{0.540}	&	0.1757	&	0.0948	&	0.2721	&	23.2\% 	\\
			&	8	&	\bf{0.491}	&	0.1840	&	0.0903	&	0.2726	&	9.1\% 	\\
			&	16	&	\bf{0.404}	&	0.1978	&	0.0800	&	0.2298	&	2.1\% 	\\
\bottomrule
\end{tabular}
\captionof{table}{{\bf Directional covariance decay model}: The geographic locations $\bfz_i$ of the individuals are simulated by sampling each coordinate according to a Beta($\beta, \beta$) distribution from the unit square, with $\beta = 4$. The other simulation settings are as described in \tref{tab:logisticDirectionalExpDecayCov_MDS_PCA_beta1}.}
\label{tab:logisticDirectionalExpDecayCov_MDS_PCA_beta4}
\end{center}

\subsection{Evaluation on the real datasets}\label{sec:evaluation_localization_real_data}
Each of the three real datasets we analyzed (\HO{}, \GLOBETROTTER{}, and \POPRES{}) contained latitude and longitude coordinates at the subpopulation level. When applying \localizationAlgorithm{} to each dataset, we used the genotype samples from a random subset of 20\% of the subpopulations along with their true sampling locations in order to pick the local genetic distance threshold $\tau$ and to rescale the inferred locations to latitude-longitude coordinates. Recall from \eqref{eq:eta_z_relation} that \localizationAlgorithm{} infers locations up to translation and the unknown $2 \times 2$ invertible matrix $J$. Similarly, when applying PCA to each dataset, we used the same random subset of subpopulations to estimate the translation and rescaling of the principal components.\footnote{From equation \eqref{eq:PCA1}, we see that we should only have to transform the PC coordinates by a translation and a $2 \times 2$ rotation/reflection matrix. However, we provide PCA the same degrees of freedom in rescaling locations to latitude-longitude coordinates as we do for \localizationAlgorithm{}.}

In particular, let $\cD$ denote the subpopulations in the full dataset and $\cT \subset \cD$ the random subset of training subpopulations, where $|\cD| = m$ and $|\cT| = m_0 = \lceil m/5 \rceil$. Let $S_j \subset \{ 1, \ldots, n \}$ be the set of individuals in subpopulation $j$ in the sample, $1 \leq j \leq m$.
Letting the inferred coordinates from \localizationAlgorithm{} or PCA be denoted by $\hat{\bfz}_i$ for individual $i$, the mean inferred coordinates $\hat{\bfy}_j$ for each subpopulation $j$ is computed by averaging the inferred coordinates over the individuals in the subpopulation,
\begin{align*}
   \hat{\bfy}_j &= \frac{1}{|S_j|} \sum_{i \in S_j} \hat{\bfz}_i \,.
\end{align*}
If the true latitude-longitude sampling coordinates of population $j$ is denoted by $\bfy_j$, we are interested in estimating the $2 \times 2$ coordinate rescaling matrix $A^*$ and translation vector $\bfb^*$ which minimizes the following objective function measuring reconstruction error,
\begin{align}\label{eq:optimize_rescaling_coordinates}
  A^*, \bfb^* &= \argmin_{\substack{A \in \reals^{2\times2}\\ \bfb \in \reals^{2}}} \,\sum_{j \in \cT} \frac{|S_j|}{n} \| \bfy_j - (A \hat{\bfy}_j + \bfb) \|^2\,.
\end{align}
We can solve \eqref{eq:optimize_rescaling_coordinates}, for example, via the weighted least squares estimator for linear regression.
To choose the value of the genetic distance parameter $\tau$ when applying \localizationAlgorithm{}, we do a grid search over $\tau$ using the leave-one-out cross-validation error over the subpopulations in the training set of subpopulations $\cT$.
The RMSE of \localizationAlgorithm{} and PCA on the full dataset is computed by transforming the inferred subpopulation coordinates using the estimated translation and rescaling in \eqref{eq:optimize_rescaling_coordinates},
\begin{align}
\text{RMSE} = \sqrt{\sum_{j \in \cD} \frac{|S_j|}{n} \| \bfy_j - (A^* \hat{\bfy}_j + \bfb^*) \|^2}\,.  \label{eq:rmse_whole_data}
\end{align}
Since the reconstruction error in \eqref{eq:rmse_whole_data} depends on the training subset $\cT$, we performed the above procedure with 100 randomly drawn subsets of training populations and report some statistics of the reconstruction RMSE in \tref{tab:reconstruction_rmse}.

\begin{table}
\begin{center}
\begin{tabular}{ccccccc}
\toprule
\multirow{2}{*}{Dataset}  &   \multicolumn{2}{c}{$\displaystyle\frac{\text{RMSE \localizationAlgorithm{}}}{\text{RMSE PCA}}$}                &   \multicolumn{2}{c}{RMSE PCA}   &   \multicolumn{2}{c}{RMSE \localizationAlgorithm{}}  \\
\vspace{-3mm} &&&&&& \\
                          &   Median   &   90\% CI           &   Median   &   90\% CI           &   Median         &   90\% CI   \\
\midrule
\HO{}                     &   0.69      &  (0.28, 0.89)         &   \ang{16.5}  &   (\ang{14.1}, \ang{44.2})          &   \ang{11.5}     &   (~\,\ang{9.7}, \ang{14.9})   \\
\GLOBETROTTER{}           &   0.90      &  (0.65, 1.23)         &   \ang{15.2}  &   (\ang{13.0}, \ang{26.1})          &   \ang{13.5}     &   (\ang{11.5}, \ang{25.8})  \\
\POPRES{}                 &   0.44      &  (0.20, 1.09)         &   \ang{18.0}  &   (~\,\ang{8.6}, \ang{41.4})        &   ~\,\ang{7.5}      &   (~\,\ang{5.3}, \ang{24.4})   \\
\POPRES{} ($\text{MAF} \geq 10\%$)               &   1.09      &  (0.60, 3.18)         &   \ang{5.2}  &   (~\,\ang{3.6}, \ang{16.2})        &   ~\,\ang{6.0}      &   (~\,\ang{4.1}, \ang{18.2})   \\
\bottomrule
\end{tabular}
\captionof{table}{{\bf Spatial assignment accuracy of PCA and \localizationAlgorithm{} on the \HO{}, \GLOBETROTTER{}, and \POPRES{} datasets.} The reconstruction RMSE statistics are based on 100 random subsamples of 20\% of the subpopulations in the full datasets. Using the known sampling locations of the subpopulations in the training sample, we rescaled the inferred coordinates from PCA and \localizationAlgorithm{} in order to learn latitude-longitude coordinates for each subpopulation in the test set. For the \POPRES{} dataset, we found that both PCA and \localizationAlgorithm{} performed significantly better at spatial reconstruction if we used only the SNPs with minor allele frequency at least $10\%$ (also see \fref{fig:POPRES_heavyLDpruned_commonSNPs}).
}
\label{tab:reconstruction_rmse}
\end{center}
\end{table}

\begin{figure}
\begin{subfigure}[]{0.5\textwidth}
\centering
\caption{}\label{fig:GLOBETROTTER_true_locations}
\includegraphics[width=0.9\textwidth]{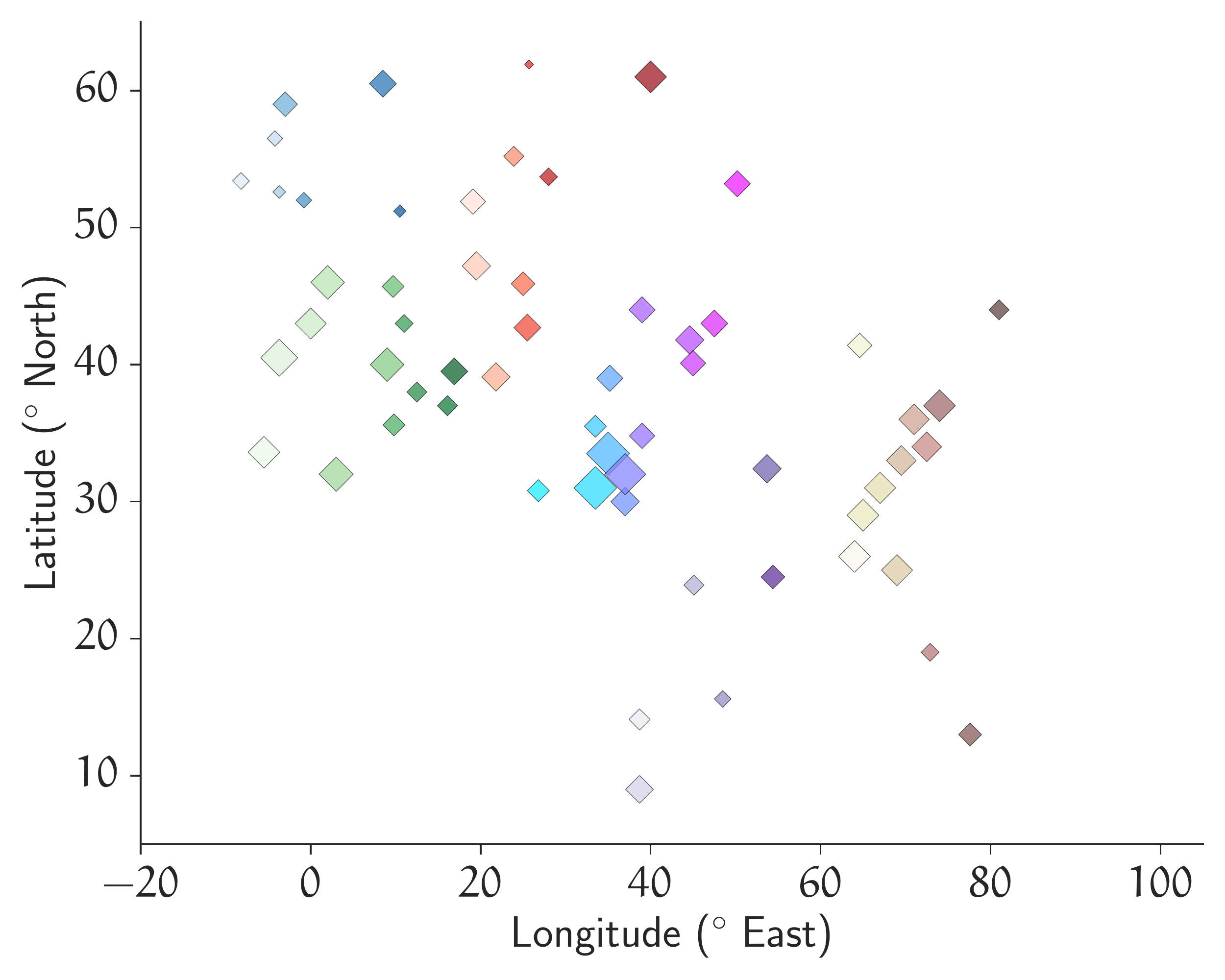}
\end{subfigure}
\begin{subfigure}[]{0.5\textwidth}
\centering
\caption{}\label{fig:GLOBETROTTER_MDS}
\includegraphics[width=0.9\textwidth]{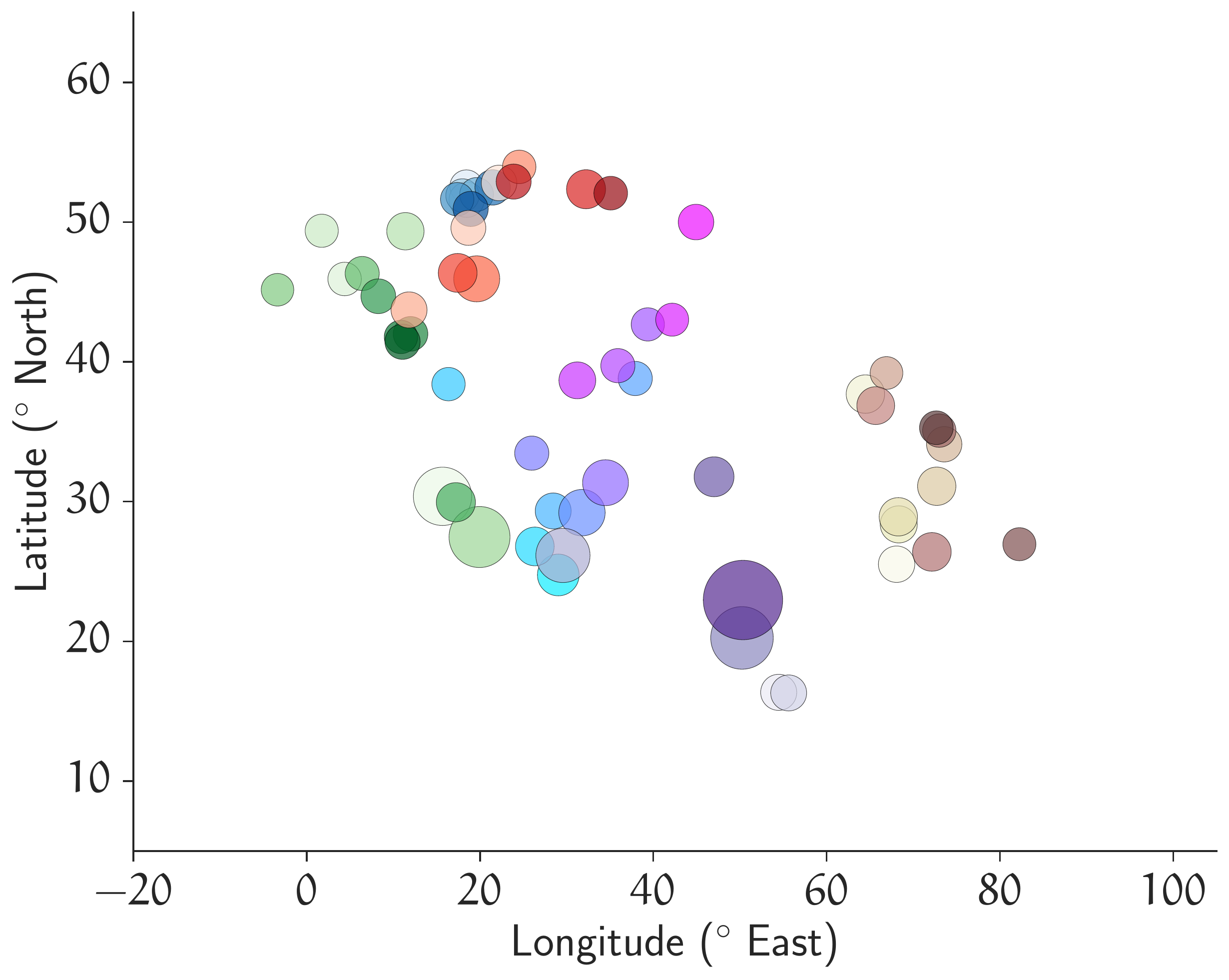}
\end{subfigure}

\begin{subfigure}[]{0.5\textwidth}
\centering
\caption{}\label{fig:GLOBETROTTER_PCA}
\includegraphics[width=0.9\textwidth]{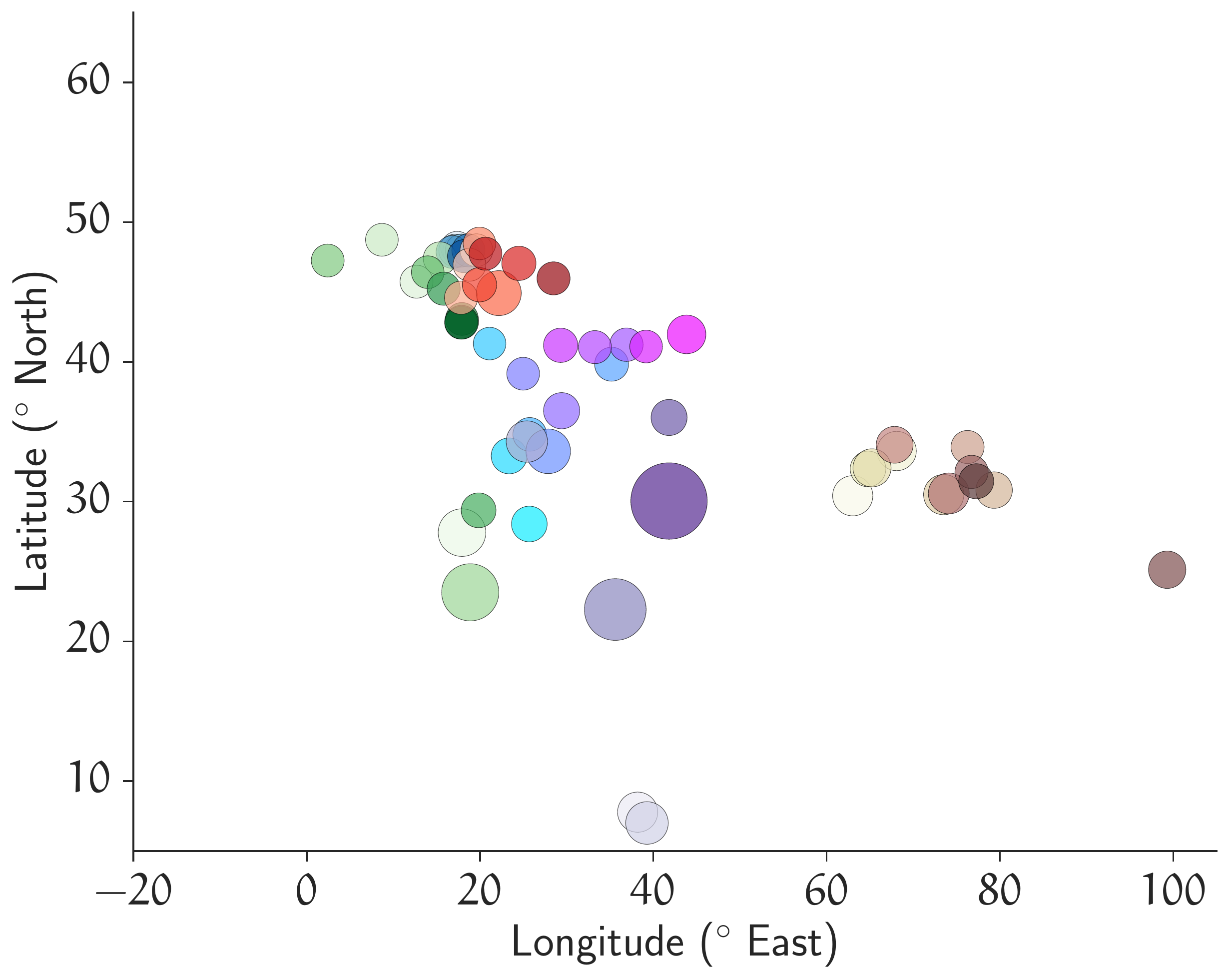}
\end{subfigure}
\begin{subfigure}[]{0.5\textwidth}
\centering
\caption{}\label{fig:GLOBETROTTER_legends}
\includegraphics[width=0.9\textwidth, trim=0 0 0 2.5cm, clip]{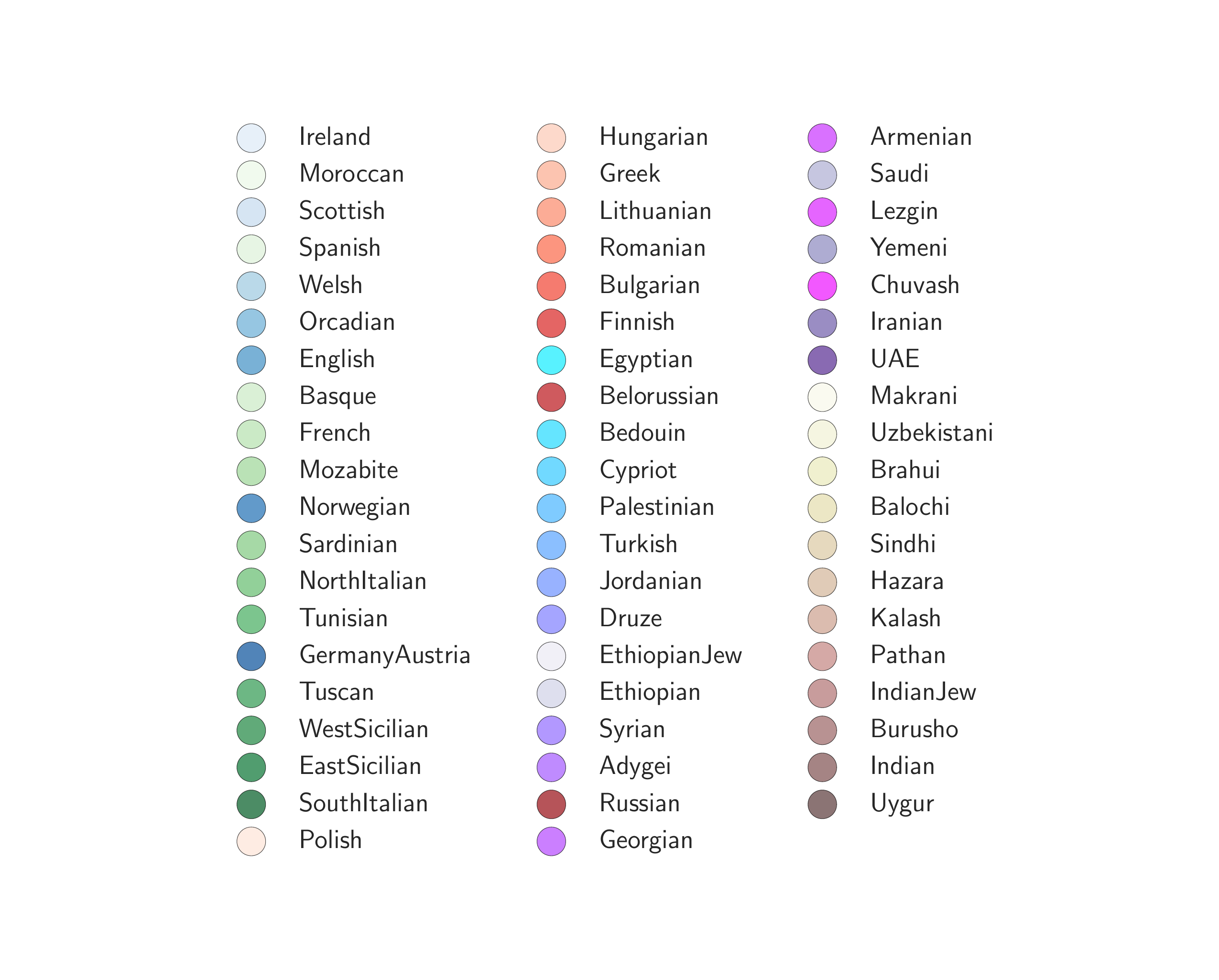}
\end{subfigure}
\caption{PCA and \localizationAlgorithm{} visualization of the populations in the \GLOBETROTTER{} dataset. We analyzed 59 populations from Europe, the Middle East, North and East Africa, and Western, Central and South Asia. \subref{fig:GLOBETROTTER_true_locations} True sampling locations, \subref{fig:GLOBETROTTER_MDS} \localizationAlgorithm{} reconstructed locations, \subref{fig:GLOBETROTTER_PCA} PCA reconstructed locations, and \subref{fig:GLOBETROTTER_legends} population legends. The diamonds are placed at the sampling locations for each population, while the circles are placed at the mean inferred location of the samples in each population. PCA tends to localize individuals from Southern Europe, North Africa, and the Middle East closer together, while \localizationAlgorithm{} is better at separating them.
}
\label{fig:GLOBETROTTER}
\end{figure}

\subsection{Application to the POPRES dataset}\label{sec:popres_localization}
The \POPRES{} dataset is an aggregation of 5{,}918 individuals with self-reported ancestry from several studies \citep{nelson:2008,preisig:2009,kooner:2008}. The dataset we analyzed contained individuals genotyped at 457{,}297 SNPs. We filtered SNPs deviating from Hardy-Weinberg equilibrium and thinned SNPs with linkage disequilibrium $r^2$ greater than 10\% in sliding windows of 50 SNPs. This left us with a dataset of 77{,}678 SNPs. We selected 1{,}217 individuals from Europe who reported all four grandparents belonging to the same country in Europe and whose reported primary language matched the country of origin of their grandparents. This filtering was performed to avoid picking individuals that might be recently admixed, and is similar to the filters applied in previous works analyzing this dataset \citep{novembre:2008,yang:2012}. Using the true sampling locations of 20\% of the subpopulations to assign spatial coordinates to a test set of subpopulations, \localizationAlgorithm{} has median RMSE \ang{7.47} while PCA has median RMSE of \ang{17.97}, where the assignment was performed using 100 random training/test splits of the full dataset.
We also analyzed the data after discarding SNPs with minor allele frequency below 10\%. In this setting, the spatial assignment accuracy of \localizationAlgorithm{} and PCA are quite similar, with median RMSE of \ang{5.99} and \ang{5.15}, respectively. The visualizations produced by PCA and \localizationAlgorithm{} are very similar (\fref{fig:POPRES_heavyLDpruned_commonSNPs}), and closely recapitulate the geography of Europe as has been previously observed by \citet{novembre:2008}. 

\begin{figure}[H]
\begin{subfigure}[]{0.5\textwidth}
\centering
\caption{}\label{fig:POPRES_true_locations}
\includegraphics[width=0.9\textwidth]{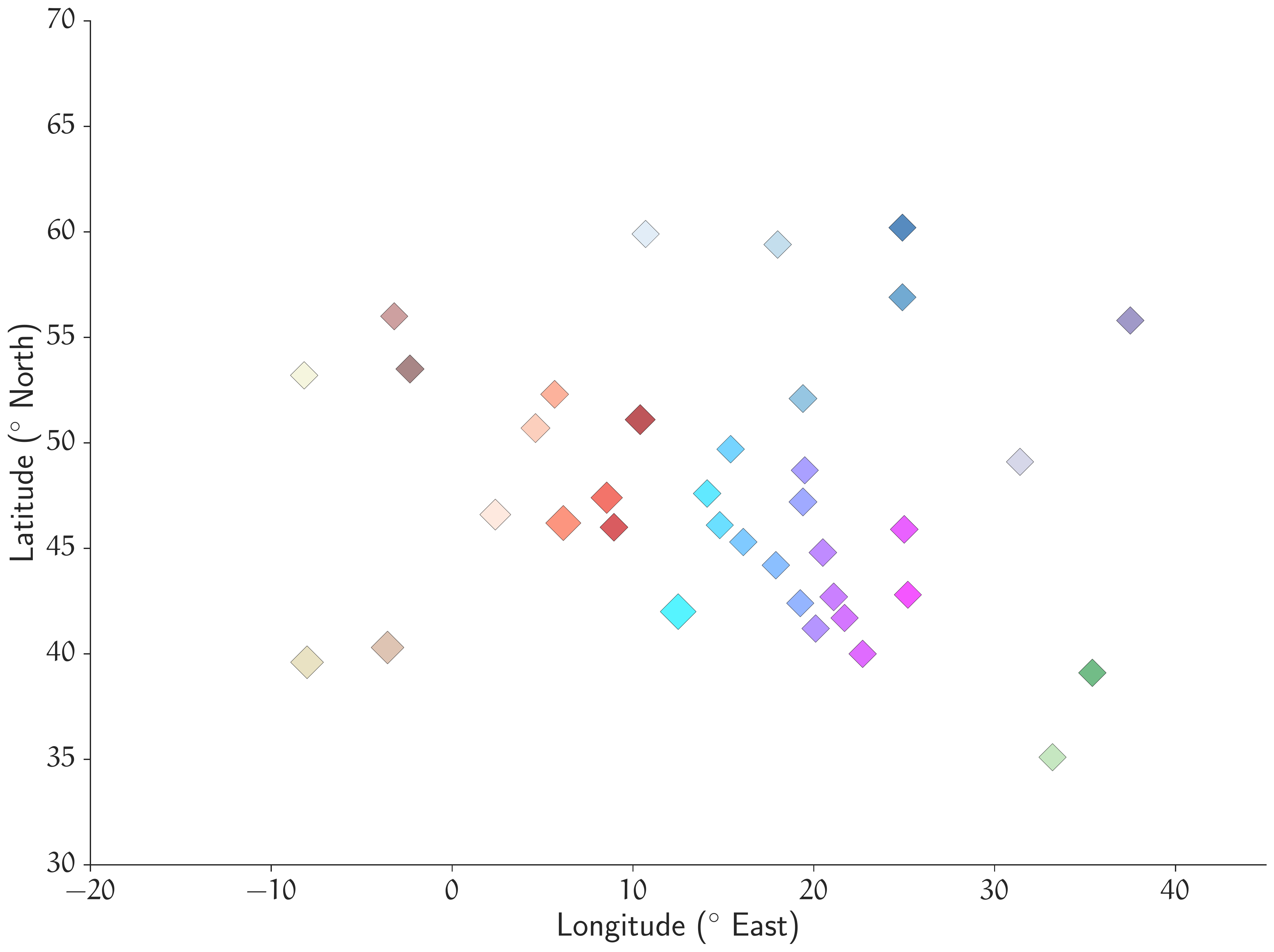}
\end{subfigure}
\begin{subfigure}[]{0.5\textwidth}
\centering
\caption{}\label{fig:POPRES_MDS_heavyLDpruned_commonSNPs}
\includegraphics[width=0.9\textwidth]{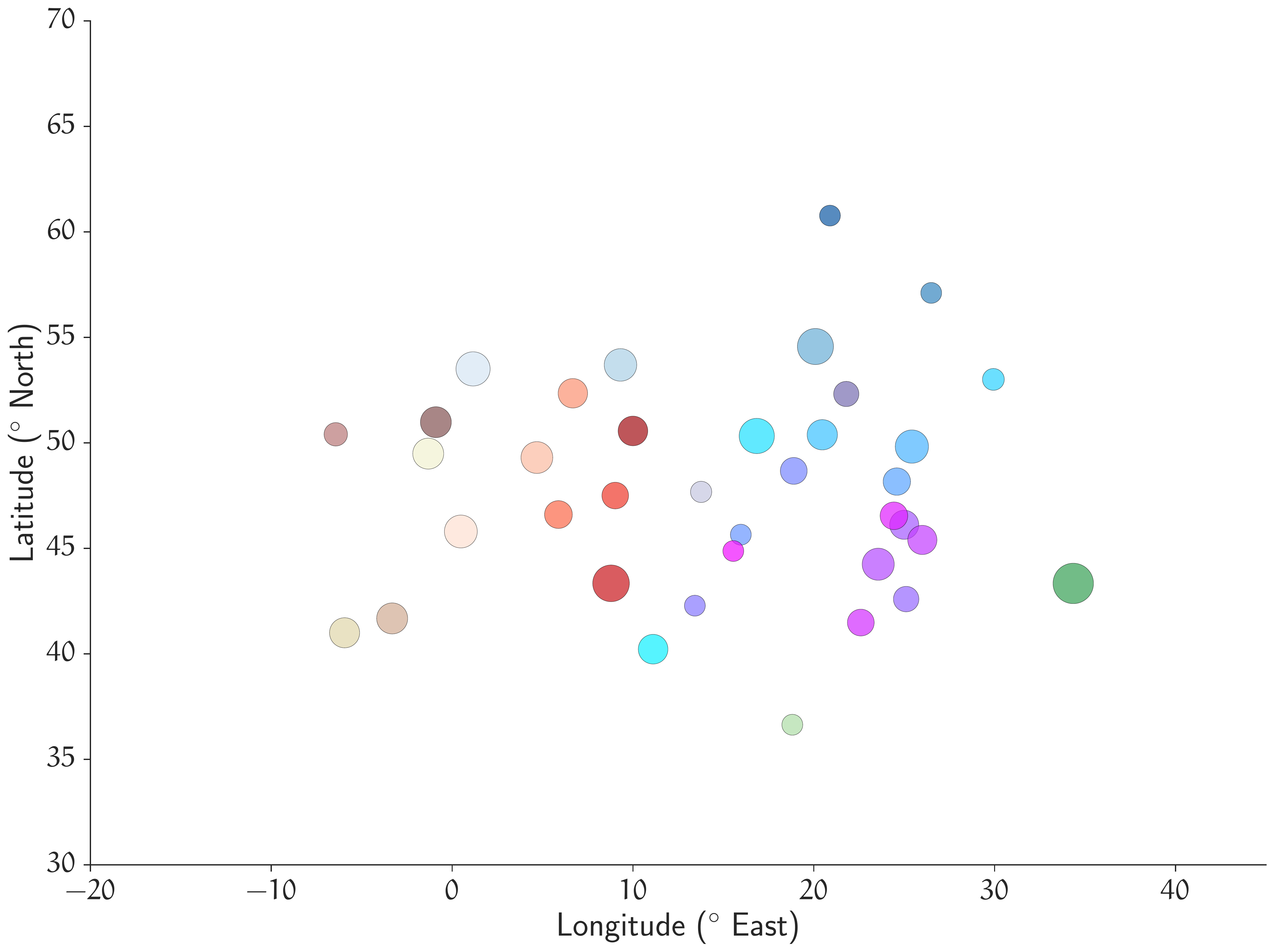}
\end{subfigure}

\begin{subfigure}[]{0.5\textwidth}
\centering
\caption{}\label{fig:POPRES_PCA_heavyLDpruned_commonSNPs}
\includegraphics[width=0.9\textwidth]{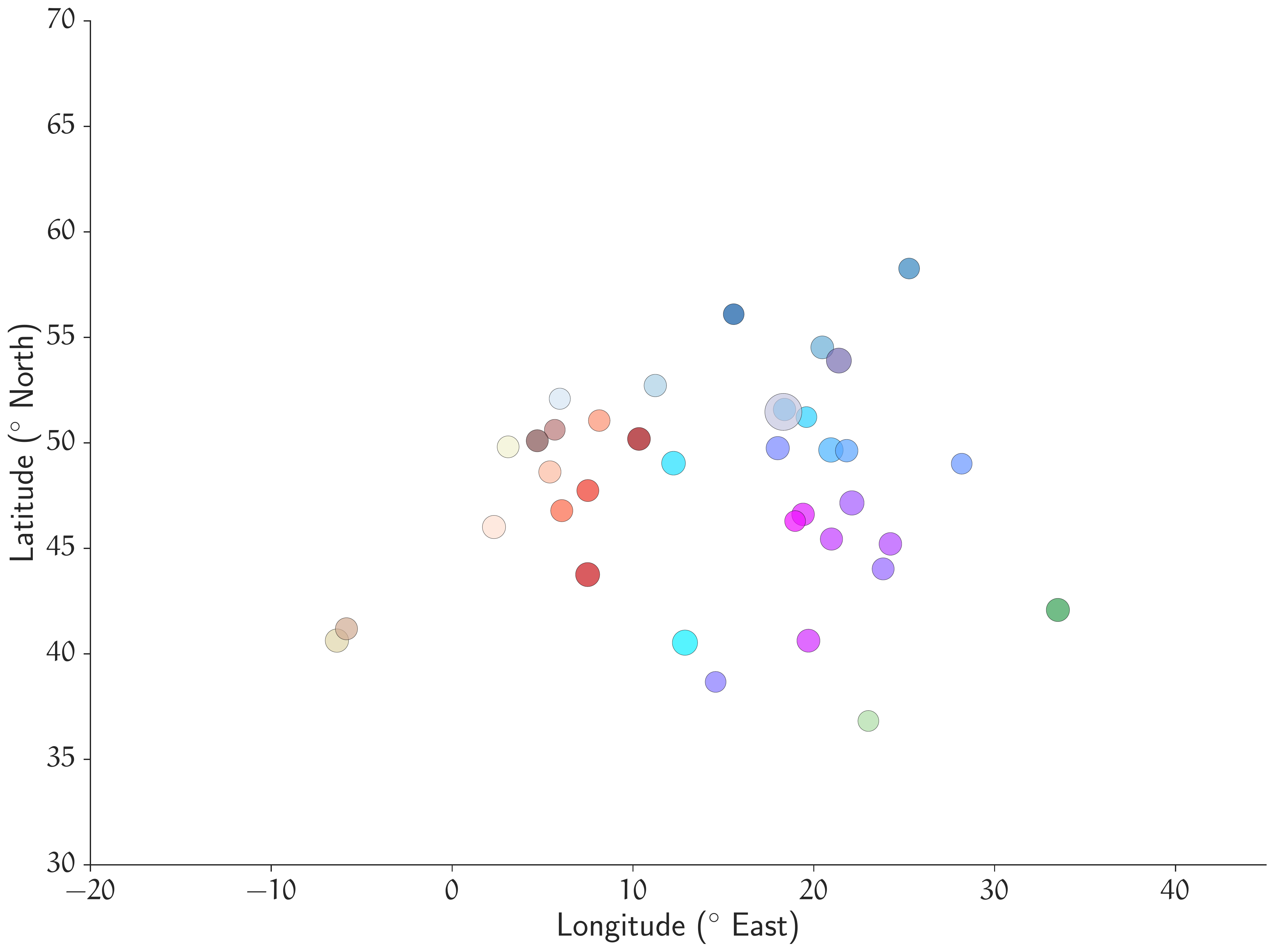}
\end{subfigure}
\begin{subfigure}[]{0.5\textwidth}
\centering
\caption{}\label{fig:POPRES_legends}
\includegraphics[width=1.8\textwidth, trim=4in 3.5in 0 4in, clip]{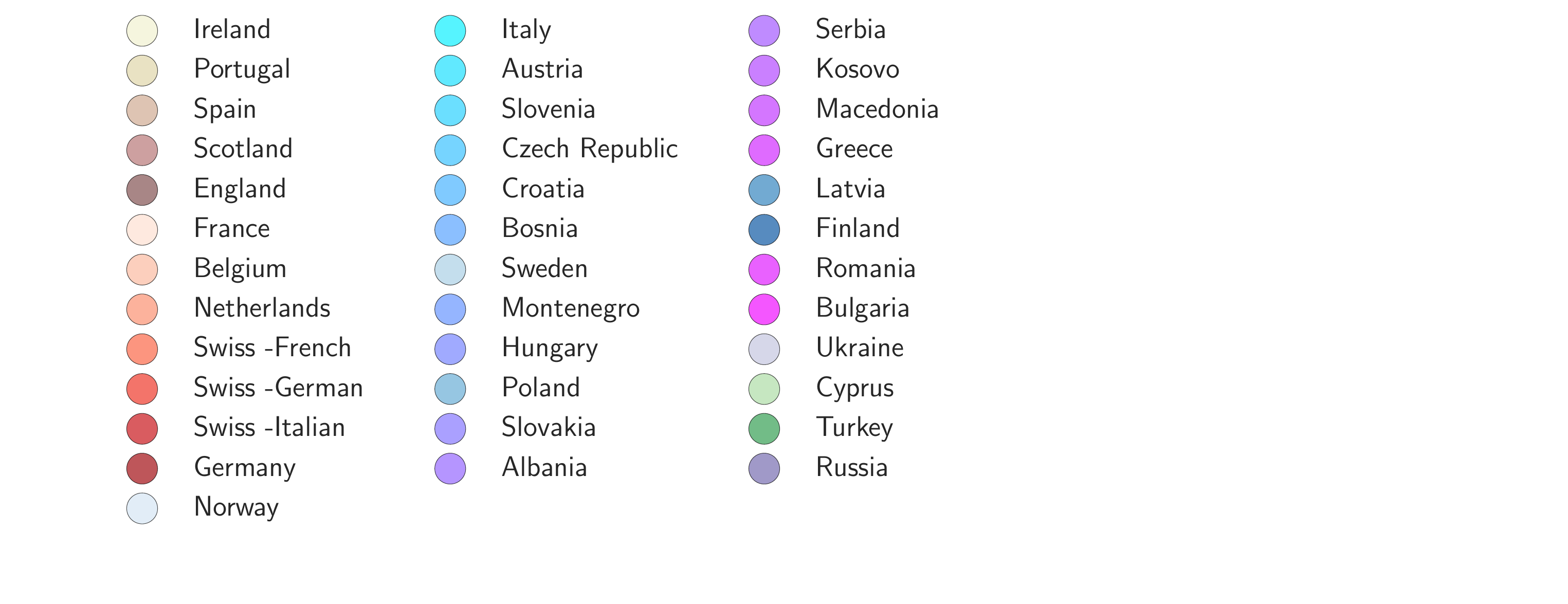}
\end{subfigure}
\caption{{\bf PCA and \localizationAlgorithm{} visualization of the POPRES dataset.} \subref{fig:POPRES_true_locations} True sampling locations, \subref{fig:POPRES_MDS_heavyLDpruned_commonSNPs} \localizationAlgorithm{} reconstructed locations, \subref{fig:POPRES_PCA_heavyLDpruned_commonSNPs} PCA reconstructed locations, and \subref{fig:POPRES_legends} population legends.
}
\label{fig:POPRES_heavyLDpruned_commonSNPs}
\end{figure}

\newpage
\section{Association testing procedure}\label{sec:association_testing_procedure}
We present a statistical framework for testing associations between genotype and trait (either binary or quantitative) in the presence of population structure. Ancestry, on account of influencing genetic variation, can induce correlations between genotypes. Furthermore, ancestry is also correlated with the phenotype, due to varying trait prevalence with geography, ancestry-biased sampling, etc. In such cases, association tests can be confounded by population stratification, and the genotype and trait will be statistically dependent even when there is no genetic basis for the trait.

We correct for population stratification by conditioning the genotype on the confounding variable, which in our setting are the ancestry coordinates $\bfz_i$. As shown schematically in \citet[Figure 1]{song:2015}, this conditioning removes the statistical dependence between genotype $x_{i\ell}$ and confounder $\bfz_i$. To be more concrete, we consider a retrospective model to describe the effect of population structure and the trait on the genotype. For a particular SNP $\ell$ and an individual $i$ with trait $y_i$, genotype $x_{i\ell}$ is generated according to the following model,
\begin{eqnarray}\label{eq:trait-model}
&x_{i\ell} \mid y_i,\bfz_i \sim \Bino(2,\theta_{i\ell}) \notag \\
&\theta_{i\ell} = \dfrac{\kappa_\ell R_\ell^{y_i}q_\ell(\bfz_i)}{1-q_\ell(\bfz_i) + \kappa_\ell R_\ell^{y_i} q_{\ell}(\bfz_i)}\,,
\end{eqnarray}
where $R_\ell$ is the genetic risk factor of the SNP $\ell$ on the trait, with the effect of population structure encoded in the allele frequency $q_{\ell}(\bfz_i)$. 
Note that for a binary trait $y_i\in \{0,1\}$, by setting $\kappa_\ell=1$ for all SNPs, the model reduces to the one studied by \citet{price:2006}. 
Further, with the change of variables $a_\ell = \log \kappa_\ell$ and $b_\ell = \log R_\ell$, the model \eqref{eq:trait-model} can be rewritten as,
\begin{eqnarray}
&x_{i\ell} \mid y_i,\bfz_i \sim \Bino(2,\theta_{i\ell}) \notag \\
&\theta_{i\ell} = \logit^{-1}(a_\ell+b_\ell y_i + \logit(q_\ell(\bfz_i)))\,,
\end{eqnarray}
with $\logit(x) = \log(x/(1-x))$ for $0<x<1$.
This inverse regression model has been put forth by \citet{song:2015} as a means of correcting for population structure and environmental confounders under fairly general assumptions.
Under model \eqref{eq:trait-model}, the genotypes $x_{i\ell}$ depend on the confounder $\bfz_i$ only through the allele frequency $q_{\ell}(\bfz_i)$.
Hence, by conditioning on the allele frequencies $q_\ell(\bfz_i)$, the genotype 
becomes independent of ancestry location, $x_{i\ell} \independent \bfz_i \mid q_\ell(\bfz_i)$. This conditioning also allows us to ignore the dependency between the trait $y_i$ and ancestry $\bfz_i$ in our model. 
In practice, we do not know the underlying allele frequencies $q_{\ell}(\bfz_i)$, and must estimate it by making some assumptions which we describe shortly.

Given a particular trait of interest, a SNP $\ell$ is considered non-associated if $R_{\ell}= 1$, and considered associated otherwise. We are thus interested in testing between the following null and alternate hypotheses,
\begin{align}
H_{0,\ell}:&\; R_{\ell} = 1\,,  \label{eq:association_null_hypothesis}\\
H_{A,\ell}:&\; R_{\ell} \neq 1\,.  \label{eq:association_alternate_hypothesis}
\end{align}

The retrospective model in \eqref{eq:trait-model} enjoys both computational and statistical advantages over the prospective model (main text, equations \eqref{eq:forward_model_quantitative} and \eqref{eq:forward_model_binary}). 
From a statistical point of view, \eqref{eq:trait-model} allows unbiased testing under fairly general assumptions about the phenotype distribution and about the ancestry and environmental confounding variables (see Theorem 1 of \citet{song:2015}). From a computational viewpoint, the association tests in \eqref{eq:trait-model} for different SNPs can be performed separately. This is in contrast to the prospective model where the trait is modeled as a linear combination of genotypes, environmental effects and random noise variation, and where a principled testing procedure might require joint estimation over all SNPs. The retrospective model thus allows conceptually simpler, statistically valid and much more efficient and parallelizable association testing algorithms than procedures based on prospective models.

We next describe our testing procedure which consists of three steps: $(1)$ estimation of ancestry coordinates $\hat{\bfz}_i$ using \localizationAlgorithm{}; $(2)$ estimation of spatial allele frequencies $q_{\ell}(\hat{\bfz}_i)$; and $(3)$ estimation of the risk factor $R_\ell$ and intercept term $\kappa_\ell$. Below, we elaborate on each of these steps.

\bigskip
\noindent{\bf Step $(1)$: estimation of ancestry coordinates $\hat{\bfz}_i$.}
We use our localization algorithm \localizationAlgorithm{} on the genotype matrix $X$ as described in \sref{sec:localization_algorithm} to estimate the ancestry coordinates $\hat{\bfz}_i$ for each individual in the sample. The threshold parameter $\tau$ for local distances used in our algorithm \localizationAlgorithm{} can be chosen using the strategies described in \sref{sec:choosing_tau}. As we will see in the next step, we do not need to rescale the ancestry coordinates by the dilation matrix $J$ of \eqref{eq:eta_z_relation}.

\bigskip
\noindent{\bf Step $(2)$: estimation of $q_\ell(\hat{\bfz}_i)$.}
We start with a naive estimate of the allele frequencies and then apply a kernel to smooth the allele frequencies over space.
Using the genotype data, we compute the initial estimate $\xi_{i\ell}$ of the unknown allele frequency $q_{\ell}(\bfz_i)$ as,
\begin{align}
  \xi_{i\ell} = \frac{x_{il}}{2}. \label{eq:allele_frequency_crude_estimate}
\end{align}
We then refine these estimates by making the assumption that the allele frequency function $q_{\ell}(\bfz)$ vary smoothly over space $\bfz$. Without such an additional assumption on the allele frequency function, the estimation problem is not well defined.
We smooth the crude estimates of allele frequencies in \eqref{eq:allele_frequency_crude_estimate} via an exponential kernel interpolation to get estimates $\hat{q}_{i\ell}$ for $q_\ell(\bfz_i)$ as follows.
\begin{eqnarray}\label{eq:q_estimate}
\hat{q}_{i\ell} =  \frac{\sum_{j=1}^n \xi_{j\ell} \exp(-\frac{1}{2} \|H^{-1}(\hat{\bfz}_i-\hat{\bfz}_j)\|^2)}{\sum_{j=1}^n \exp(-\frac{1}{2} \|H^{-1}(\hat{\bfz}_i-\hat{\bfz}_j)\|^2)} \,,
\end{eqnarray}
where $H$ is a $2\times 2$ bandwidth matrix.
We use Scott's rule \citep{scott:1979} for selecting the bandwidth, where $H$ is chosen so that $H H^T = n^{-1/3} \Sigma$ and $\Sigma$ is the $2 \times 2$ covariance matrix of the estimated locations $\hat{\bfz}_i$.
Finally, we threshold $\hat{q}_{i\ell}$ by one to ensure that they are valid probabilities.
Note that with the choice of the bandwidth matrix in \eqref{eq:q_estimate}, we only need to estimate the ancestry coordinates $\hat{\bfz}$ up to the invertible $2 \times 2$ linear transformation $J$ in \eqref{eq:eta_z_relation}. 

\bigskip
\noindent{\bf Step $(3)$: estimation of $R_\ell$ and $\kappa_\ell$.}
Given an estimate of the allele frequencies $\hat{q}_{i\ell}$ from Step (2), we use Newton's method to estimate the risk factor $R_\ell$ (under the alternate hypothesis $H_{A,\ell}$) and intercept term $\kappa_\ell$ (under both the null and alternate hypotheses).

Given $y_i$ and $x_{i\ell}$, the log-likelihood of the model~\eqref{eq:trait-model}, in terms of $\theta_{i\ell}$, is given by
\begin{align}
\cL = \sum_{i=1}^n x_{i\ell} \log \theta_{i\ell} + (2-x_{i\ell}) \log(1-\theta_{i\ell})\,.
\end{align}
Define the entries of the scaled gradient of the log-likelihood function, $F_1(R_\ell, \kappa_\ell) \defeq R_\ell {\partial \cL}/{\partial R_\ell}$, and $F_2(\kappa_\ell, \kappa_\ell) \defeq \kappa_\ell {\partial \cL}/{\partial \kappa_\ell}$.

After some algebraic manipulation we get,
\begin{align}
F_1(R_\ell, \kappa_\ell) &=  \sum_{i=1}^n \Big(x_{i\ell} y_i - \frac{2y_i q_{i\ell} \kappa_\ell R_\ell^{y_i}}{1-q_{i\ell} + \kappa R_\ell^{y_i} q_{i\ell}}\Big)\,, \\
F_2(R_\ell, \kappa_\ell) &= \sum_{i=1}^n \Big(x_{i\ell} - \frac{2\kappa_\ell R_\ell^{y_i} q_{i\ell}}{1-q_{i\ell} + \kappa_\ell R_\ell^{y_i} q_{i\ell}} \Big)\,.
\end{align}
Under the alternate hypothesis $H_{A,\ell}$, we obtain the maximum-likelihood estimate of $R_\ell$ by simultaneously solving $F_1(R_\ell, \kappa_\ell) = F_2(R_\ell, \kappa_\ell) = 0$ using Newton's method. The initial value we used for Newton's method is $R_\ell = \kappa_\ell = 1$ which corresponds to a non-associated SNP. We similarly compute the maximum likelihood estimate for $\kappa_\ell$ under the null hypothesis, and compute a $p$-value from the log-likelihood ratio using the $\chi^2_1$ distribution.

\subsection{Association test simulations}\label{sec:association_test_simulations}
We simulated genotype data for $n=2{,}000$ individuals at $p=50{,}000$ SNPs sampled uniformly from the unit square using the isotropic and direction-dependent allele frequency covariance decay models described in \sref{sec:localization_simulations}. We used the same set of parameter combinations as we did for the ancestry localization simulations. We then generated quantitative phenotype data using the following linear model,
\begin{align*}
  y_i = \alpha + \sum_{\ell=1}^{p} \beta_\ell x_{i\ell} + \lambda_i + \varepsilon_i,
\end{align*}
where $y_i$ is the phenotype and $x_{i\ell} \in \{0, 1,2\}$ is the genotype at SNP $\ell$ for individual $i$, $\alpha$ is an intercept term, $\beta_\ell$ is the effect size of SNP $\ell$, and $\lambda_i$ and $\varepsilon_i$ are the ancestral and random environmental/noise contributions respectively to the phenotype of individual $i$. 
We randomly selected 10 SNPs to be causal, with the effect sizes $\beta_\ell$ drawn from a standard normal distribution. The ancestry contribution $\lambda_i$ to the phenotype was set to the first component ($x$-coordinate) of the sampling location $\bfz_i$, and the environmental/noise contribution was drawn from a standard normal distribution. We then rescaled the genotypic (i.e. $\sum_{\ell=1}^{p}\beta_\ell x_{i\ell}$), ancestry (i.e. $\lambda_i$), and environment/noise contributions ($\varepsilon_i$) so that they accounted for 20\%, 10\%, and 70\% respectively of the variance of the phenotype in the sample.

To perform our association test, we first inferred the ancestry coordinates using \localizationAlgorithm{}, where we used 20\% of the data points as anchors with known sampling locations.
To choose the threshold parameter $\tau$ for estimating local distances from the genetic distances, we swept over a range of $\tau$ which minimized the localization error over the subset of anchor data points. We then applied our association test described in the previous section with the inferred locations of all data points. Note that our association test only needs the ancestry coordinates up to the coordinate transformation matrix $J$ given in \eqref{eq:eta_z_relation}, and hence this matrix does not need to be estimated.

We find that for parameter combinations where \localizationAlgorithm{} has lower reconstruction RMSE in inferring ancestry coordinates compared to PCA, there is a concomitant increase in power to detect associations. Moreover, using the ancestry coordinates inferred from \localizationAlgorithm{} in our association test performs almost as well as an oracle that has the true ancestry coordinates (Figures \ref{fig:logisticExpDecayCov_n2000_p50k_pc10_beta1_gv0.20_lv0.10_nv0.70_alpha20.5}--\ref{fig:logisticDirectionalExpDecayCov_n2000_p50k_pc10_beta1_gv0.20_lv0.10_nv0.70_vkappa10} and Tables \ref{tab:association_logisticExpDecayCov_nv70_auc_fp_conditional_1e-3}--\ref{tab:association_logisticDirectionalExpDecayCov_nv70_auc_fp_conditional_1e-3}).

\newpage
\begin{center}
\begin{tabular}{cccccc}
\toprule
\multirow{2}{0.5cm}{$\alpha_2$} &	\multirow{2}{0.5cm}{{$\alpha_1$}} &	\multirow{2}{1.5cm}{\centering \associationTestingAlgorithm{}}	&	\multirow{2}{1.7cm}{\centering PCA coordinates}	&	\multirow{2}{1.7cm}{\centering True coordinates}	&	\multirow{2}{1.5cm}{\centering GCAT ($d=6$)} \\
&&&&&\\
\midrule
\multirow{5}{*}{0.5}	&	1	&	0.5937	&	0.5989	&	0.6031	&	0.5958	\\
			&	2	&	0.6266	&	0.6268	&	0.6313	&	0.6233	\\
			&	4	&	0.6556	&	0.6535	&	0.6535	&	0.6545	\\
			&	8	&	0.5759	&	0.5650	&	0.5788	&	0.5574	\\
			&	16	&	0.5867	&	0.5755	&	0.5840	&	0.5833	\\
\midrule
\multirow{5}{*}{1}	&	1	&	0.6139	&	0.6143	&	0.6205	&	0.6066	\\
			&	2	&	0.6046	&	0.6045	&	0.6152	&	0.6080	\\
			&	4	&	0.5881	&	0.5669	&	0.5929	&	0.5621	\\
			&	8	&	0.6100	&	0.5802	&	0.6104	&	0.5568	\\
			&	16	&	0.6345	&	0.6015	&	0.6354	&	0.5820	\\
\midrule
\multirow{5}{*}{1.5}	&	1	&	0.6345	&	0.6348	&	0.6369	&	0.6310	\\
			&	2	&	0.5521	&	0.5474	&	0.5613	&	0.5517	\\
			&	4	&	0.6122	&	0.5489	&	0.6117	&	0.5458	\\
			&	8	&	0.5841	&	0.5483	&	0.5842	&	0.5250	\\
			&	16	&	0.6043	&	0.5446	&	0.6072	&	0.5443	\\
\bottomrule
\end{tabular}
\captionof{table}{{\bf Association tests on data simulated under the isotropic covariance decay model.} Area under the ROC curve conditional on the FP rate being $\leq 10^{-3}$ for \associationTestingAlgorithm{} and for our allele frequency estimation procedure applied to ancestry coordinates inferred by PCA (column 3) and to the true ancestry coordinates (column 4). For comparison, we also show the performance of \GCAT{} using $d=6$ latent factors used for estimating the allele frequencies. The genotype data were simulated according to the isotropic covariance decay model (see \fref{fig:logisticExpDecayCovAlleleFreqFnExample} for an example) with the same parameter combinations as in \tref{tab:logisticExpDecayCov_MDS_PCA_beta1}. Each parameter combination row corresponds to 40 simulated datasets with $n = 2{,}000$ individuals and $p = 50{,}000$ SNPs, where 10 SNPs were chosen to have non-zero effects, with their effect sizes drawn from a standard normal distribution. The genotypic, ancestry, and environmental contribution to the phenotypic variance were set to 20\%, 10\% and 70\% respectively.}
\label{tab:association_logisticExpDecayCov_nv70_auc_fp_conditional_1e-3}
\end{center}

\newpage
\begin{center}
\begin{tabular}{cccccc}
\toprule
\multirow{2}{0.5cm}{$\kappa$} &	\multirow{2}{0.5cm}{{$\alpha_1$}} &	\multirow{2}{1.5cm}{\centering \associationTestingAlgorithm{}}	&	\multirow{2}{1.7cm}{\centering PCA coordinates}	&	\multirow{2}{1.7cm}{\centering True coordinates}	&	\multirow{2}{1.5cm}{\centering GCAT ($d=6$)} \\
&&&&&\\
\midrule
\multirow{5}{*}{0.1}	&	1	&	0.6309	&	0.6311	&	0.6318	&	0.6329	\\
			&	2	&	0.5968	&	0.5962	&	0.6044	&	0.5974	\\
			&	4	&	0.5551	&	0.5477	&	0.5664	&	0.5372	\\
			&	8	&	0.5867	&	0.5659	&	0.5863	&	0.5683	\\
			&	16	&	0.5931	&	0.5676	&	0.5920	&	0.5569	\\
\midrule
\multirow{5}{*}{1}	&	1	&	0.5920	&	0.5922	&	0.5972	&	0.5995	\\
			&	2	&	0.6150	&	0.6158	&	0.6263	&	0.6164	\\
			&	4	&	0.5678	&	0.5565	&	0.5800	&	0.5471	\\
			&	8	&	0.5970	&	0.5705	&	0.6015	&	0.5467	\\
			&	16	&	0.5873	&	0.5679	&	0.5907	&	0.5514	\\
\midrule
\multirow{5}{*}{10}	&	1	&	0.6130	&	0.6097	&	0.6144	&	0.6131	\\
			&	2	&	0.5677	&	0.5754	&	0.5732	&	0.5787	\\
			&	4	&	0.5825	&	0.5687	&	0.5863	&	0.5780	\\
			&	8	&	0.5858	&	0.5807	&	0.5918	&	0.5772	\\
			&	16	&	0.5929	&	0.5807	&	0.5920	&	0.5791	\\
\bottomrule
\end{tabular}
\captionof{table}{{\bf Association tests on data simulated under the directional covariance decay model.} Area under the ROC curve conditional on the FP rate being $\leq 10^{-3}$ for \associationTestingAlgorithm{} and for our allele frequency estimation procedure applied to ancestry coordinates inferred by PCA (column 3) and to the true ancestry coordinates (column 4). For comparison, we also show the performance of \GCAT{} using $d=6$ latent factors used for estimating the allele frequencies. The genotype data were simulated according to the directional covariance decay model (see \fref{fig:logisticDirectionalExpDecayCovAlleleFreqFnExample} for an example) with the same parameter combinations as in \tref{tab:logisticDirectionalExpDecayCov_MDS_PCA_beta1}. Each parameter combination row corresponds to 40 simulated datasets with $n = 2{,}000$ individuals and $p = 50{,}000$ SNPs, where 10 SNPs were chosen to have non-zero effects, with their effect sizes drawn from a standard normal distribution. The genotypic, ancestry, and environmental contribution to the phenotypic variance were set to 20\%, 10\% and 70\% respectively.}
\label{tab:association_logisticDirectionalExpDecayCov_nv70_auc_fp_conditional_1e-3}
\end{center}

\begin{figure}[h]
\begin{subfigure}[]{0.5\linewidth}
\centering
\caption{\hspace{0.75cm}$\alpha_1=1$}
\includegraphics[width=3.3in]{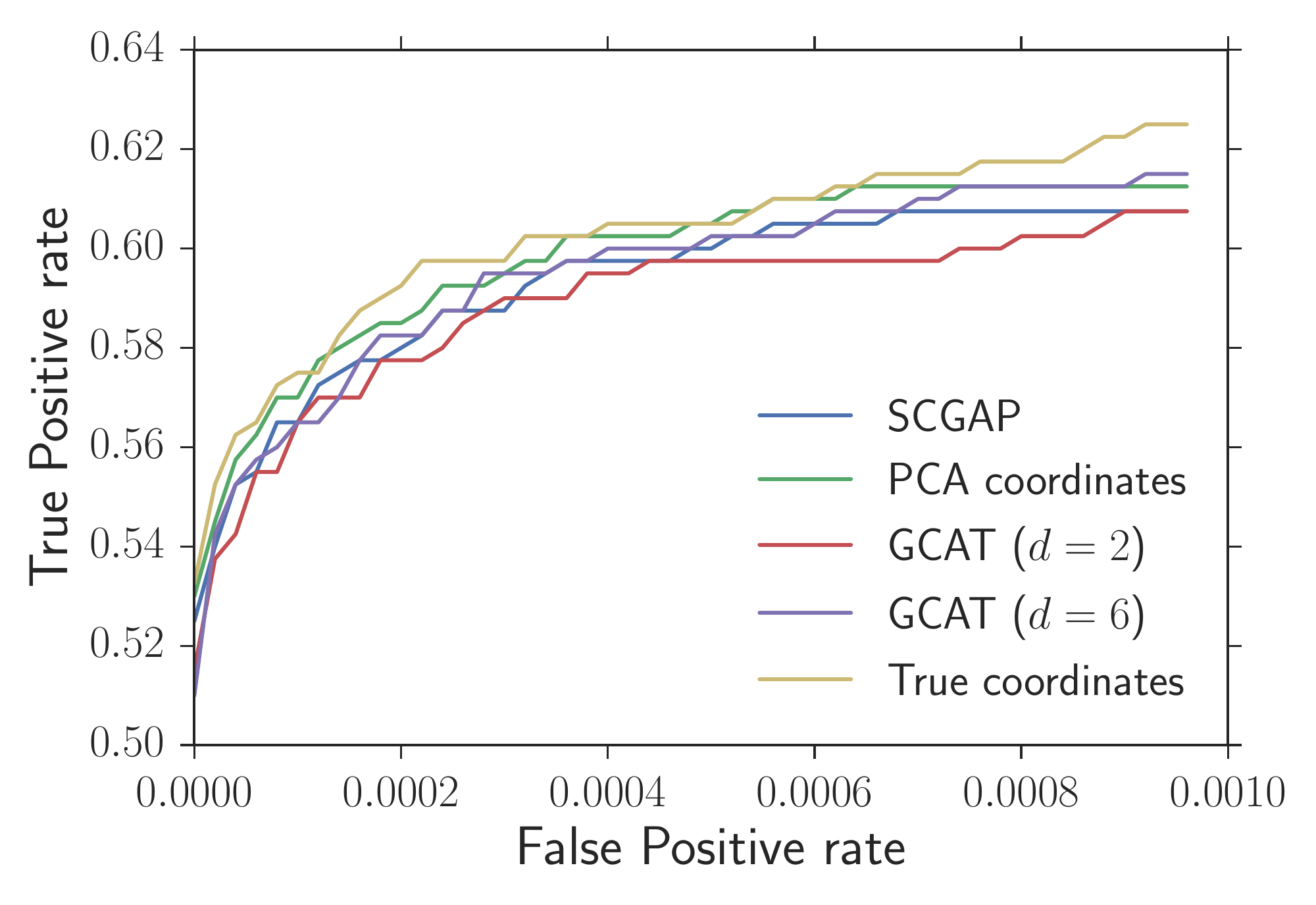}
\end{subfigure}
\begin{subfigure}[]{0.5\linewidth}
\centering
\caption{\hspace{0.75cm}$\alpha_1=2$}
\includegraphics[width=3.3in]{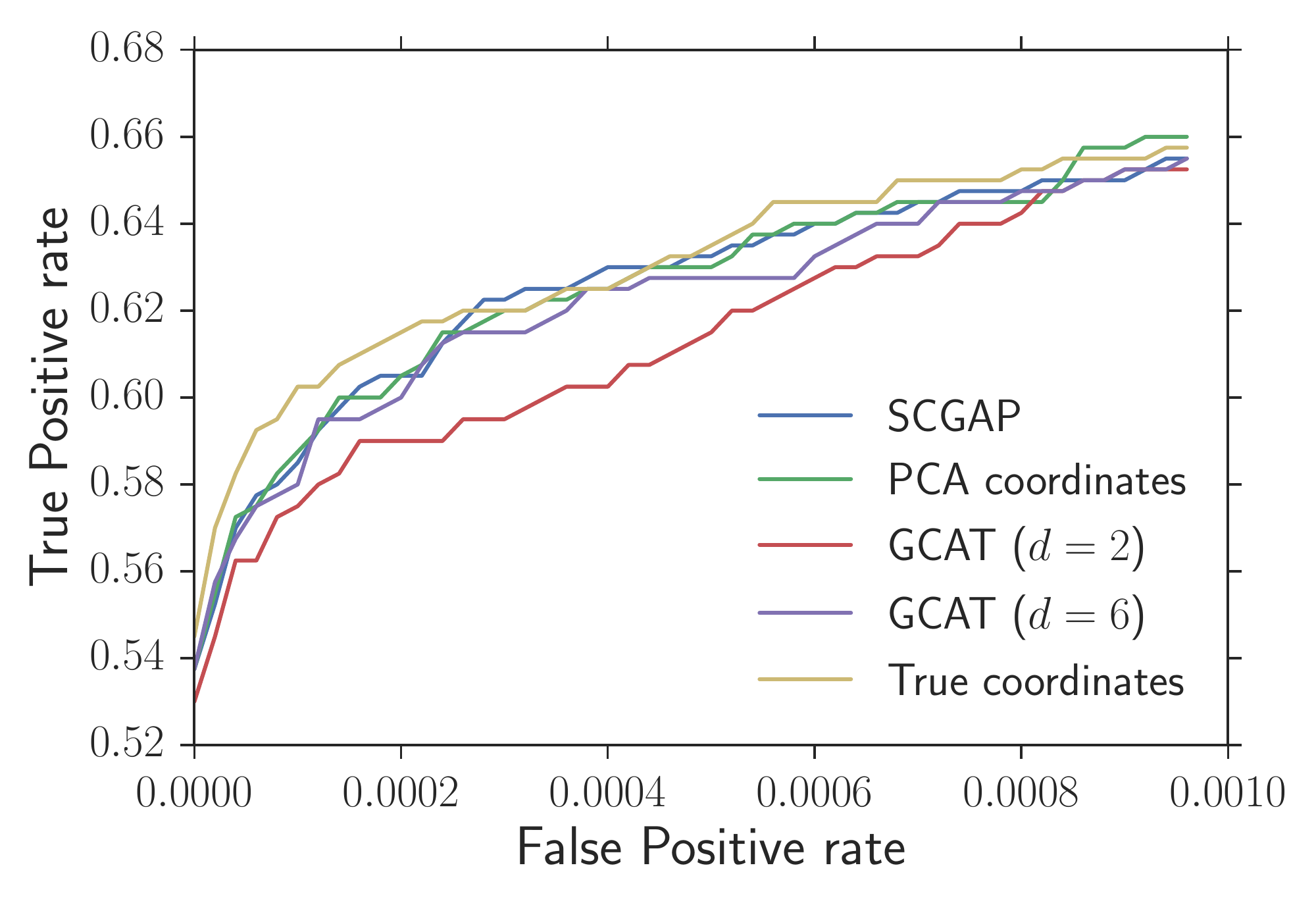}
\end{subfigure}
\begin{subfigure}[]{0.5\linewidth}
\centering
\caption{\hspace{0.75cm}$\alpha_1=4$}
\includegraphics[width=3.3in]{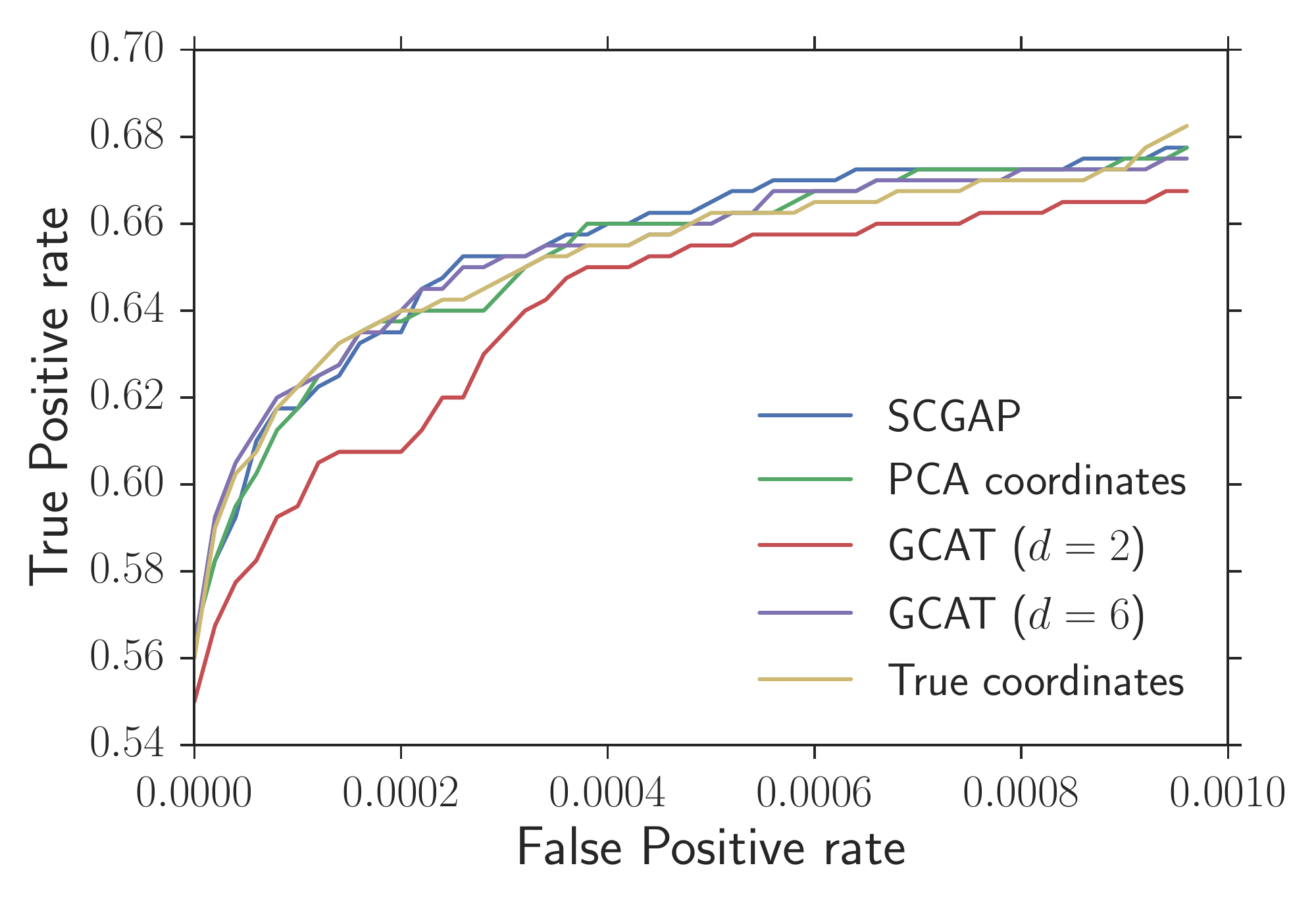}
\end{subfigure}
\begin{subfigure}[]{0.5\linewidth}
\centering
\caption{\hspace{0.75cm}$\alpha_1=8$}
\includegraphics[width=3.3in]{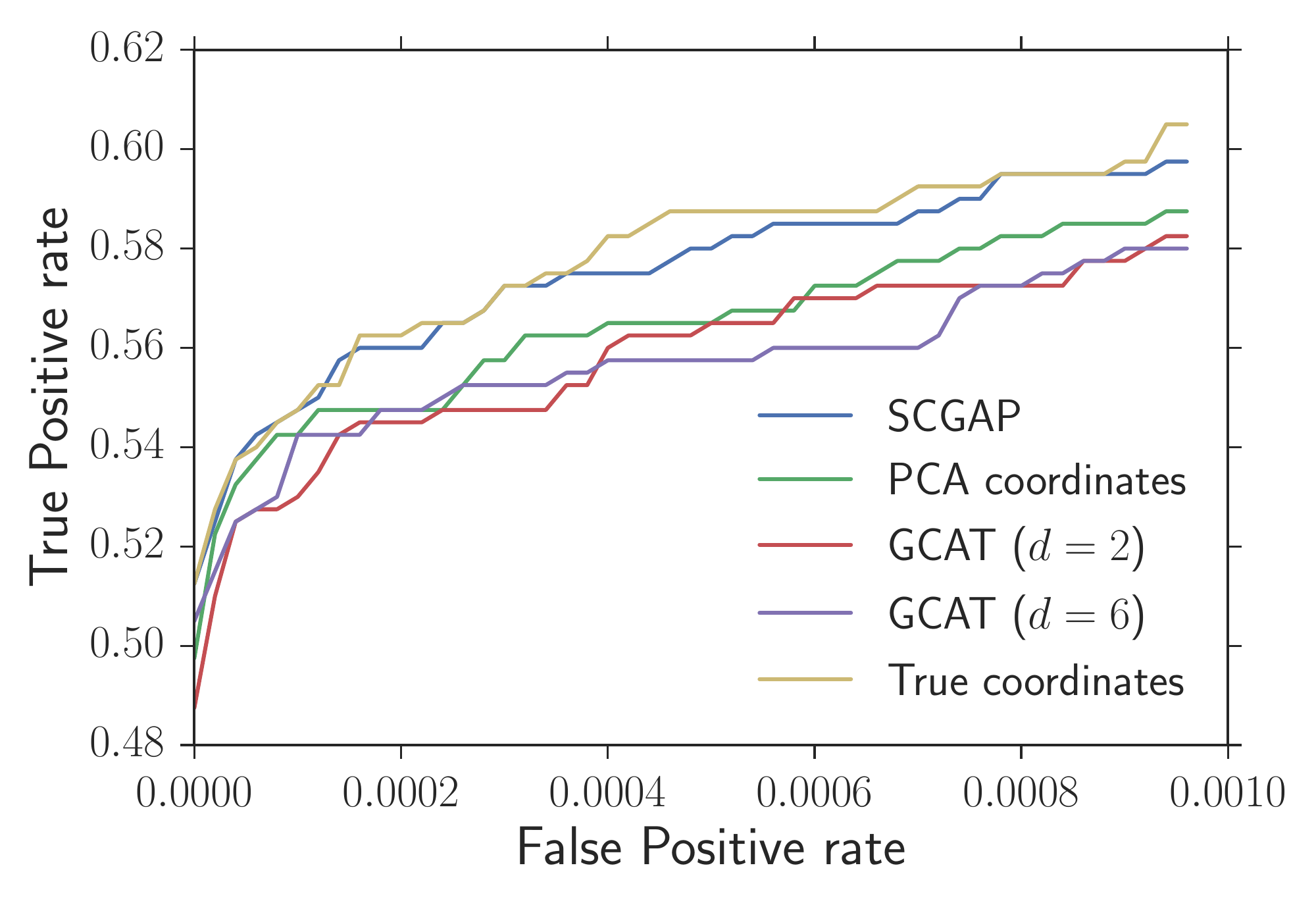}
\end{subfigure}
\begin{subfigure}[]{\linewidth}
\centering
\caption{\hspace{0.75cm}$\alpha_1=16$}
\includegraphics[width=3.3in]{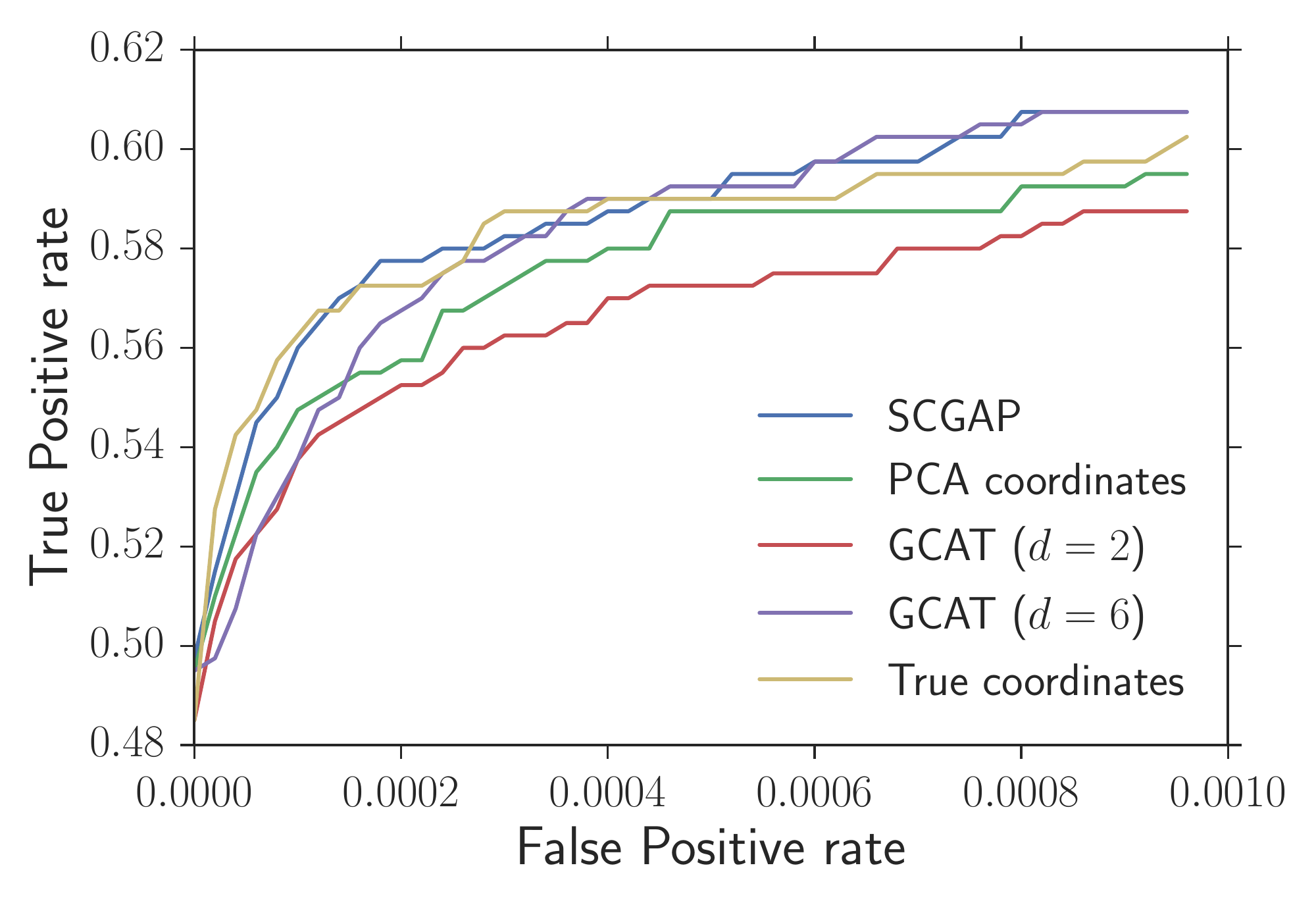}
\end{subfigure}
\caption{{\bf Isotropic covariance decay.} ROC curves for our association testing procedure with ancestral locations inferred using \localizationAlgorithm{}, PCA, or using the true locations. We also compared our results with the \GCAT{} method, which uses a latent factor model with $d$ factors to estimate the allele frequencies for each individual at each locus. Genotypes were drawn according to the isotropic covariance decay model with $\alpha_0 = 1$ and $\alpha_2 = 0.5$, with the different choices of $\alpha_1$ given in the panel captions. These are the same simulation parameters used in \tref{tab:logisticExpDecayCov_MDS_PCA_beta1}.}
\label{fig:logisticExpDecayCov_n2000_p50k_pc10_beta1_gv0.20_lv0.10_nv0.70_alpha20.5}
\end{figure}

\newpage
\begin{figure}[h]
\begin{subfigure}[]{0.5\linewidth}
\centering
\caption{\hspace{0.75cm}$\alpha_1=1$}
\includegraphics[width=3.3in]{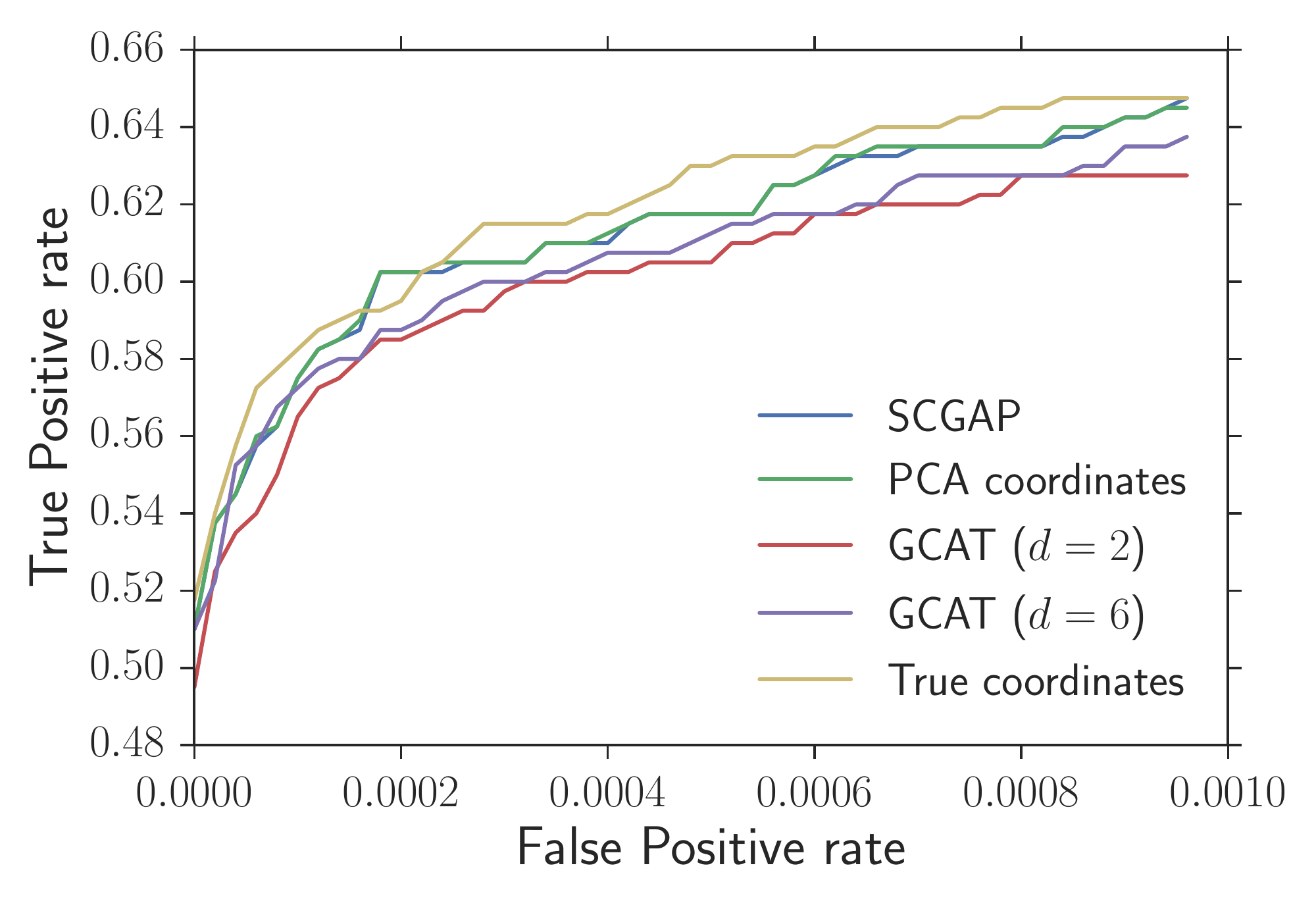}
\end{subfigure}
\begin{subfigure}[]{0.5\linewidth}
\centering
\caption{\hspace{0.75cm}$\alpha_1=2$}
\includegraphics[width=3.3in]{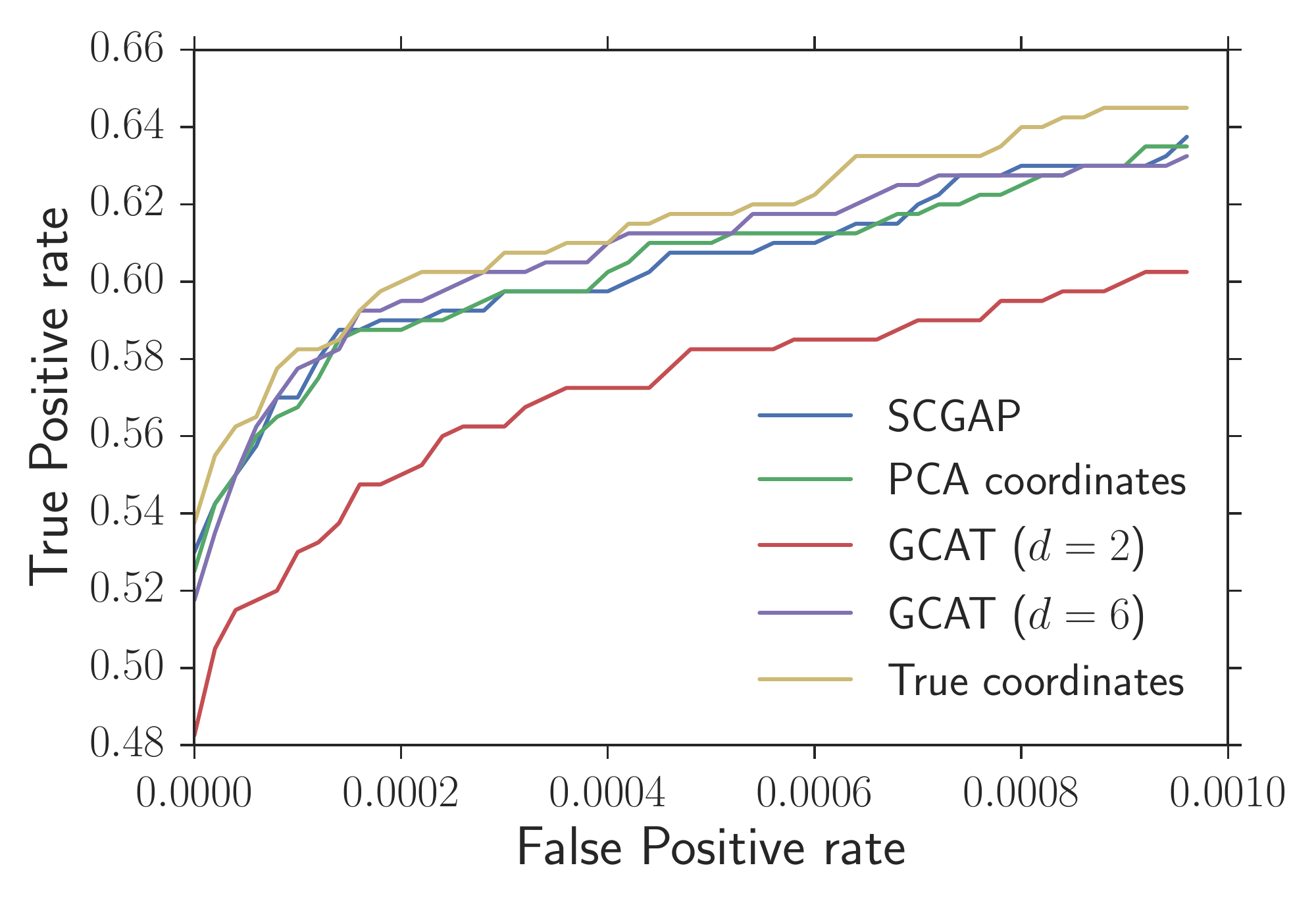}
\end{subfigure}
\begin{subfigure}[]{0.5\linewidth}
\centering
\caption{\hspace{0.75cm}$\alpha_1=4$}
\includegraphics[width=3.3in]{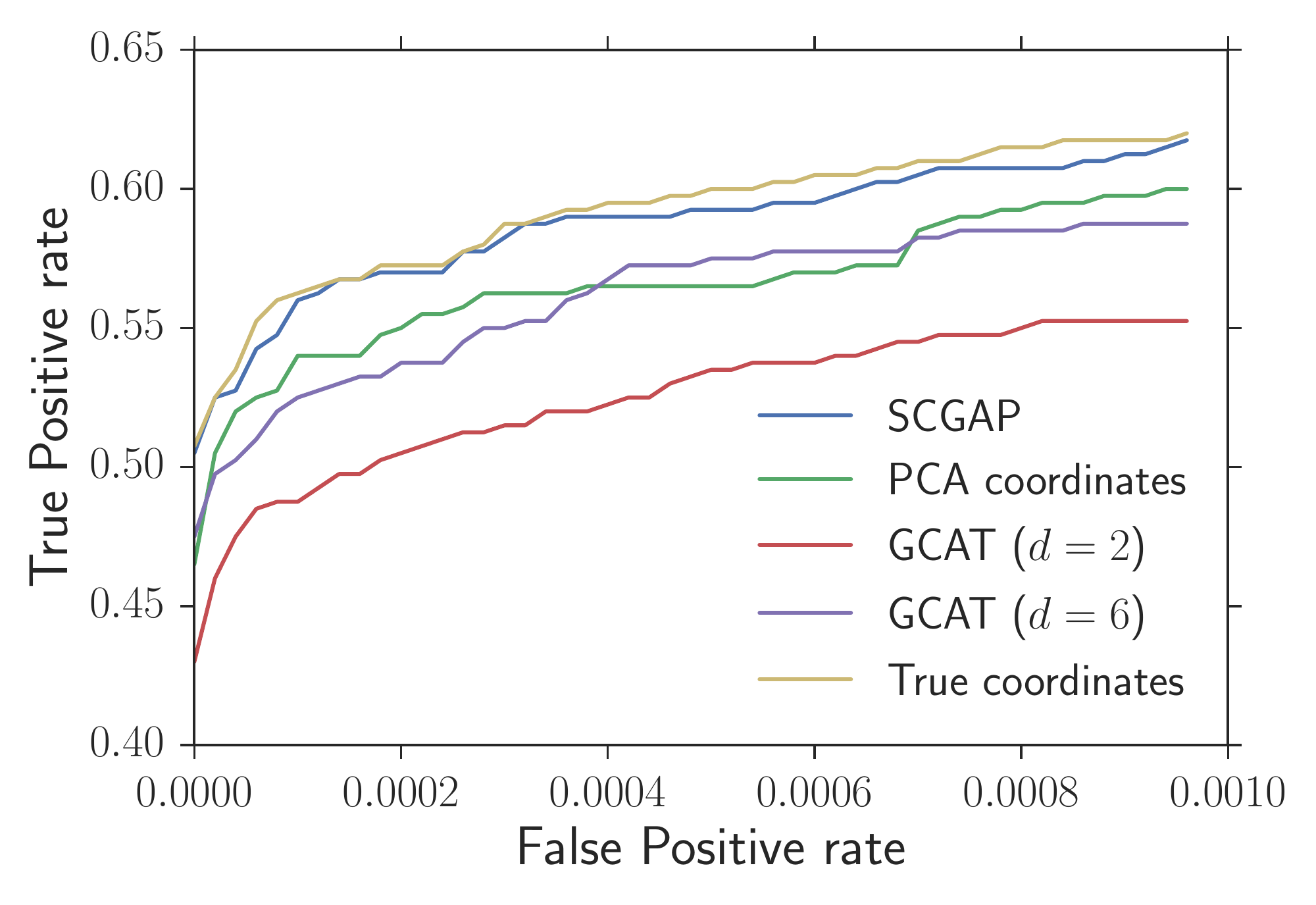}
\end{subfigure}
\begin{subfigure}[]{0.5\linewidth}
\centering
\caption{\hspace{0.75cm}$\alpha_1=8$}
\includegraphics[width=3.3in]{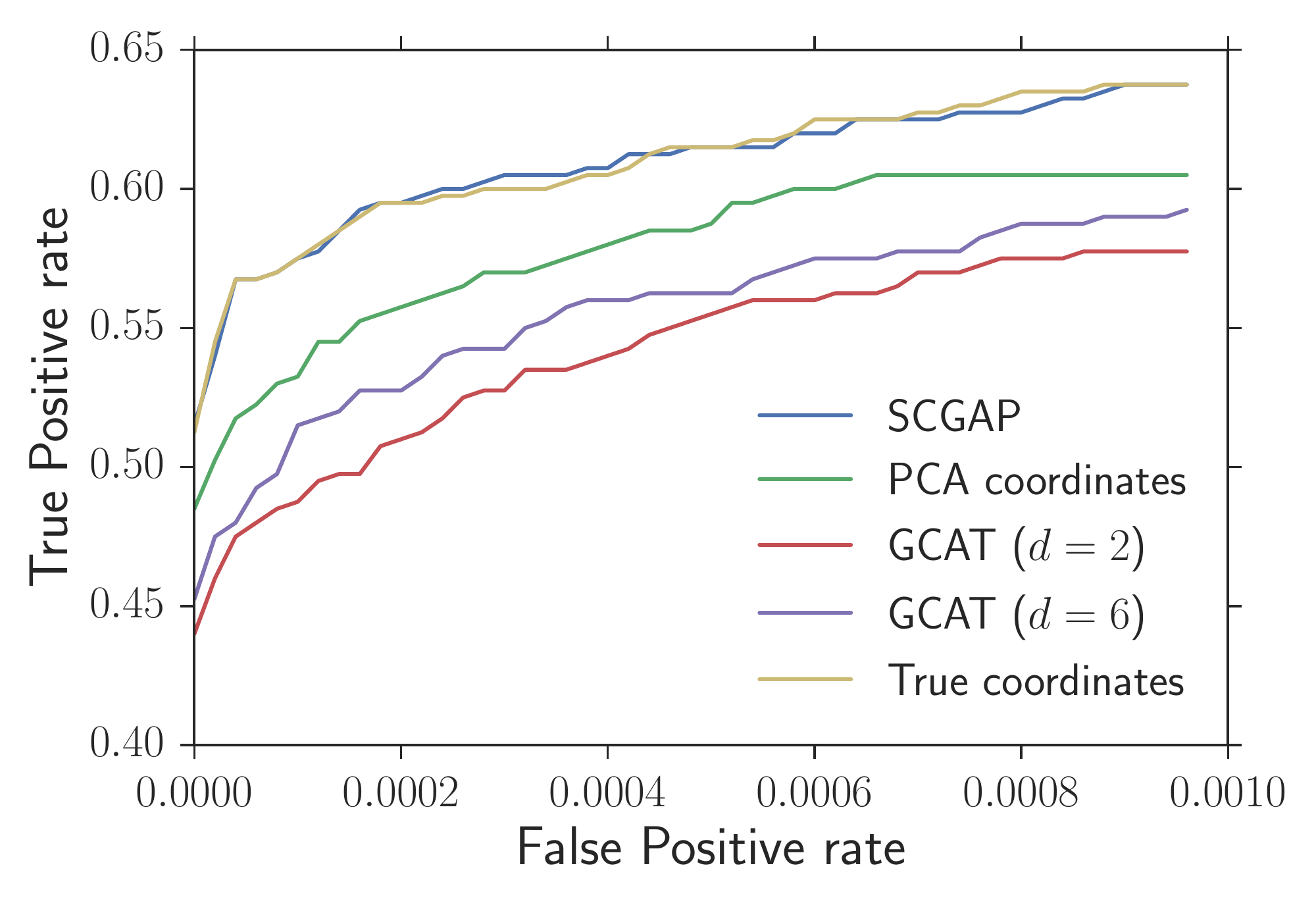}
\end{subfigure}
\begin{subfigure}[]{\linewidth}
\centering
\caption{\hspace{0.75cm}$\alpha_1=16$}
\includegraphics[width=3.3in]{{logisticExpDecayCov_n2000_p50k_pc10_beta1_gv0.20_lv0.10_nv0.70_alpha01_alpha116_alpha21}.pdf}
\end{subfigure}
\caption{Same simulation scenario as in \fref{fig:logisticExpDecayCov_n2000_p50k_pc10_beta1_gv0.20_lv0.10_nv0.70_alpha20.5}, with $\alpha_2 = 1$.}
\label{fig:logisticExpDecayCov_n2000_p50k_pc10_beta1_gv0.20_lv0.10_nv0.70_alpha21}
\end{figure}

\newpage
\begin{figure}[h]
\begin{subfigure}[]{0.5\linewidth}
\centering
\caption{\hspace{0.75cm}$\alpha_1=1$}
\includegraphics[width=3.3in]{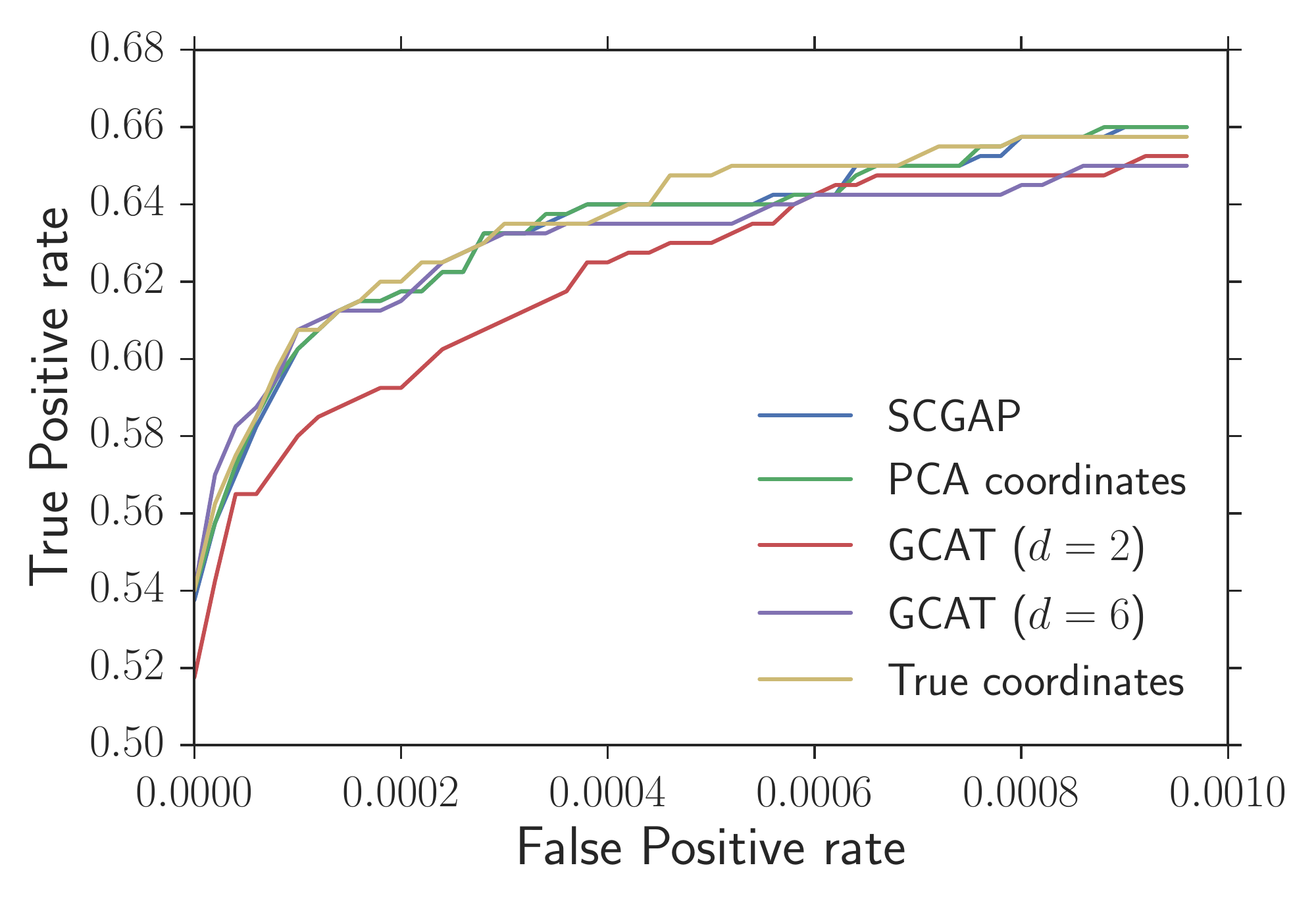}
\end{subfigure}
\begin{subfigure}[]{0.5\linewidth}
\centering
\caption{\hspace{0.75cm}$\alpha_1=2$}
\includegraphics[width=3.3in]{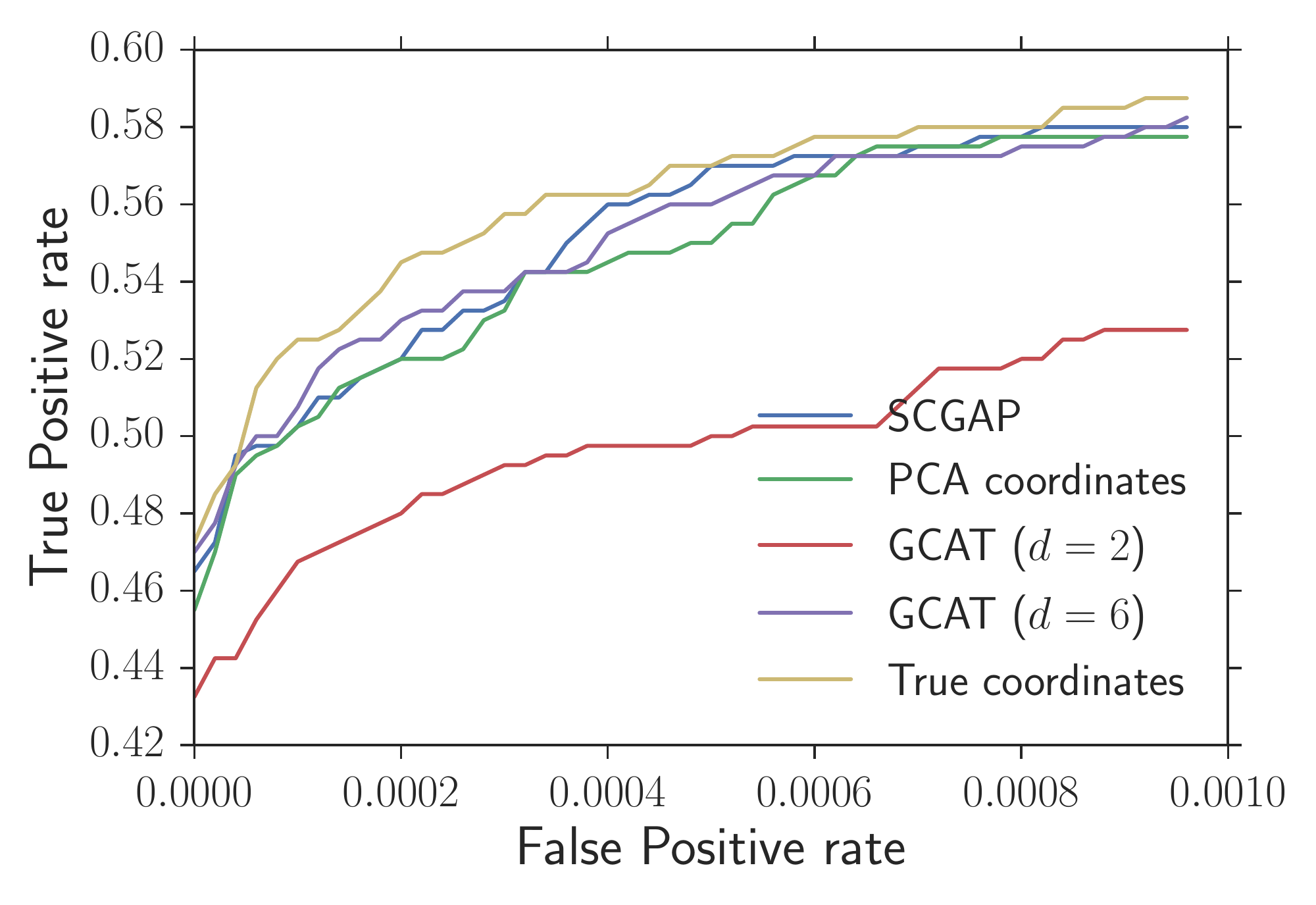}
\end{subfigure}
\begin{subfigure}[]{0.5\linewidth}
\centering
\caption{\hspace{0.75cm}$\alpha_1=4$}
\includegraphics[width=3.3in]{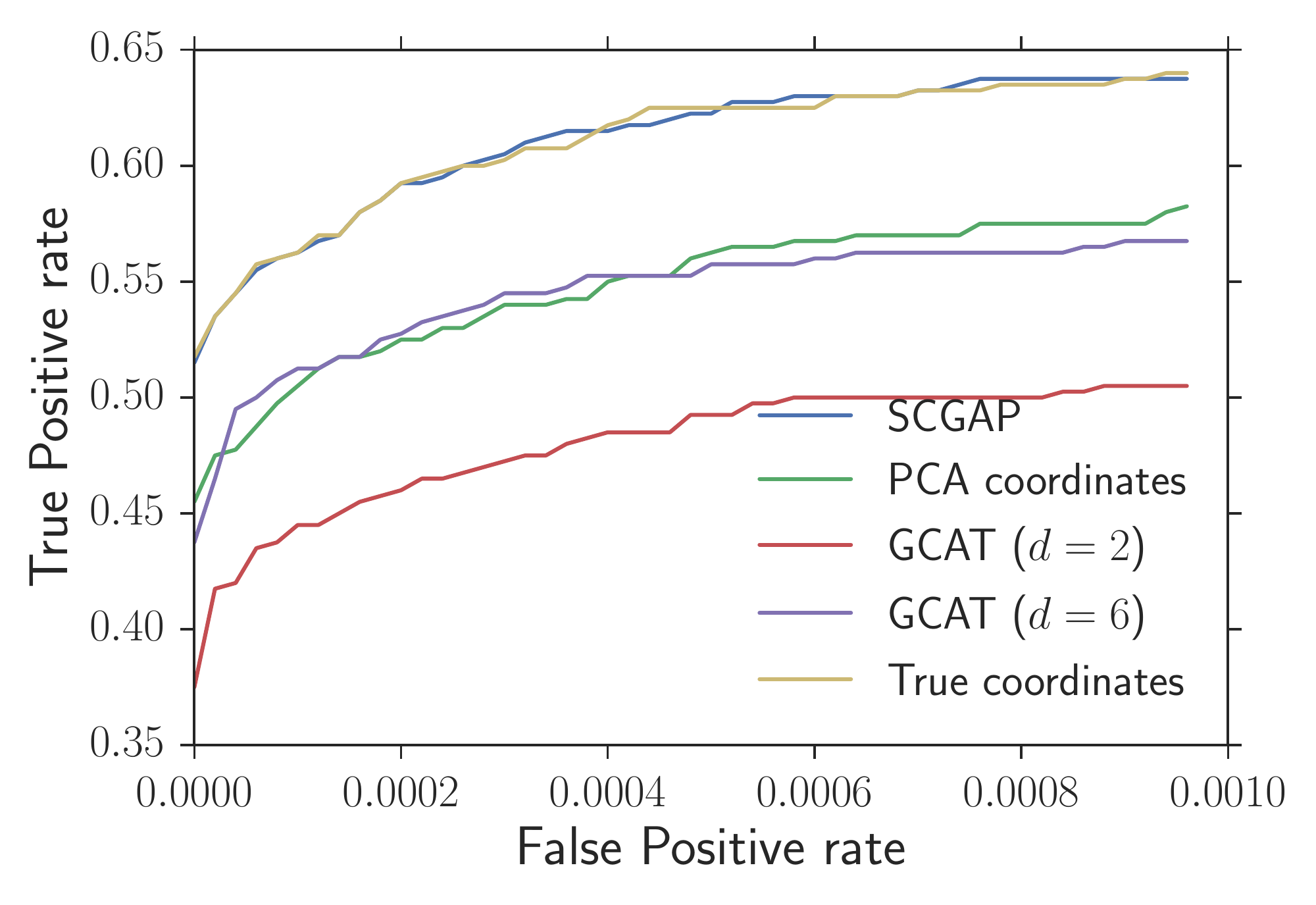}
\end{subfigure}
\begin{subfigure}[]{0.5\linewidth}
\centering
\caption{\hspace{0.75cm}$\alpha_1=8$}
\includegraphics[width=3.3in]{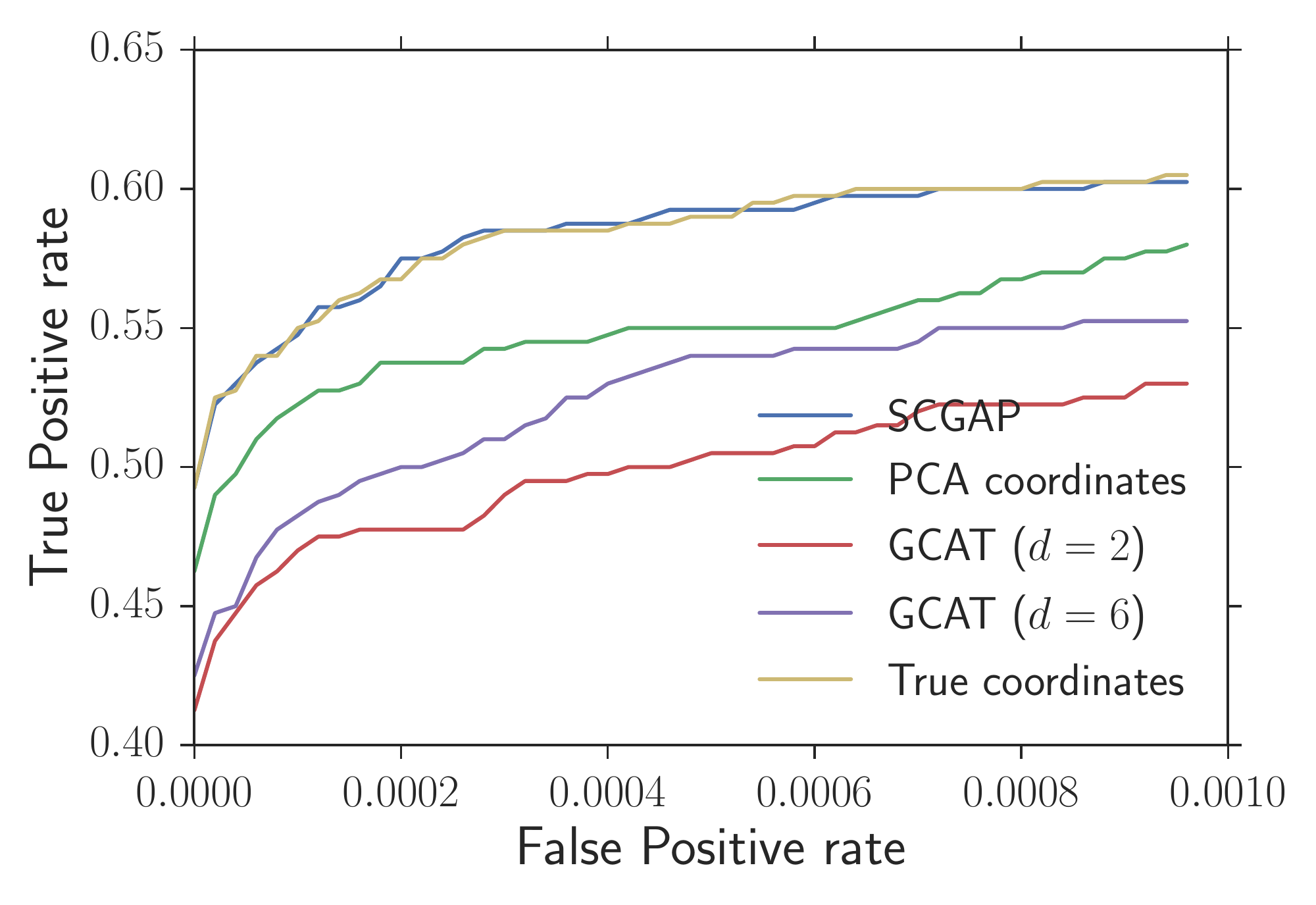}
\end{subfigure}
\begin{subfigure}[]{\linewidth}
\centering
\caption{\hspace{0.75cm}$\alpha_1=16$}
\includegraphics[width=3.3in]{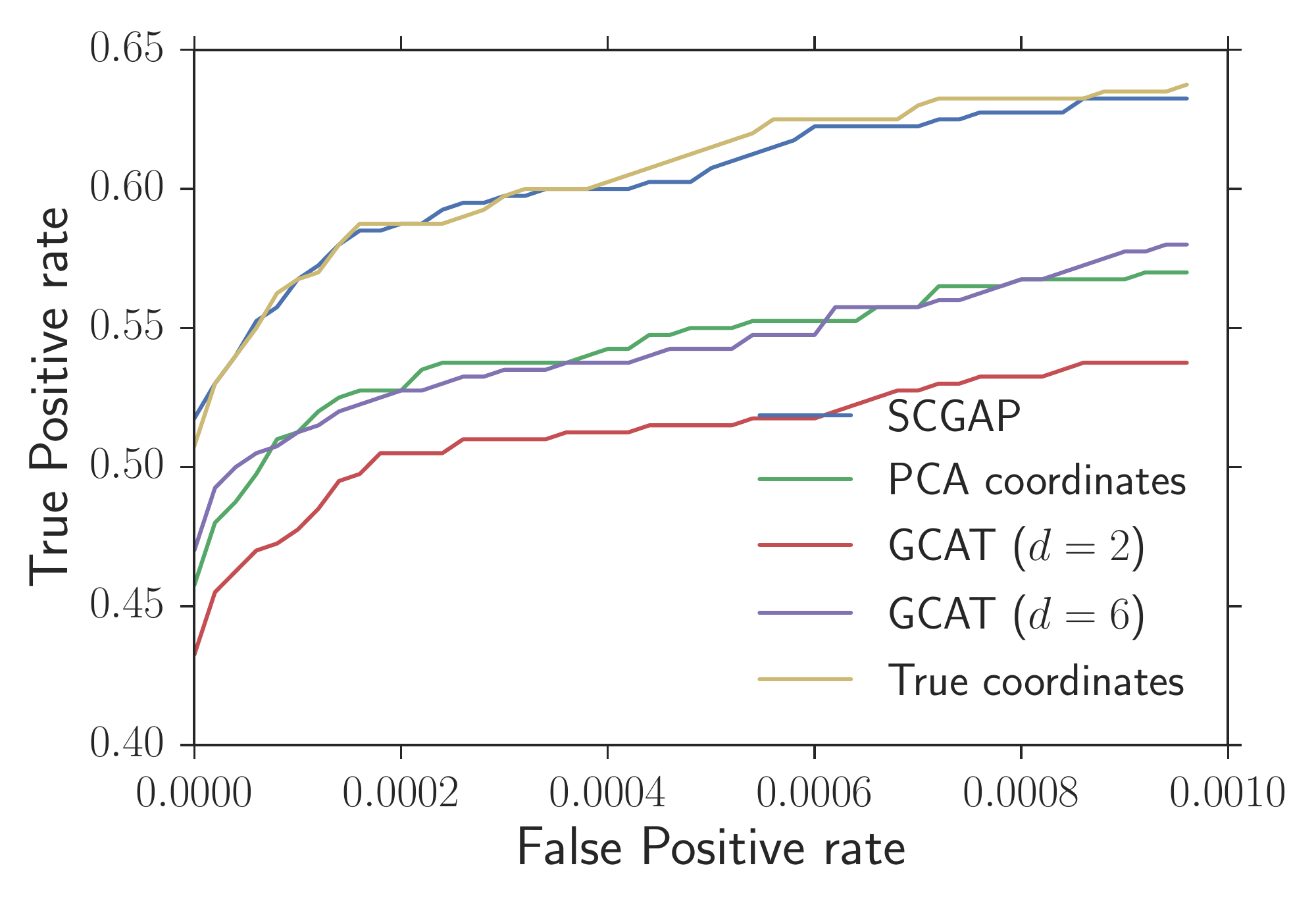}
\end{subfigure}
\caption{Same simulation scenario as in \fref{fig:logisticExpDecayCov_n2000_p50k_pc10_beta1_gv0.20_lv0.10_nv0.70_alpha20.5}, with $\alpha_2 = 1.5$.}
\label{fig:logisticExpDecayCov_n2000_p50k_pc10_beta1_gv0.20_lv0.10_nv0.70_alpha21.5}
\end{figure}

\newpage
\begin{figure}[h]
\begin{subfigure}[]{0.5\linewidth}
\centering
\caption{\hspace{0.75cm}$\alpha_1=1$}
\includegraphics[width=3.3in]{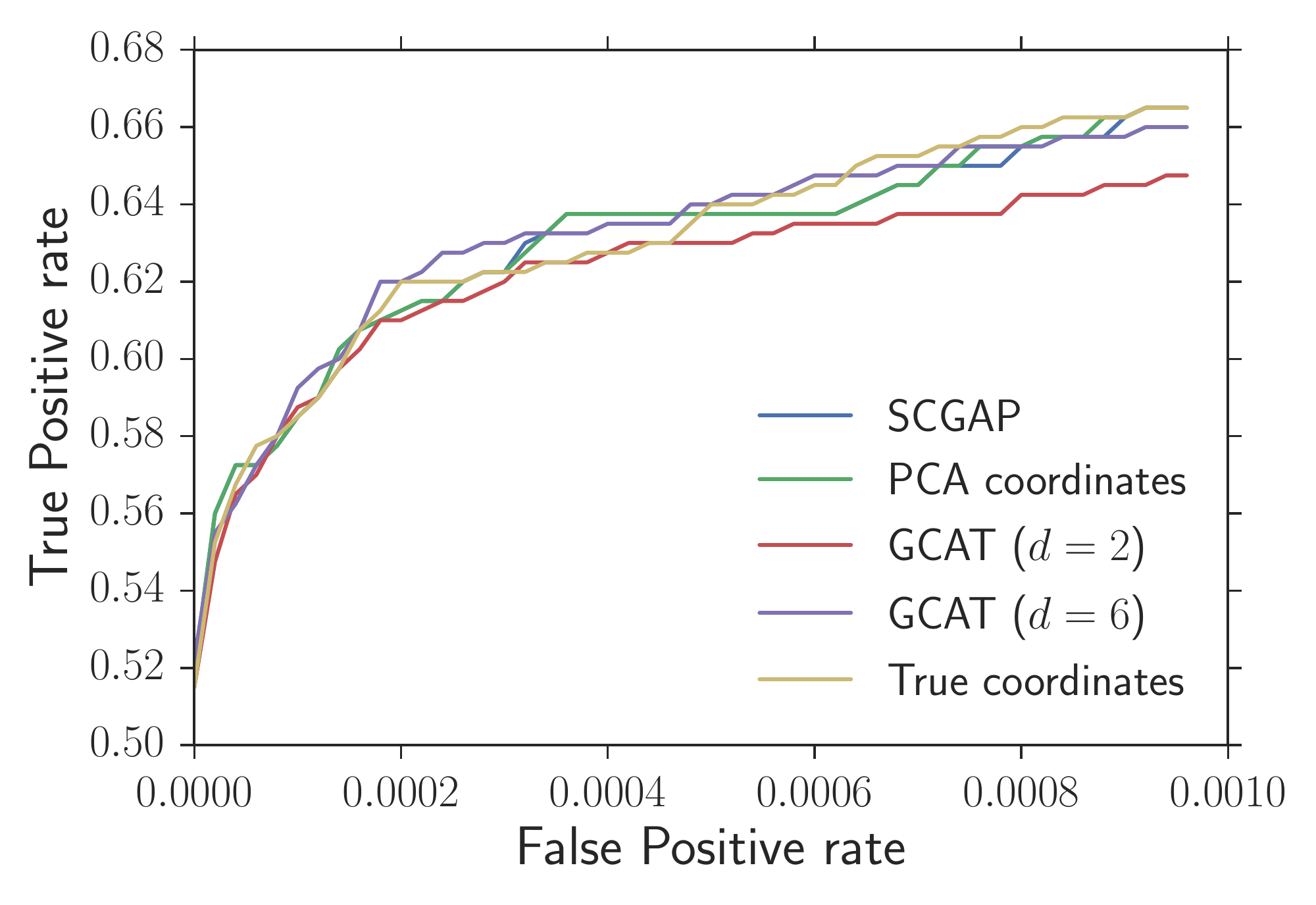}
\end{subfigure}
\begin{subfigure}[]{0.5\linewidth}
\centering
\caption{\hspace{0.75cm}$\alpha_1=2$}
\includegraphics[width=3.3in]{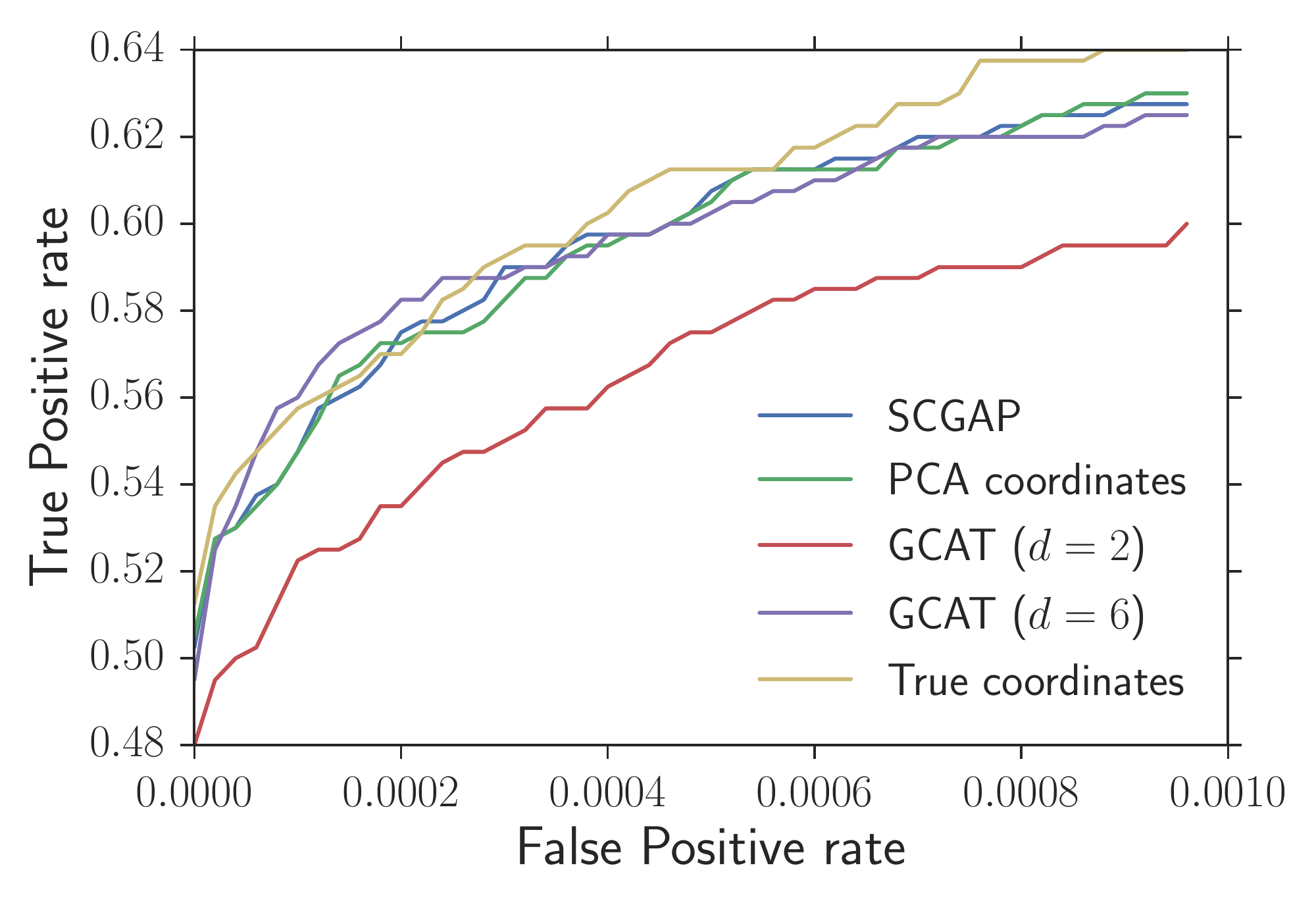}
\end{subfigure}
\begin{subfigure}[]{0.5\linewidth}
\centering
\caption{\hspace{0.75cm}$\alpha_1=4$}
\includegraphics[width=3.3in]{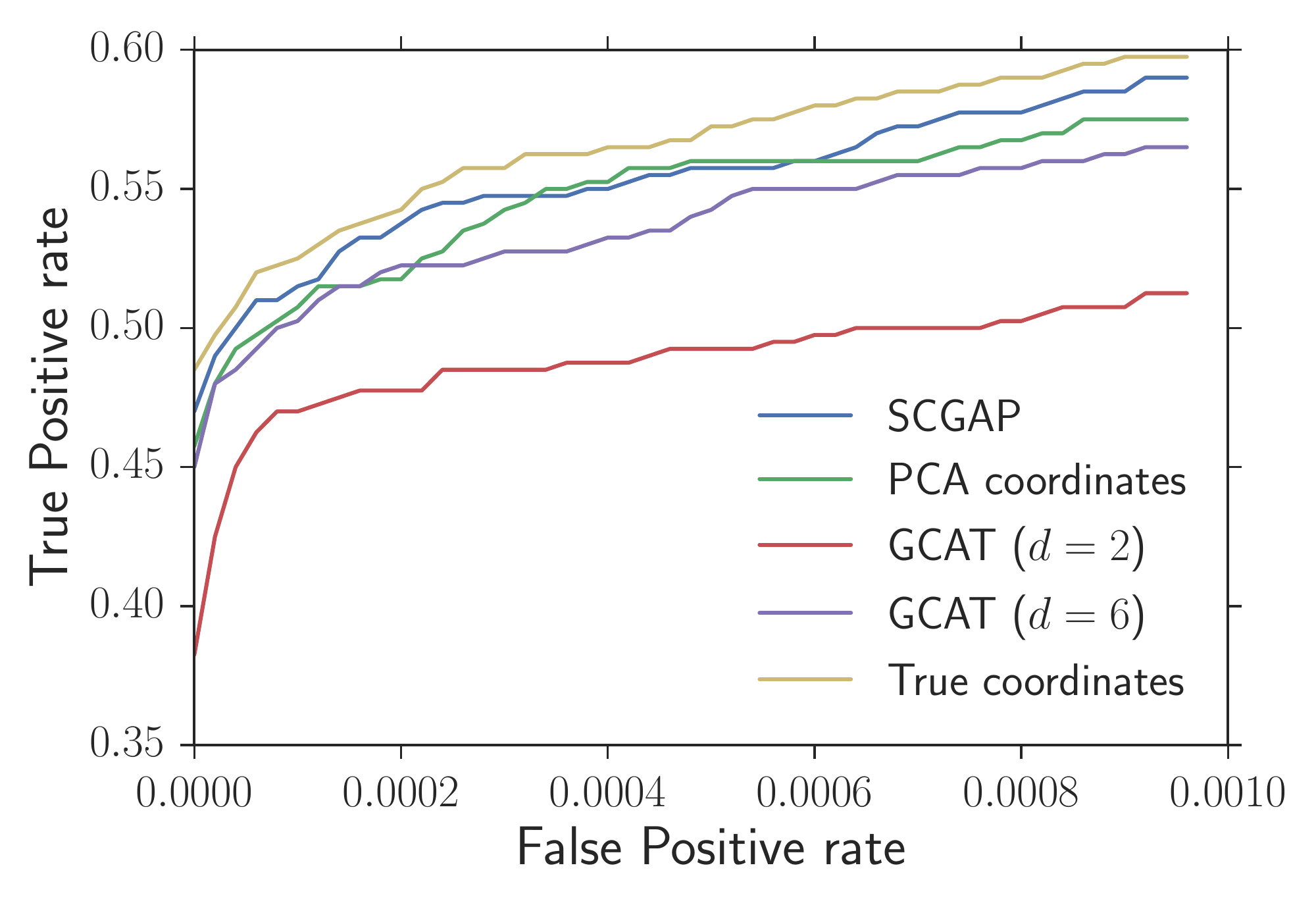}
\end{subfigure}
\begin{subfigure}[]{0.5\linewidth}
\centering
\caption{\hspace{0.75cm}$\alpha_1=8$}
\includegraphics[width=3.3in]{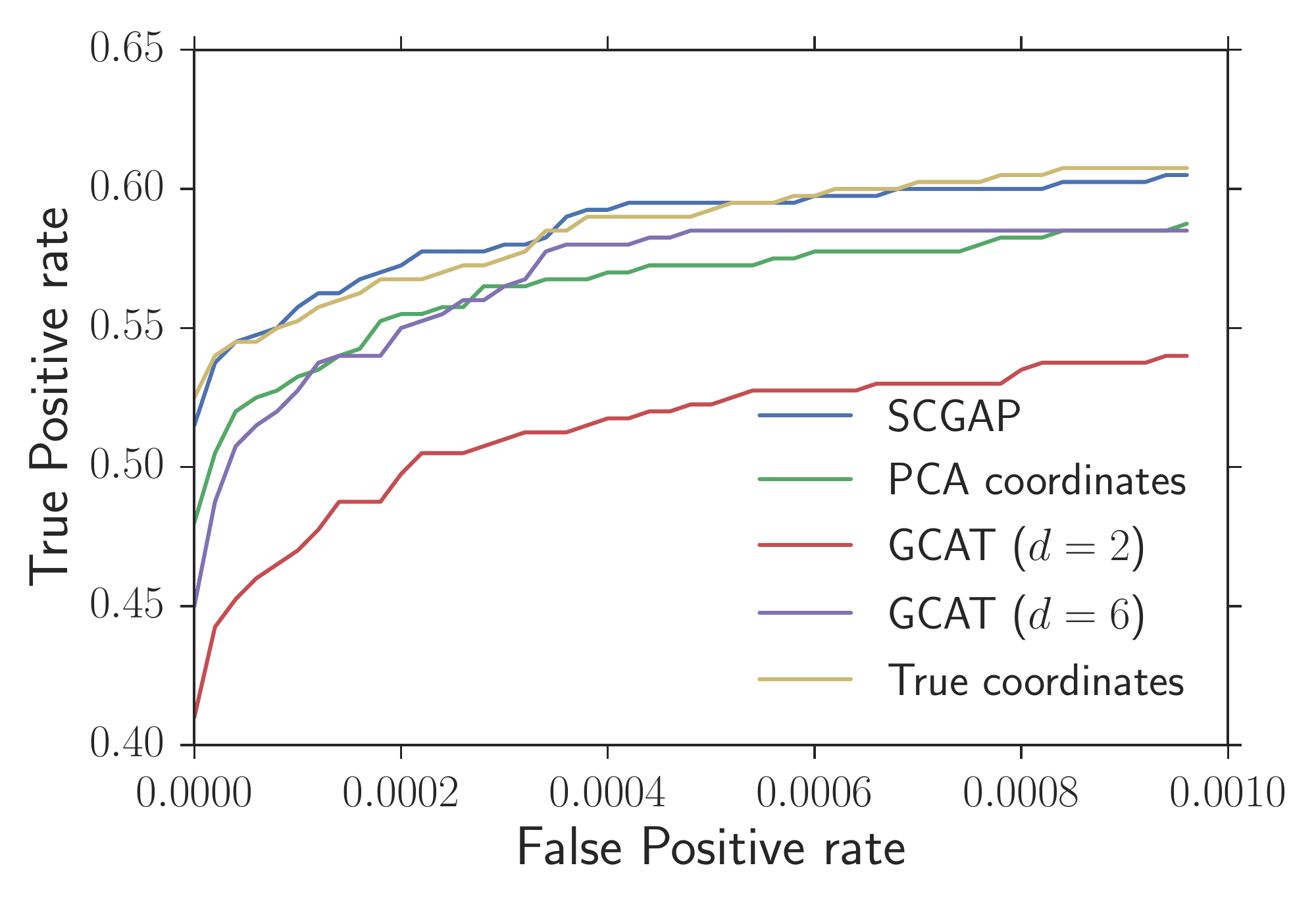}
\end{subfigure}
\begin{subfigure}[]{\linewidth}
\centering
\caption{\hspace{0.75cm}$\alpha_1=16$}
\includegraphics[width=3.3in]{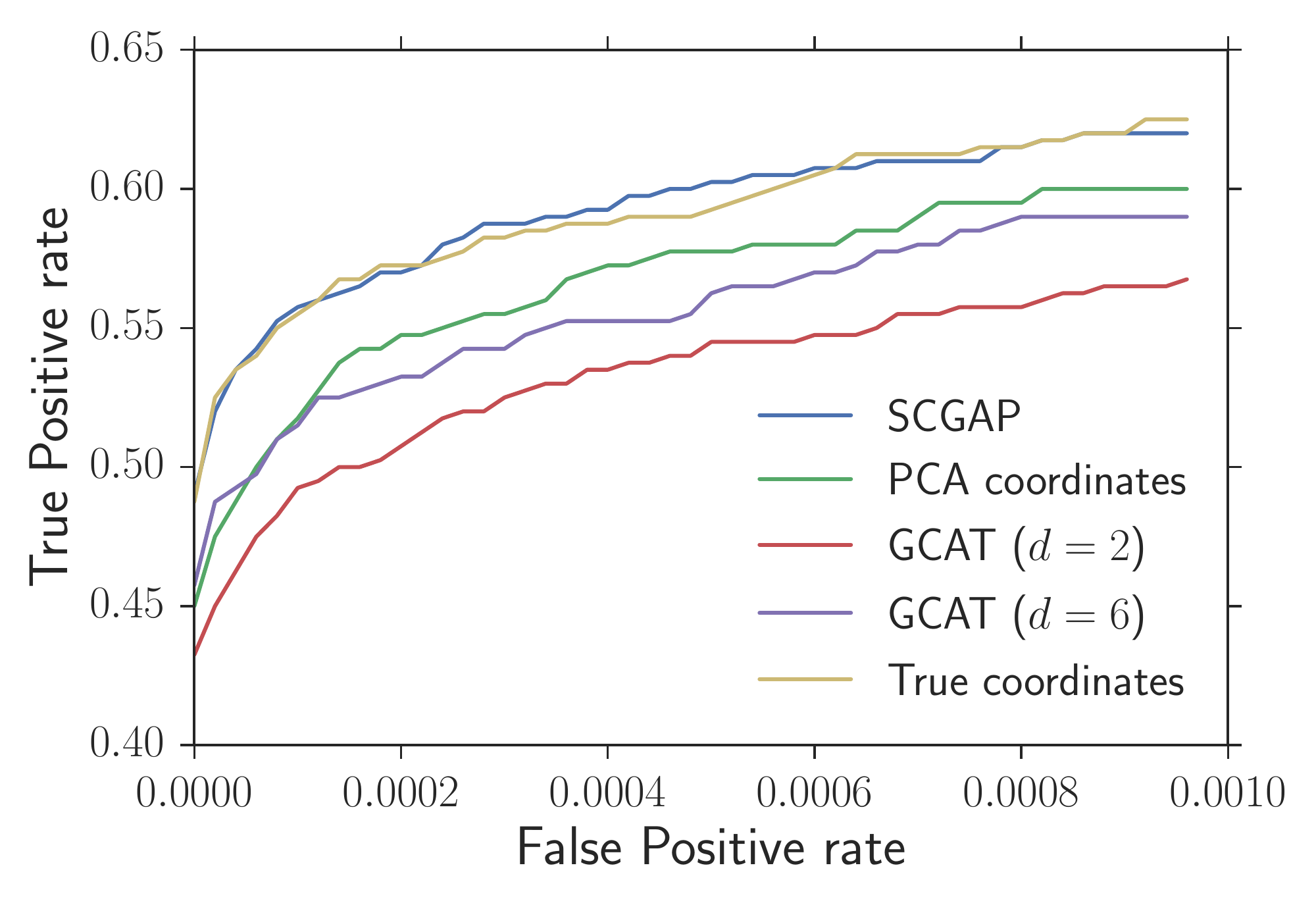}
\end{subfigure}
\caption{{\bf Directional covariance decay.} ROC curves for our association testing procedure with ancestral locations inferred using \localizationAlgorithm{}, PCA, or using the true locations. We also compared our results with the \GCAT{} method, which uses a latent factor model with $d$ factors to estimate the allele frequencies for each individual at each locus. Genotypes were drawn according to the directional covariance decay model with $\alpha_0 = \alpha_2 = 1$ and $\kappa = 0.1$, with the different choices of $\alpha_1$ given in the panel captions. These are the same simulation parameters used in \tref{tab:logisticDirectionalExpDecayCov_MDS_PCA_beta1}.}
\label{fig:logisticDirectionalExpDecayCov_n2000_p50k_pc10_beta1_gv0.20_lv0.10_nv0.70_vkappa0.1}
\end{figure}

\newpage
\begin{figure}[h]
\begin{subfigure}[]{0.5\linewidth}
\centering
\caption{\hspace{0.75cm}$\alpha_1=1$}
\includegraphics[width=3.3in]{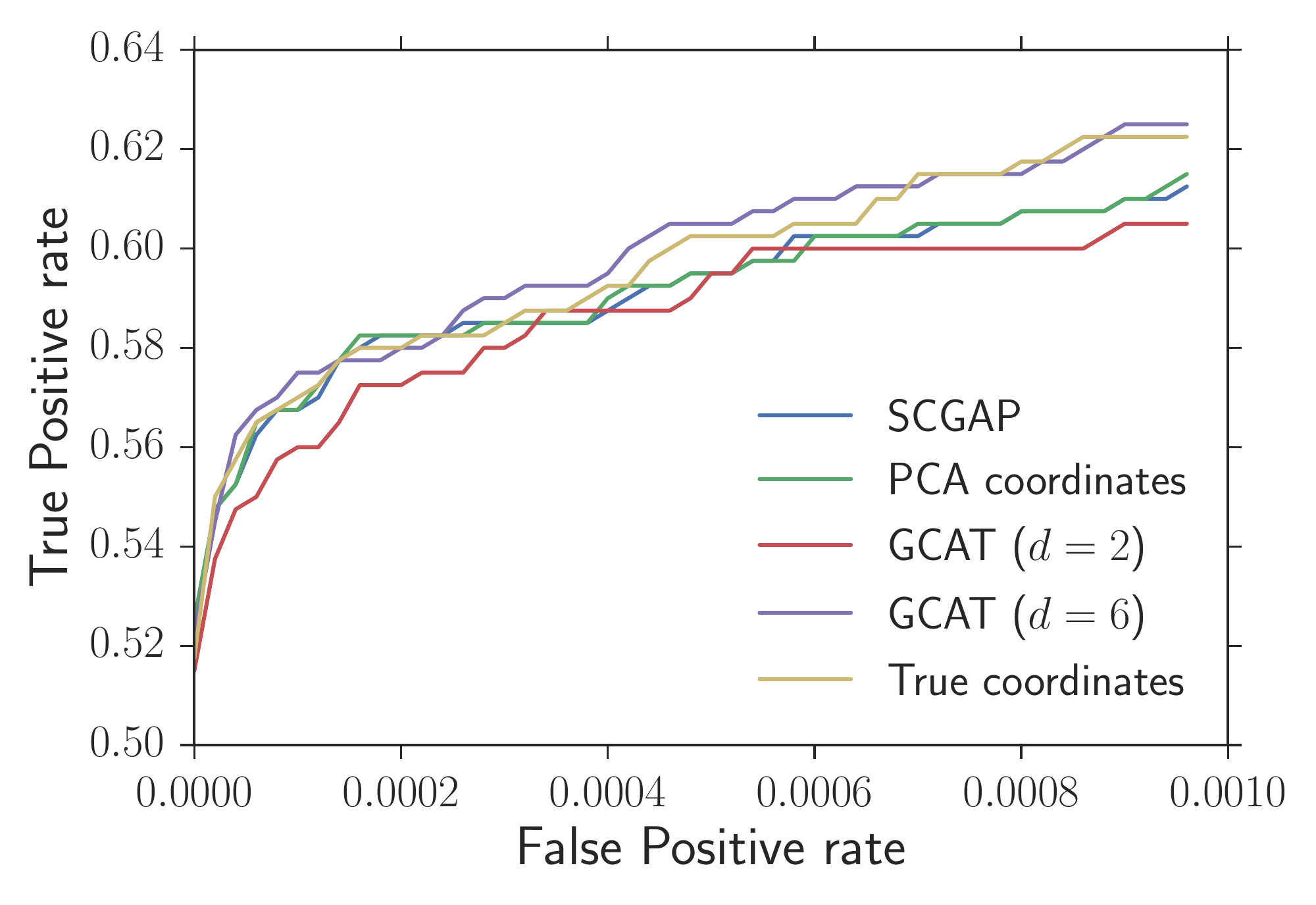}
\end{subfigure}
\begin{subfigure}[]{0.5\linewidth}
\centering
\caption{\hspace{0.75cm}$\alpha_1=2$}
\includegraphics[width=3.3in]{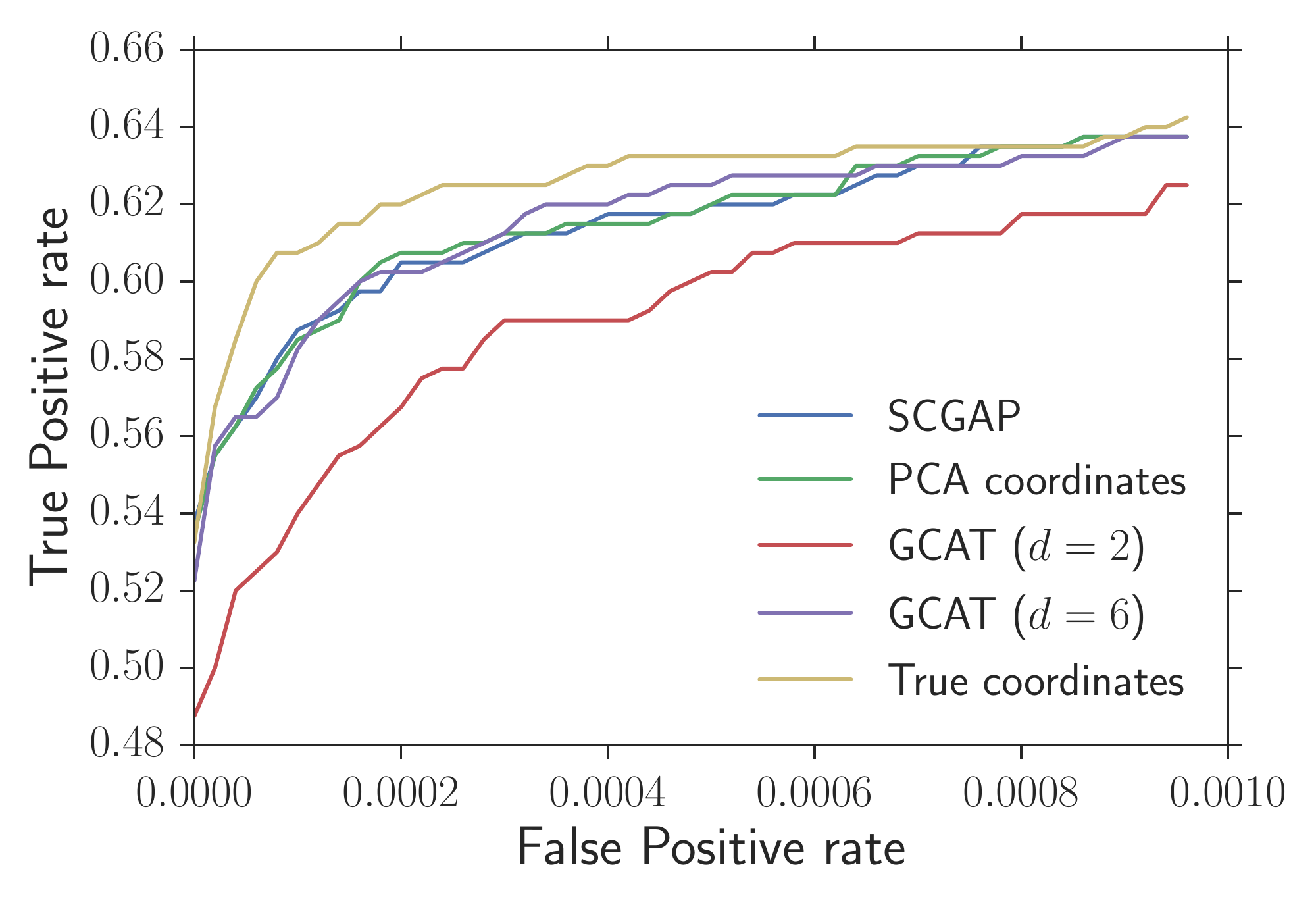}
\end{subfigure}
\begin{subfigure}[]{0.5\linewidth}
\centering
\caption{\hspace{0.75cm}$\alpha_1=4$}
\includegraphics[width=3.3in]{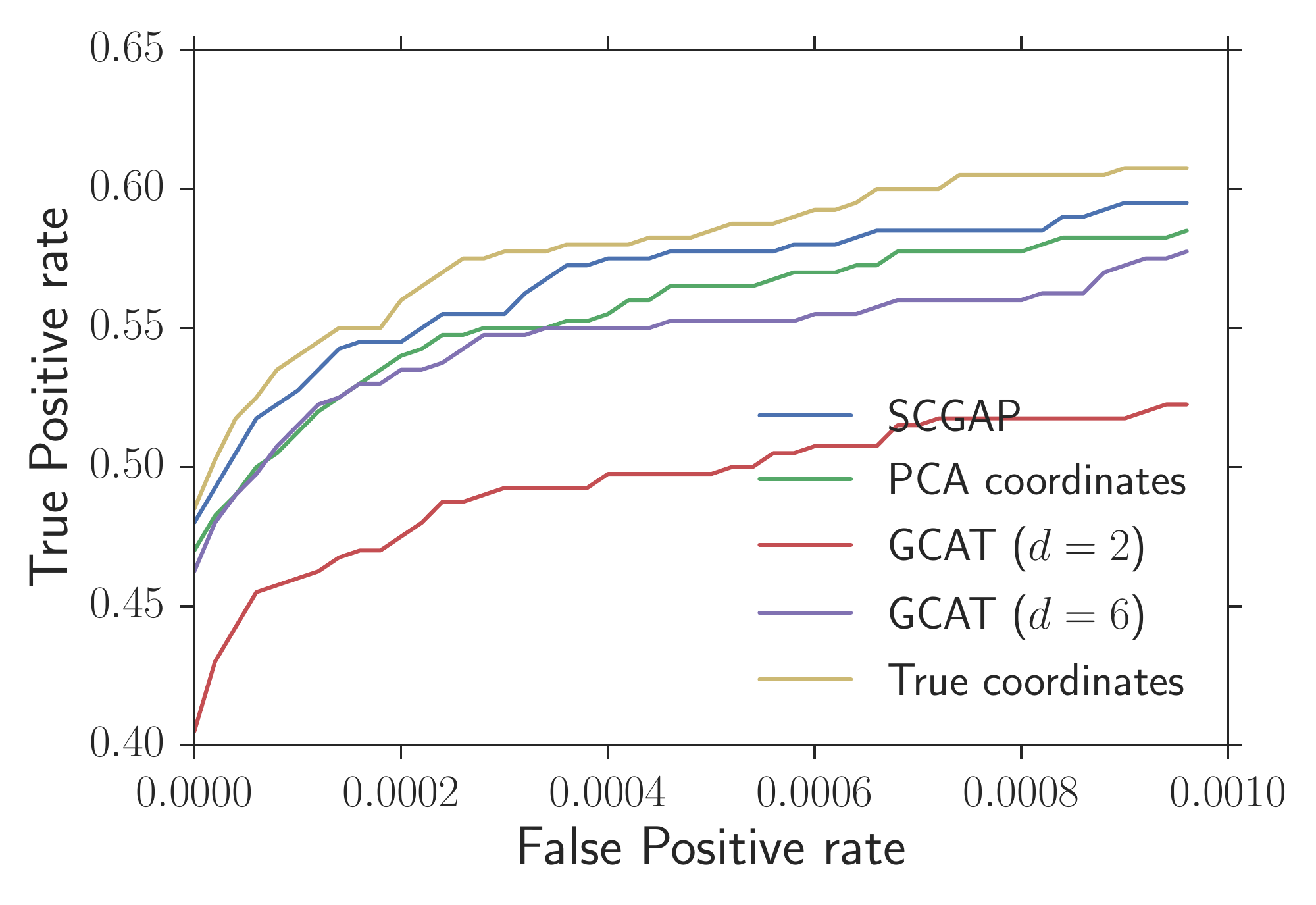}
\end{subfigure}
\begin{subfigure}[]{0.5\linewidth}
\centering
\caption{\hspace{0.75cm}$\alpha_1=8$}
\includegraphics[width=3.3in]{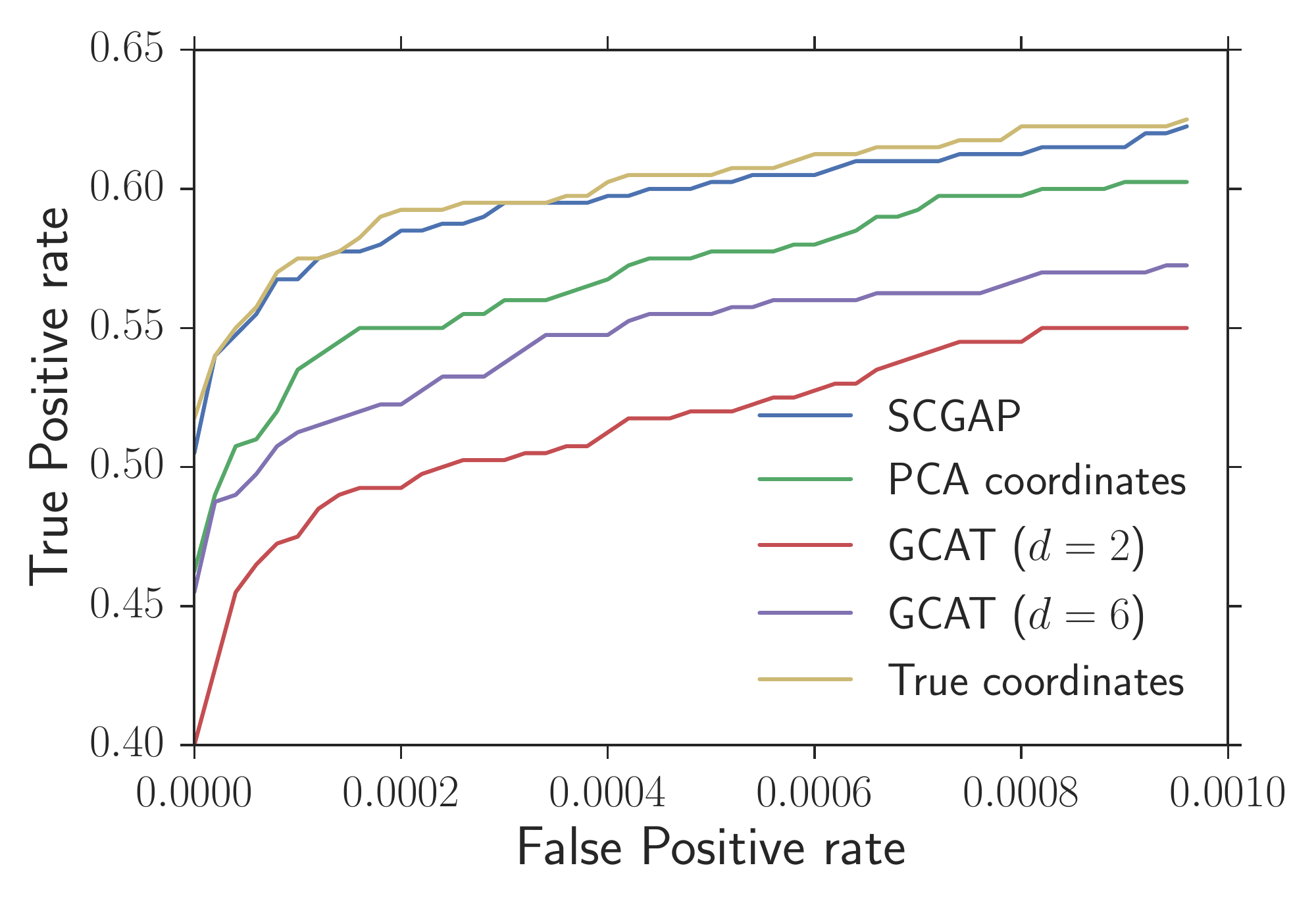}
\end{subfigure}
\begin{subfigure}[]{\linewidth}
\centering
\caption{\hspace{0.75cm}$\alpha_1=16$}
\includegraphics[width=3.3in]{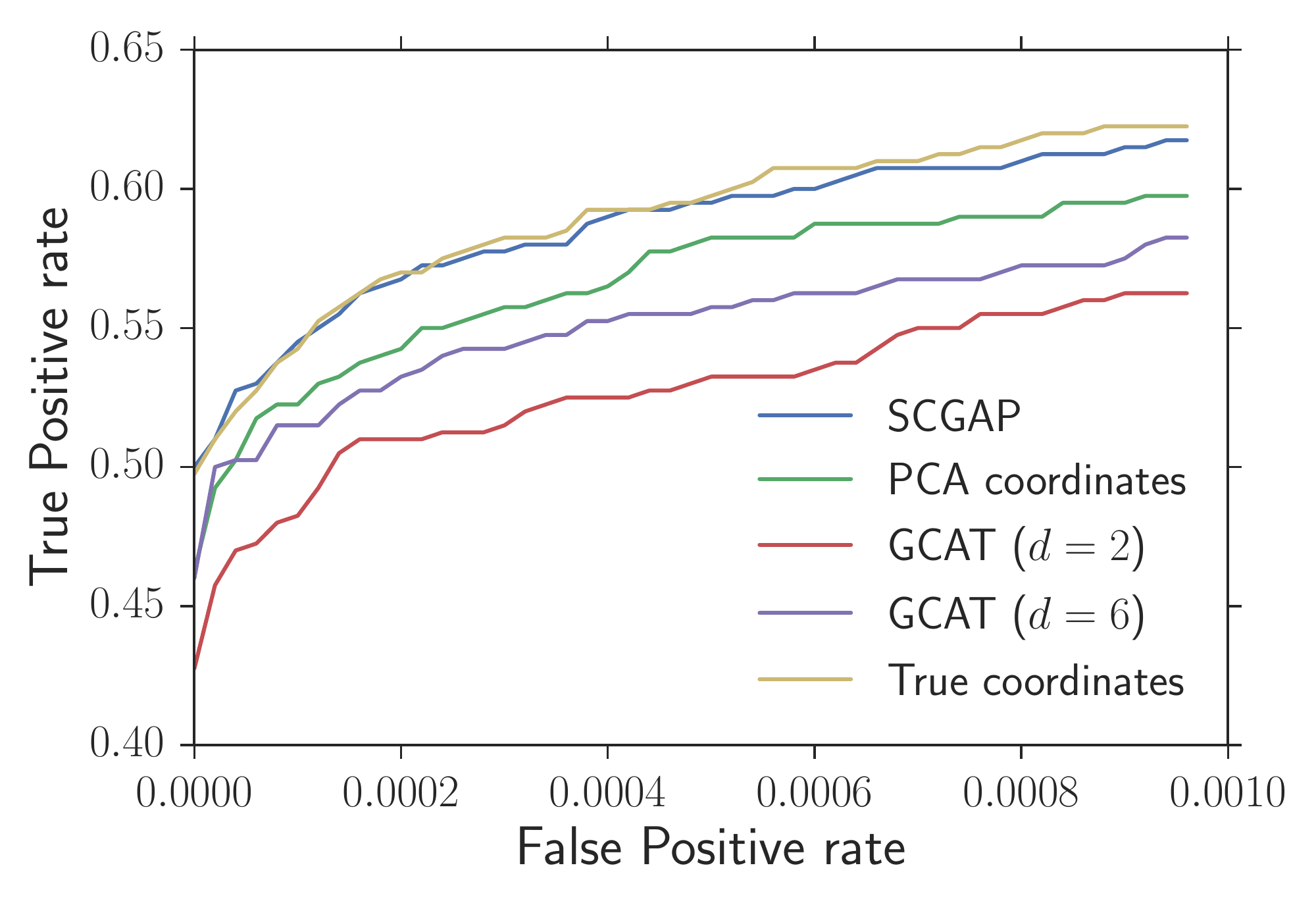}
\end{subfigure}
\caption{Same simulation scenario as in \fref{fig:logisticDirectionalExpDecayCov_n2000_p50k_pc10_beta1_gv0.20_lv0.10_nv0.70_vkappa0.1}, with $\kappa = 1$.}
\label{fig:logisticDirectionalExpDecayCov_n2000_p50k_pc10_beta1_gv0.20_lv0.10_nv0.70_vkappa1}
\end{figure}

\newpage
\begin{figure}[h]
\begin{subfigure}[]{0.5\linewidth}
\centering
\caption{\hspace{0.75cm}$\alpha_1=1$}
\includegraphics[width=3.3in]{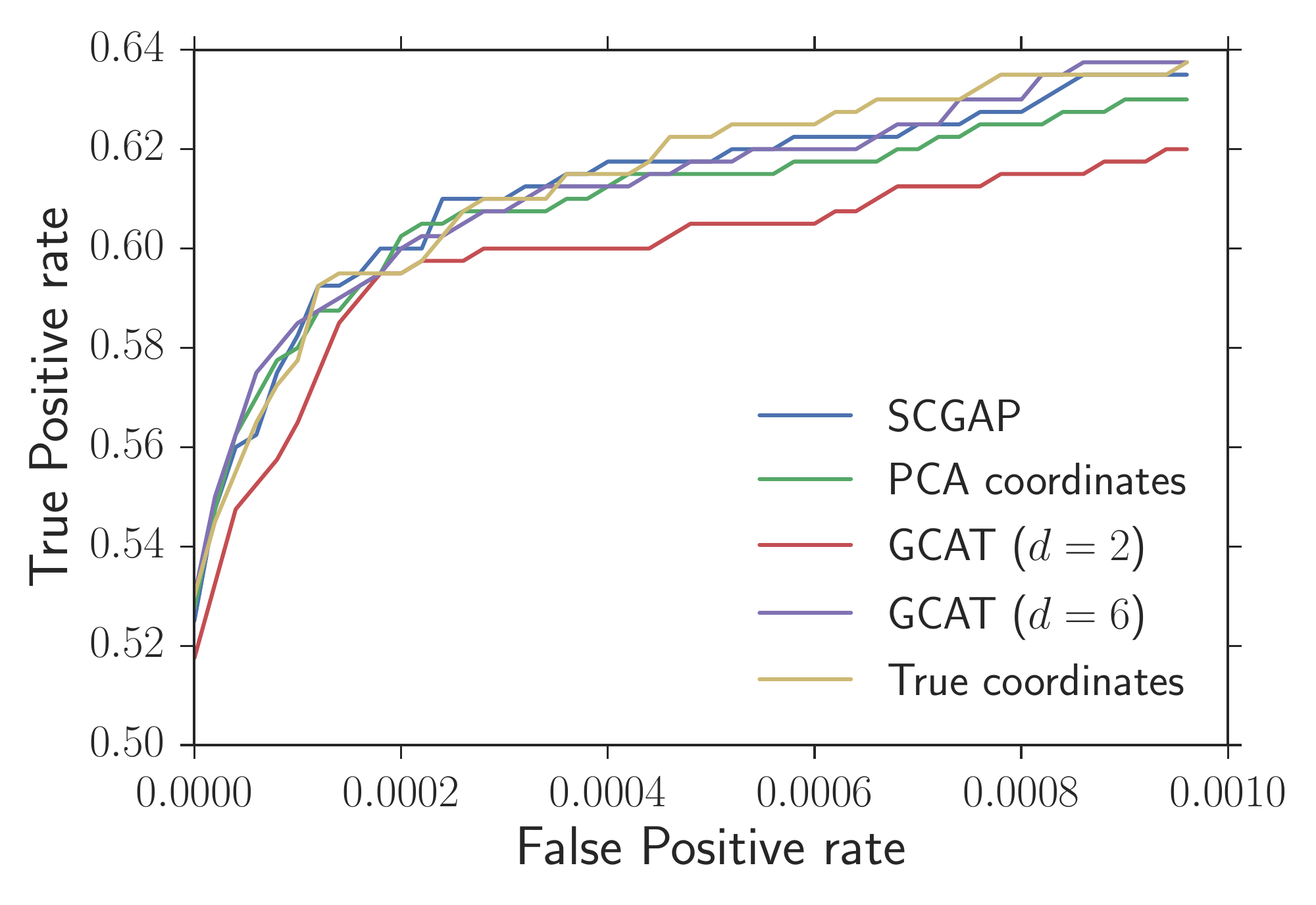}
\end{subfigure}
\begin{subfigure}[]{0.5\linewidth}
\centering
\caption{\hspace{0.75cm}$\alpha_1=2$}
\includegraphics[width=3.3in]{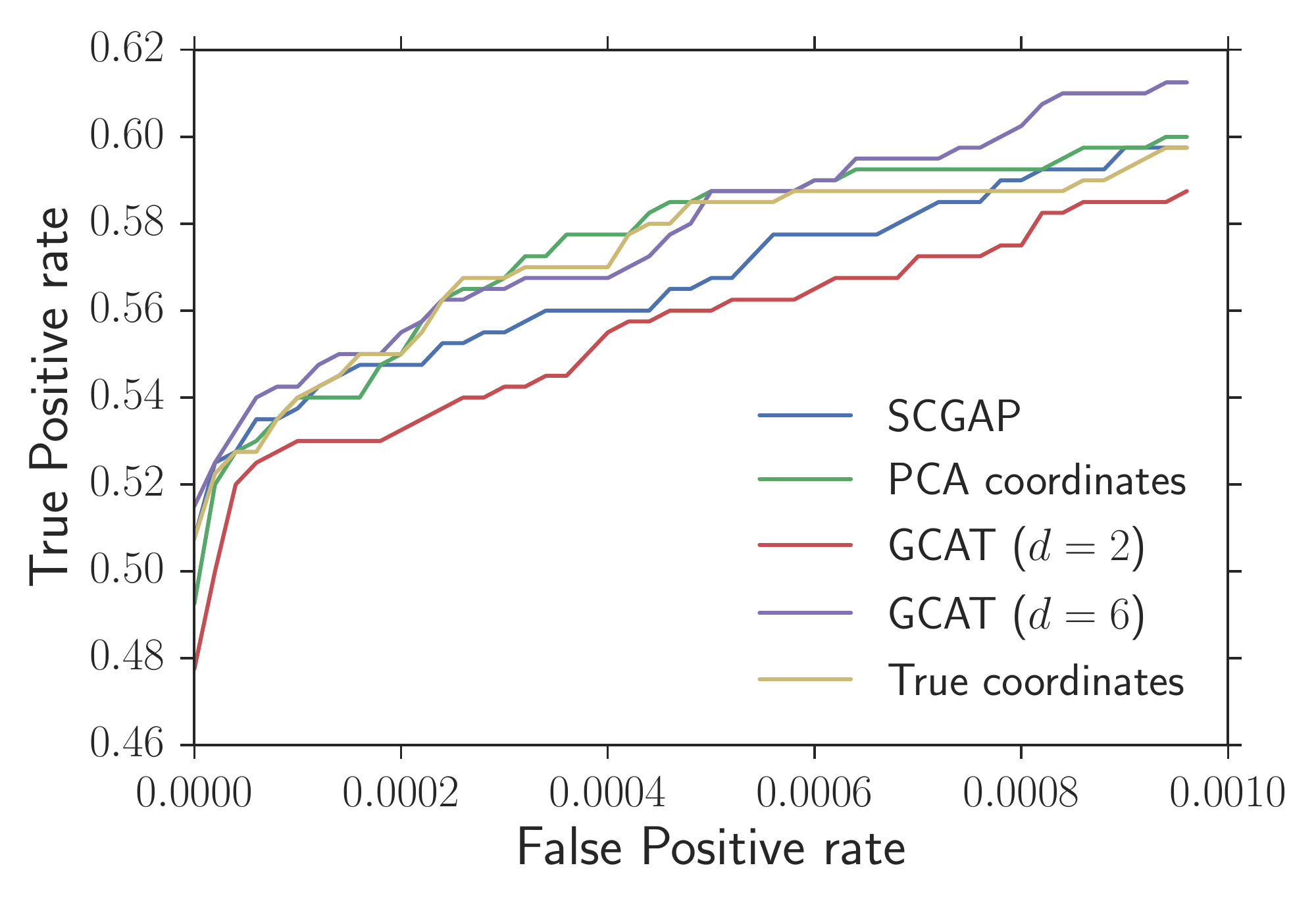}
\end{subfigure}
\begin{subfigure}[]{0.5\linewidth}
\centering
\caption{\hspace{0.75cm}$\alpha_1=4$}
\includegraphics[width=3.3in]{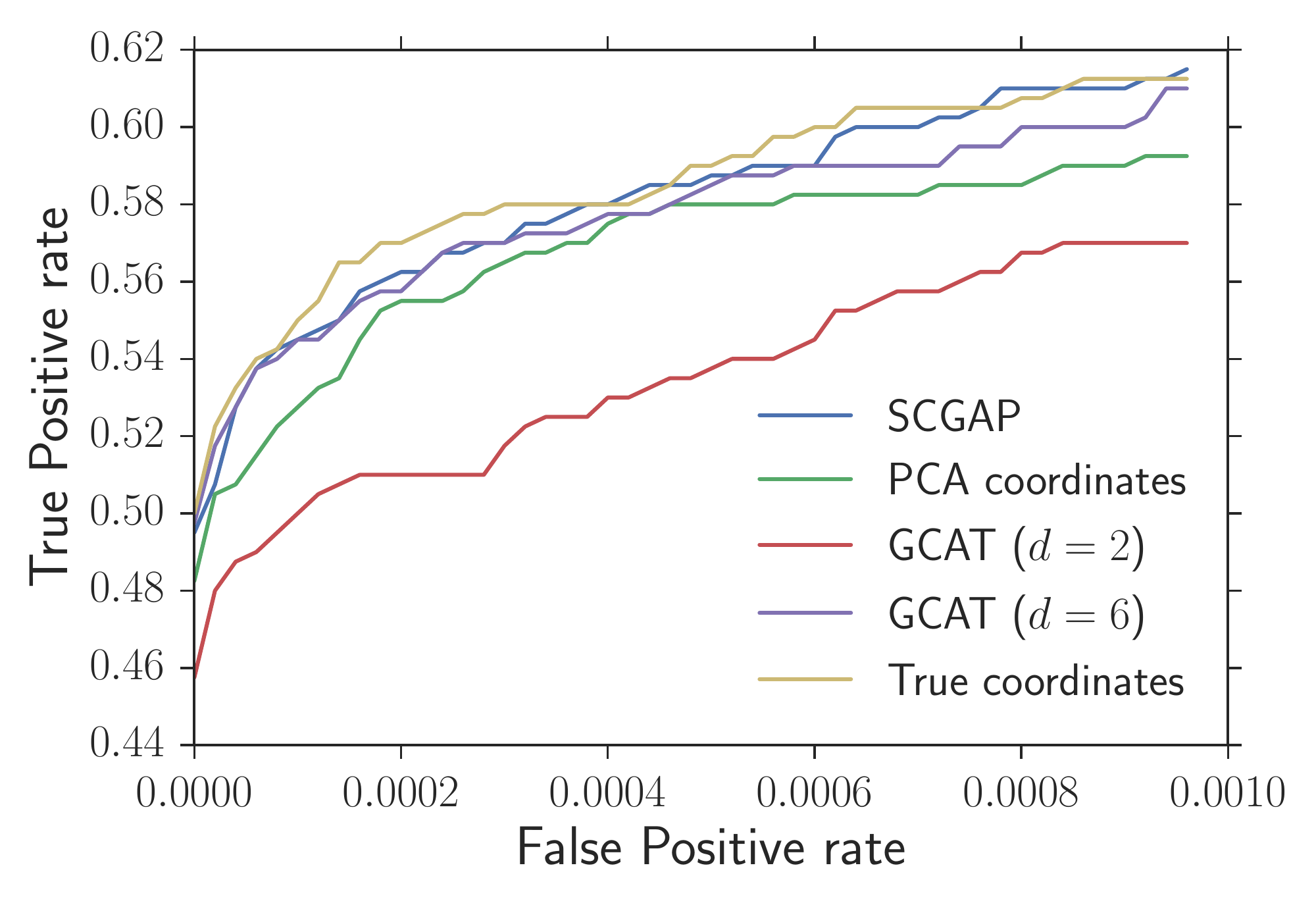}
\end{subfigure}
\begin{subfigure}[]{0.5\linewidth}
\centering
\caption{\hspace{0.75cm}$\alpha_1=8$}
\includegraphics[width=3.3in]{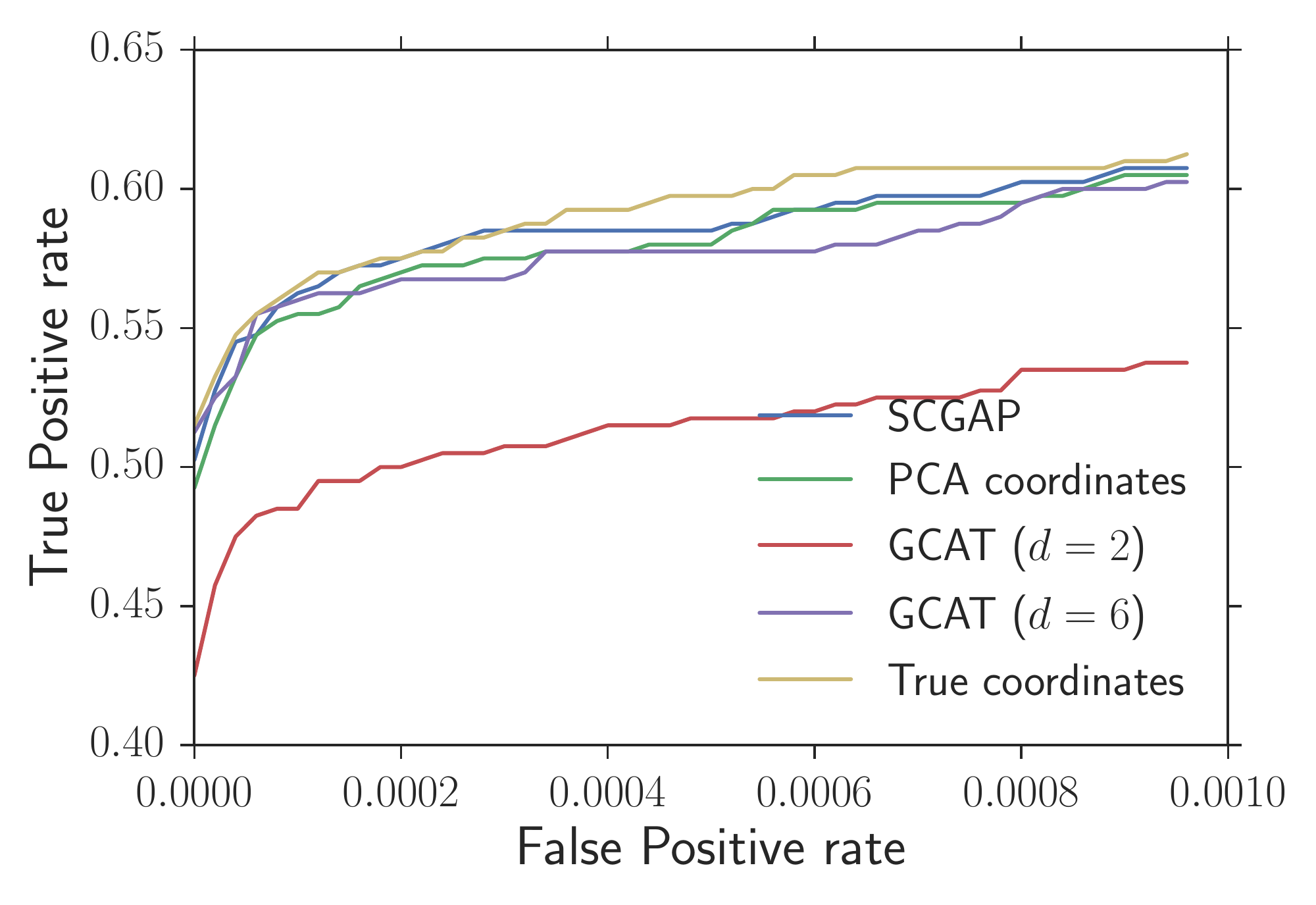}
\end{subfigure}
\begin{subfigure}[]{\linewidth}
\centering
\caption{\hspace{0.75cm}$\alpha_1=16$}
\includegraphics[width=3.3in]{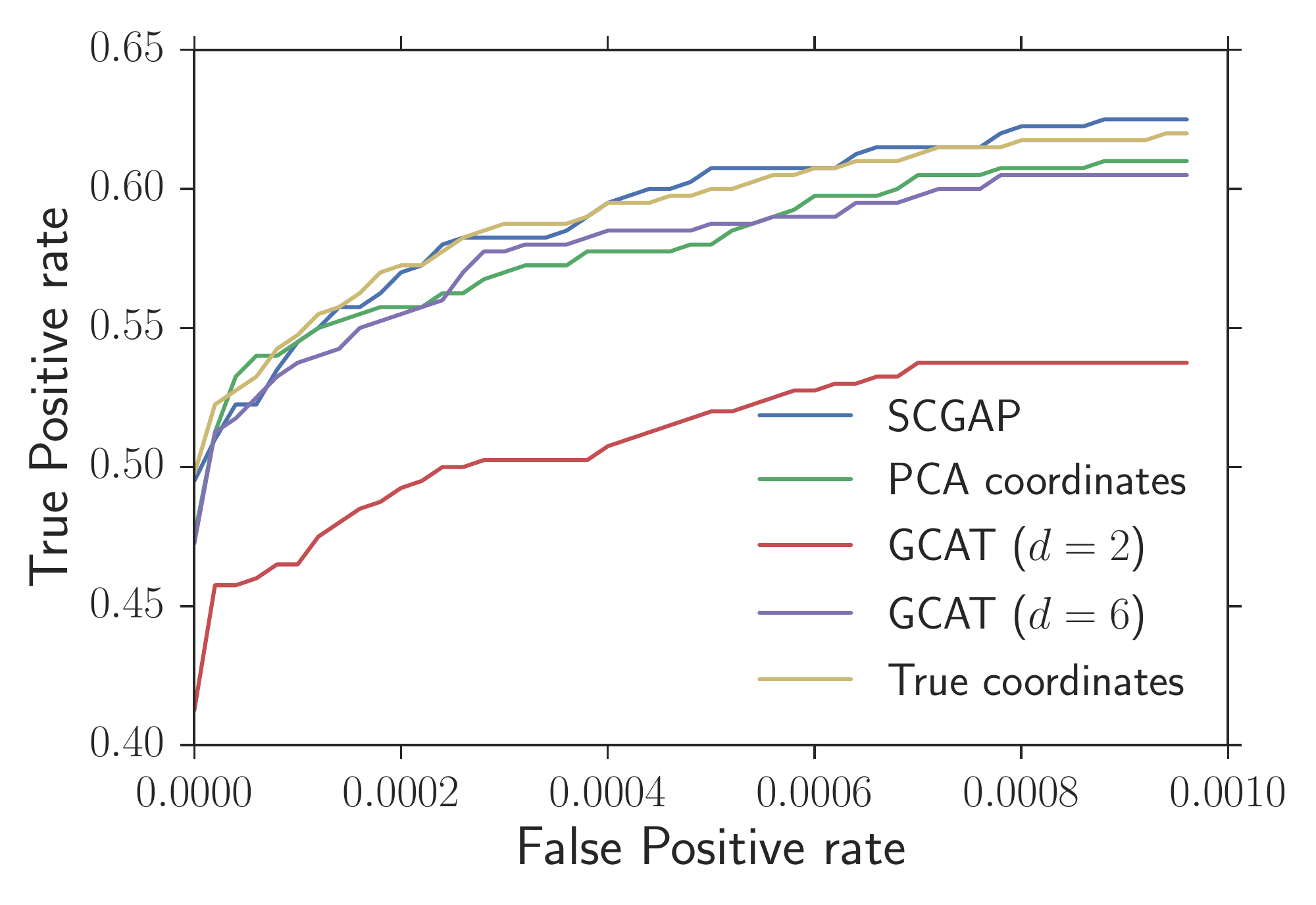}
\end{subfigure}
\caption{Same simulation scenario as in \fref{fig:logisticDirectionalExpDecayCov_n2000_p50k_pc10_beta1_gv0.20_lv0.10_nv0.70_vkappa0.1}, with $\kappa = 10$.}
\label{fig:logisticDirectionalExpDecayCov_n2000_p50k_pc10_beta1_gv0.20_lv0.10_nv0.70_vkappa10}
\end{figure}

\newpage
\subsection{NFBC dataset}\label{sec:nfbc}
The original dataset contained 364{,}590 SNPs from 5{,}402 individuals. After filtering individuals and SNPs using the same criteria for missing genotypes and deviation from Hardy-Weinberg equilibrium as described in \cite{song:2015}, we were left with 335{,}143 SNPs and 5{,}246 individuals. We added features for known confounders such as sex, oral contraceptive use, pregnancy status, and fasting status according to the procedure described in the first analysis of this dataset by \cite{sabatti:2009}. We performed a Box-Cox transform on the median 95\% of trait values to make the distribution of traits as close to a normal distribution as possible. 

We applied our localization algorithm \localizationAlgorithm{} on the genotype data to estimate two spatial ancestry coordinates for each individual. Since we did not have the ancestral or birth locations of the individuals in the sample, we could not optimize the threshold $\tau$ for estimating local spatial distances from the genetic distances as we had done in the simulations. Instead, we picked the threshold $\tau$ as described in \sref{sec:choosing_tau}. 

\vspace{1cm}
\begin{table}[H]
{
\scriptsize
\centering
\begin{tabular}{lccccccc} 
    \toprule
    Trait &   Abbreviation  &   \associationTestingAlgorithm{}+GC  &   GCAT+GC     &   LMM+GC   & PCA+GC  &   Uncorrected+GC  \\
    \midrule
    Height                  &   Height  &   1  &   1   &   0   &   0   &   0 \\
    Body mass index         &   BMI     &   0  &   0   &   0   &   0   &   0 \\
    HDL cholesterol levels  &   HDL     &   5  &   4   &   4   &   2   &   4 \\
    LDL cholesterol levels  &   LDL     &   4  &   4   &   3   &   3   &   3 \\
    Triglyceride levels     &   TG      &   $~\,2^{*}$  &   2   &   3   &   2   &   2 \\
    C-reactive protein      &   CRP     &   $~\,2^{*}$  &   $~\,2^{*}$   &   2   &   2   &   2 \\
    Glucose levels          &   GLU     &   3  &   3   &   2   &   2   &   2 \\
    Insulin levels          &   INS     &   0  &   0   &   0   &   0   &   0 \\
    Diastolic blood pressure&   DBP     &   0  &   0   &   0   &   0   &   0 \\
    Systolic blood pressure &   SBP     &   0  &   0   &   0   &   0   &   0 \\
    \bottomrule    
\end{tabular}
\caption{{\bf Number of significant loci discovered by \associationTestingAlgorithm{} and several other association-testing approaches on the Northern Finland Birth Cohorts (\NFBC{}) dataset.} The log-likelihood ratios from each method were corrected using genomic control, denoted by ``+GC".The genome-wide significance level was set to $7.2 \times 10^{-8}$. \\
\footnotesize{*Result when the trait was not transformed using the Box-Cox transformation. Under the transformation, one locus was significant.}}
\label{tab:nfbc_association_summary}
}
\end{table}

\begin{figure}[H]
\begin{subfigure}[]{0.65\textwidth}
\centering
\caption{}\label{fig:nfbc_ppplot}
\includegraphics[width=\textwidth, trim=0 1cm 0 2cm, clip]{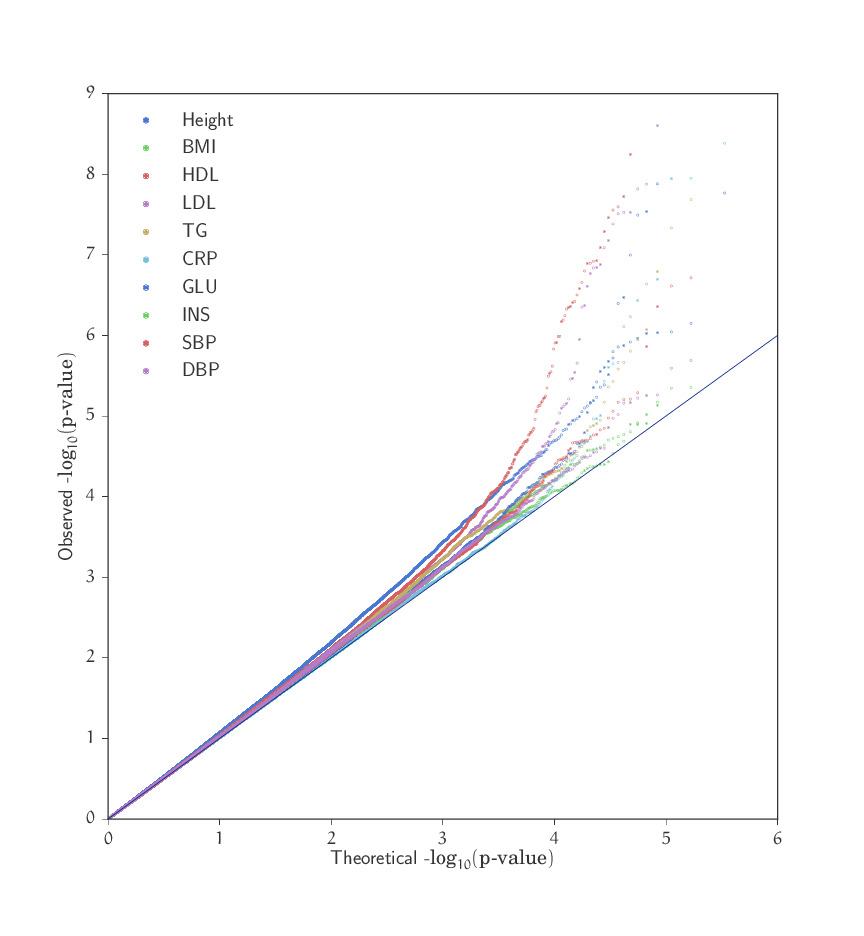}
\end{subfigure}
\begin{subfigure}[]{0.35\textwidth}
\centering
\caption{}\label{tab:nfbc_gcif_values}
\vspace{2.4cm}
\begin{tabular}{ll}
\toprule
Phenotype &    GCIF \\
\midrule
   Height &  1.0794 \\
      BMI &  1.0521 \\
      HDL &  1.0527 \\
      LDL &  1.0515 \\
       TG &  0.9983 \\
      CRP &  1.0614 \\
      GLU &  1.0049 \\
      INS &  1.0462 \\
      SBP &  1.0312 \\
      DBP &  1.0485 \\
\bottomrule
\end{tabular}
\vspace{2.5cm}
\end{subfigure}
\caption{\subref{fig:nfbc_ppplot} Plot of the observed versus expected log $p$-values of \associationTestingAlgorithm{} for the 10 metabolic traits in the Northern Finland Birth Cohorts (NFBC) dataset. See \tref{tab:nfbc_significant_hits} for the most significant p-values for each trait. \subref{tab:nfbc_gcif_values} Genomic control inflaction factor estimated for each of the 10 quantitative traits in the NFBC dataset using SNPs spaced 250kb apart.
}
\label{fig:nfbc_ppplot_gcif}
\end{figure}

\begin{table}
{\scriptsize
\parbox{.45\linewidth}{
\centering
{\bf Height} \vspace{2mm} \\
\begin{tabular}{lrrcc}
\toprule
\multicolumn{1}{c}{RSID}  & \multicolumn{1}{c}{Chr} & \multicolumn{1}{c}{Pos} & \multicolumn{1}{c}{\associationTestingAlgorithm{}}  & \multicolumn{1}{c}{\associationTestingAlgorithm{}+GC} \\
\midrule
 rs2814982 &    6 &   34654538 & 1.697e-08 & 7.101e-08 \\
 rs6719545 &    2 &  218160079 & 7.032e-07 & 2.147e-06 \\
 rs2815005 &    6 &   34746825 & 8.971e-07 & 2.684e-06 \\
 rs2744972 &    6 &   34767032 & 9.203e-07 & 2.748e-06 \\
 rs2814983 &    6 &   34699185 & 9.344e-07 & 2.786e-06 \\
 rs2814993 &    6 &   34726871 & 1.078e-06 & 3.175e-06 \\
 rs2814985 &    6 &   34656274 & 1.197e-06 & 3.496e-06 \\
 rs4911494 &   20 &   33435328 & 1.324e-06 & 3.833e-06 \\
 rs6088813 &   20 &   33438595 & 1.380e-06 & 3.981e-06 \\
 rs6058154 &   20 &   33049495 & 1.892e-06 & 5.316e-06 \\
\bottomrule
\end{tabular}
\vspace{4mm}

{\bf HDL cholestorol levels (HDL)}  \vspace{2mm} \\
\begin{tabular}{lrrcc}
\toprule
      RSID &  Chr &       Pos &     SCGAP &  SCGAP+GC \\
\midrule
 rs1532624 &   16 &  55562980 & 0.000e+00 & 0.000e+00 \\
 rs7499892 &   16 &  55564091 & 1.110e-16 & 2.220e-16 \\
 rs1532085 &   15 &  56470658 & 2.351e-12 & 5.686e-12 \\
 rs9989419 &   16 &  55542640 & 8.050e-11 & 1.723e-10 \\
 rs1800961 &   20 &  42475778 & 5.643e-09 & 1.044e-08 \\
 rs7120118 &   11 &  47242866 & 1.882e-08 & 3.341e-08 \\
 rs2167079 &   11 &  47226831 & 2.525e-08 & 4.436e-08 \\
  rs255049 &   16 &  66570972 & 2.759e-08 & 4.833e-08 \\
  rs415799 &   15 &  56478046 & 3.438e-08 & 5.977e-08 \\
  rs255052 &   16 &  66582496 & 5.116e-08 & 8.774e-08 \\
\bottomrule
\end{tabular}
\vspace{4mm}

{\bf Triglyceride levels (TG) (untransformed)}  \vspace{2mm} \\
\begin{tabular}{lrrcc}
\toprule
 \multicolumn{1}{c}{RSID}  & \multicolumn{1}{c}{Chr} & \multicolumn{1}{c}{Pos} & \multicolumn{1}{c}{\associationTestingAlgorithm{}}  & \multicolumn{1}{c}{\associationTestingAlgorithm{}+GC} \\
\midrule
  rs1260326 &    2 &   27584444 & 2.053e-08 & 2.318e-08 \\
 rs10096633 &    8 &   19875201 & 4.597e-08 & 5.160e-08 \\
   rs780094 &    2 &   27594741 & 1.608e-07 & 1.788e-07 \\
  rs5939593 &   23 &   92156237 & 8.433e-07 & 9.268e-07 \\
   rs673548 &    2 &   21091049 & 1.125e-06 & 1.233e-06 \\
   rs676210 &    2 &   21085029 & 1.551e-06 & 1.697e-06 \\
  rs6728178 &    2 &   21047434 & 2.140e-06 & 2.337e-06 \\
  rs3923037 &    2 &   21011755 & 2.592e-06 & 2.825e-06 \\
  rs4825619 &   23 &  117987171 & 3.688e-06 & 4.010e-06 \\
  rs6754295 &    2 &   21059688 & 4.327e-06 & 4.699e-06 \\
\bottomrule
\end{tabular}
}
\hfill
\parbox{.45\linewidth}{
\centering
{\bf Body mass index (BMI)} \vspace{2mm} \\
\begin{tabular}{lrrcc}
\toprule
 \multicolumn{1}{c}{RSID}  & \multicolumn{1}{c}{Chr} & \multicolumn{1}{c}{Pos} & \multicolumn{1}{c}{\associationTestingAlgorithm{}}  & \multicolumn{1}{c}{\associationTestingAlgorithm{}+GC} \\
\midrule
  rs5957365 &   23 &  119412593 & 4.390e-06 & 6.394e-06 \\
  rs6567030 &   18 &   54679876 & 4.476e-06 & 6.515e-06 \\
 rs10247383 &    7 &   63536549 & 6.667e-06 & 9.579e-06 \\
  rs1001729 &    6 &    2540477 & 1.223e-05 & 1.722e-05 \\
 rs12658762 &    5 &   18615363 & 1.263e-05 & 1.777e-05 \\
  rs6638764 &   23 &    8171611 & 1.558e-05 & 2.177e-05 \\
  rs2408165 &   16 &   61259627 & 1.691e-05 & 2.356e-05 \\
  rs4953198 &    2 &   45248172 & 1.792e-05 & 2.493e-05 \\
 rs17207196 &    7 &   74939001 & 1.876e-05 & 2.606e-05 \\
  rs8050136 &   16 &   52373776 & 1.971e-05 & 2.733e-05 \\
\bottomrule
\end{tabular}
\vspace{4mm}

{\bf LDL cholestorol levels (LDL)}  \vspace{2mm} \\
\begin{tabular}{lrrcc}
\toprule
 \multicolumn{1}{c}{RSID}  & \multicolumn{1}{c}{Chr} & \multicolumn{1}{c}{Pos} & \multicolumn{1}{c}{\associationTestingAlgorithm{}}  & \multicolumn{1}{c}{\associationTestingAlgorithm{}+GC} \\
\midrule
   rs646776 &    1 &  109620053 & 1.419e-12 & 4.361e-12 \\
      rs693 &    2 &   21085700 & 2.010e-11 & 5.512e-11 \\
   rs754524 &    2 &   21165046 & 2.477e-09 & 5.525e-09 \\
  rs6754295 &    2 &   21059688 & 1.317e-08 & 2.734e-08 \\
  rs6728178 &    2 &   21047434 & 1.509e-08 & 3.115e-08 \\
   rs207150 &    1 &   55579053 & 2.946e-08 & 5.910e-08 \\
 rs11668477 &   19 &   11056030 & 2.948e-08 & 5.913e-08 \\
  rs3923037 &    2 &   21011755 & 3.078e-08 & 6.164e-08 \\
  rs4844614 &    1 &  205941798 & 4.107e-08 & 8.123e-08 \\
   rs754523 &    2 &   21165196 & 6.584e-08 & 1.276e-07 \\
\bottomrule
\end{tabular}
\vspace{4mm}

{\bf C-reactive protein (CRP) (untransformed)}  \vspace{2mm} \\
\begin{tabular}{lrrcc}
\toprule
 \multicolumn{1}{c}{RSID}  & \multicolumn{1}{c}{Chr} & \multicolumn{1}{c}{Pos} & \multicolumn{1}{c}{\associationTestingAlgorithm{}}  & \multicolumn{1}{c}{\associationTestingAlgorithm{}+GC} \\
\midrule
  rs2794520 &    1 &  157945440 & 4.086e-09 & 5.607e-09 \\
  rs1169300 &   12 &  119915608 & 1.115e-08 & 1.503e-08 \\
  rs2464196 &   12 &  119919810 & 1.131e-08 & 1.525e-08 \\
  rs2592887 &    1 &  157919563 & 1.998e-07 & 2.563e-07 \\
  rs2650000 &   12 &  119873345 & 2.277e-07 & 2.915e-07 \\
  rs2098930 &    3 &  153371624 & 3.664e-07 & 4.652e-07 \\
  rs6624381 &   23 &   68197466 & 5.836e-07 & 7.349e-07 \\
   rs735396 &   12 &  119923227 & 7.723e-07 & 9.679e-07 \\
 rs10035541 &    5 &    7592712 & 2.147e-06 & 2.644e-06 \\
 rs12093699 &    1 &  157914612 & 2.236e-06 & 2.752e-06 \\
\bottomrule
\end{tabular}
}
}
\vspace{5mm}
\caption{The top 10 most significant SNPs found by \associationTestingAlgorithm{} for each of the 10 traits in the Northern Finland Birth Cohort dataset. We report the $p$-values both before and after genomic control adjustment.}
\label{tab:nfbc_significant_hits}
\end{table}

\begin{table}
{\scriptsize
\parbox{.45\linewidth}{
\centering
{\bf Glucose levels (GLU)}  \vspace{2mm} \\
\begin{tabular}{lrrcc}
\toprule
\multicolumn{1}{c}{RSID}  & \multicolumn{1}{c}{Chr} & \multicolumn{1}{c}{Pos} & \multicolumn{1}{c}{\associationTestingAlgorithm{}}  & \multicolumn{1}{c}{\associationTestingAlgorithm{}+GC} \\
\midrule
  rs560887 &    2 &  169471394 & 3.279e-11 & 5.947e-11 \\
 rs3847554 &   11 &   92308474 & 1.575e-10 & 2.742e-10 \\
 rs2971671 &    7 &   44177862 & 3.367e-10 & 5.749e-10 \\
 rs2908290 &    7 &   44182662 & 1.306e-08 & 2.029e-08 \\
 rs1387153 &   11 &   92313476 & 2.882e-08 & 4.387e-08 \\
  rs563694 &    2 &  169482317 & 3.188e-08 & 4.840e-08 \\
 rs2166706 &   11 &   92331180 & 9.990e-08 & 1.473e-07 \\
 rs1447352 &   11 &   92362409 & 3.354e-07 & 4.793e-07 \\
 rs7121092 &   11 &   92363999 & 3.982e-07 & 5.666e-07 \\
  rs758989 &    7 &   44169531 & 1.584e-06 & 2.176e-06 \\
\bottomrule
\end{tabular}
\vspace{4mm}

{\bf Triglyceride levels (TG)} \vspace{2mm} \\
\begin{tabular}{lrrcc}
\toprule
\multicolumn{1}{c}{RSID}  & \multicolumn{1}{c}{Chr} & \multicolumn{1}{c}{Pos} & \multicolumn{1}{c}{\associationTestingAlgorithm{}}  & \multicolumn{1}{c}{\associationTestingAlgorithm{}+GC} \\
\midrule
  rs1260326 &    2 &   27584444 & 4.484e-09 & 5.903e-09 \\
   rs780094 &    2 &   27594741 & 6.890e-08 & 8.704e-08 \\
 rs10096633 &    8 &   19875201 & 1.174e-07 & 1.471e-07 \\
   rs673548 &    2 &   21091049 & 6.148e-07 & 7.515e-07 \\
   rs676210 &    2 &   21085029 & 7.149e-07 & 8.719e-07 \\
  rs6728178 &    2 &   21047434 & 9.045e-07 & 1.099e-06 \\
  rs2304130 &   19 &   19650528 & 9.208e-07 & 1.119e-06 \\
  rs6754295 &    2 &   21059688 & 1.726e-06 & 2.077e-06 \\
  rs3923037 &    2 &   21011755 & 6.887e-06 & 8.118e-06 \\
 rs12805061 &   11 &  116058235 & 7.972e-06 & 9.377e-06 \\
\bottomrule
\end{tabular}
\vspace{4mm}

{\bf Systolic blood pressure (SBP)} \vspace{2mm} \\
\begin{tabular}{lrrcc}
\toprule
\multicolumn{1}{c}{RSID}  & \multicolumn{1}{c}{Chr} & \multicolumn{1}{c}{Pos} & \multicolumn{1}{c}{\associationTestingAlgorithm{}}  & \multicolumn{1}{c}{\associationTestingAlgorithm{}+GC} \\
\midrule
   rs782588 &    2 &   55695144 & 1.916e-07 & 2.523e-07 \\
   rs782586 &    2 &   55689669 & 2.414e-07 & 3.164e-07 \\
   rs782602 &    2 &   55702813 & 4.374e-07 & 5.669e-07 \\
  rs2627759 &    2 &   55706845 & 1.363e-06 & 1.729e-06 \\
 rs10496050 &    2 &   55659817 & 5.122e-06 & 6.337e-06 \\
  rs1754154 &    1 &   43243353 & 6.118e-06 & 7.544e-06 \\
  rs9656787 &    8 &  104301120 & 6.223e-06 & 7.671e-06 \\
   rs782606 &    2 &   55740106 & 6.696e-06 & 8.243e-06 \\
   rs782652 &    2 &   55716279 & 8.643e-06 & 1.059e-05 \\
   rs782637 &    2 &   55747751 & 1.049e-05 & 1.281e-05 \\
\bottomrule
\end{tabular}
}
\hfill
\parbox{.45\linewidth}{
\centering
{\bf Insulin levels (INS)}  \vspace{2mm} \\
\begin{tabular}{lrrcc}
\toprule
      RSID &  Chr &       Pos &     SCGAP &  SCGAP+GC \\
\midrule
  rs521184 &    8 &  41720842 & 7.293e-06 & 9.437e-06 \\
 rs5985850 &   23 &  28409093 & 9.537e-06 & 1.226e-05 \\
 rs5943445 &   23 &  28411095 & 1.205e-05 & 1.541e-05 \\
 rs6502762 &   17 &   3819013 & 1.257e-05 & 1.606e-05 \\
 rs7241379 &   18 &  64306982 & 2.042e-05 & 2.579e-05 \\
  rs998223 &    2 &  64824633 & 2.260e-05 & 2.847e-05 \\
 rs6126645 &   20 &  50745422 & 2.895e-05 & 3.626e-05 \\
 rs6526679 &   23 &  27468439 & 3.672e-05 & 4.574e-05 \\
  rs932052 &   12 &  62081496 & 3.968e-05 & 4.934e-05 \\
 rs2037206 &   18 &  64323734 & 3.974e-05 & 4.942e-05 \\
\bottomrule
\end{tabular}
\vspace{4mm}

{\bf C-reactive protein (CRP)} \vspace{2mm} \\
\begin{tabular}{lrrcc}
\toprule
\multicolumn{1}{c}{RSID}  & \multicolumn{1}{c}{Chr} & \multicolumn{1}{c}{Pos} & \multicolumn{1}{c}{\associationTestingAlgorithm{}}  & \multicolumn{1}{c}{\associationTestingAlgorithm{}+GC} \\
\midrule
  rs2794520 &    1 &  157945440 & 6.284e-14 & 8.360e-14 \\
 rs12093699 &    1 &  157914612 & 8.348e-12 & 1.058e-11 \\
  rs2592887 &    1 &  157919563 & 1.260e-08 & 1.487e-08 \\
  rs1811472 &    1 &  157908973 & 4.596e-08 & 5.358e-08 \\
   rs402681 &    4 &  104634397 & 9.352e-07 & 1.059e-06 \\
  rs7694802 &    4 &  104621696 & 1.973e-06 & 2.219e-06 \\
  rs7178765 &   15 &   23672266 & 4.585e-06 & 5.114e-06 \\
 rs10107791 &    8 &  101040128 & 5.874e-06 & 6.535e-06 \\
   rs340468 &    4 &  104637688 & 7.626e-06 & 8.464e-06 \\
  rs6701469 &    1 &  199265442 & 9.789e-06 & 1.084e-05 \\
\bottomrule
\end{tabular}
\vspace{4mm}

{\bf Diastolic blood pressure (DBP)} \vspace{2mm} \\
\begin{tabular}{lrrcc}
\toprule
\multicolumn{1}{c}{RSID}  & \multicolumn{1}{c}{Chr} & \multicolumn{1}{c}{Pos} & \multicolumn{1}{c}{\associationTestingAlgorithm{}}  & \multicolumn{1}{c}{\associationTestingAlgorithm{}+GC} \\
\midrule
 rs5928929 &   23 &   35549454 & 2.031e-06 & 2.894e-06 \\
 rs6942973 &    7 &    3134277 & 2.528e-06 & 3.580e-06 \\
  rs472594 &    1 &  226668261 & 5.430e-06 & 7.521e-06 \\
 rs5927821 &   23 &   31656866 & 5.510e-06 & 7.629e-06 \\
  rs952061 &   12 &  100502356 & 5.948e-06 & 8.217e-06 \\
 rs1079199 &   11 &    6384682 & 6.771e-06 & 9.320e-06 \\
 rs2094147 &   23 &   33025497 & 6.873e-06 & 9.455e-06 \\
 rs7783562 &    7 &  106704674 & 9.493e-06 & 1.294e-05 \\
  rs808127 &   23 &    8458979 & 1.057e-05 & 1.436e-05 \\
 rs4548444 &    1 &  204956761 & 1.381e-05 & 1.861e-05 \\
\bottomrule
\end{tabular}
}
}
\vspace{5mm} \\
\tref{tab:nfbc_significant_hits} continued
\end{table}

\begin{center}
\begin{table}
\begin{tabular}{llrrlccc}
\toprule
\multicolumn{1}{c}{Phenotype}  &  \multicolumn{1}{l}{RSID}  & \multicolumn{1}{r}{Chr} & \multicolumn{1}{c}{Pos} & \multicolumn{1}{l}{Nearest gene}  & \multicolumn{1}{c}{\associationTestingAlgorithm{}+GC} & \multicolumn{1}{c}{Replication Study}\\
\midrule
Height  & rs2814982   &    6 &  34654538    &C6orf106 &   7.101e-08 &   \cite{weedon:2008}*\\
\midrule
HDL     & rs1532624   &   16 &  55562980    &   CETP  &   0.000e+00 &   \cite{aulchenko:2009} \\
HDL     & rs1532085   &   15 &  56470658    &   LIPC  &   5.686e-12 &   \cite{willer:2013} \\
HDL     & rs1800961   &   20 &  42475778    &   HNF4A &   1.044e-08 &   \cite{willer:2013} \\
HDL     & rs7120118   &   11 &  47242866    &   NR1H3 &   3.341e-08 &   --- \\
HDL     &  rs255049   &   16 &  66570972    &   LCAT  &   4.833e-08 &   \cite{willer:2013} \\ 
\midrule
TG      &  rs1260326  &    2 &   27584444   &   GCKR  &   2.318e-08 &   \cite{willer:2013} \\
TG      & rs10096633  &    8 &   19875201   &   LPL   &   5.160e-08 &   \cite{aulchenko:2009} \\
\midrule
LDL     &   rs646776  &    1 &  109620053   &  CELSR2 &   4.361e-12 &   \cite{aulchenko:2009} \\
LDL     &      rs693  &    2 &   21085700   &   APOB  &   5.512e-11 &   \cite{aulchenko:2009} \\
LDL     &   rs207150  &    1 &   55579053   &   USP24 &   5.910e-08 &   --- \\
LDL     & rs11668477  &   19 &   11056030   &   LDLR  &   5.913e-08 &   \cite{willer:2013} \\
\midrule
CRP     &  rs2794520  &    1 &  157945440   &   CRP   &   5.607e-09 &   \cite{dehghan:2011} \\
CRP     &  rs1169300  &   12 &  119915608   &   HNF1A &   1.503e-08 &   \cite{dehghan:2011} \\
\midrule
GLU     &  rs560887   &    2 &  169471394   &   G6PC2 &   5.947e-11 &   \cite{dupuis:2010} \\
GLU     & rs3847554   &   11 &   92308474   &   MTNR1B&   2.742e-10 &   \cite{dupuis:2010} \\
GLU     & rs2971671   &    7 &   44177862   &   GCK   &   5.749e-10 &   \cite{dupuis:2010} \\
\bottomrule \\
\end{tabular}
\footnotesize{* SNP rs2814993 is reported to be associated with height, and is 72kb from rs2814982 with LD $r^2 = 0.56$ in the 1000 Genomes CEU samples (see also \citet{song:2015}).}
\vspace{2mm}
\caption{Most significantly associated SNPs at each locus that were detected by \associationTestingAlgorithm{} with genomic control, and the replication studies on different datasets which have also reported these associations.}
\label{tab:nfbc_loci}
\end{table}
\end{center}

\newpage
\section{Proofs}\label{sec:proofs}
\subsection{Proof of Theorem \ref{thm:mu-eta}}
We use the shorthand $q_{i\ell}:=q_\ell(\bfz_i)$ to lighten the notation. By applying the triangle inequality, for $1\le i<j \le n$,
\begin{eqnarray}\label{etaij-dec}
\begin{split}
|\hat{\eta}_{i,j} - \eta(\bfz_i -\bfz_j)|\le &~\bigg| \frac{1}{p}  \sum_{\ell=1}^p \left(\frac{x_{i\ell}}{2} - \mu_\ell\right)\left(\frac{x_{j\ell}}{2} - \mu_\ell\right)
-\E[(q_{i\ell}-\mu_\ell)(q_{j\ell}-\mu_\ell)] \bigg|\\
&+ \frac{1}{2p}\sum_{\ell=1}^p (x_{i\ell}+x_{j\ell})\, |\hmu_\ell-\mu_\ell |\\
&+ \frac{1}{p}\sum_{\ell=1}^p (\hmu_\ell+\mu_\ell) |\hmu_\ell-\mu_\ell|\,.
\end{split}
\end{eqnarray}
Define the following events $\cA_{i,j}$ for $1\le i<j\le n$,
\begin{align}\label{eq:Aij}
\cA_{i,j}: \quad \bigg| \frac{1}{p}  \sum_{\ell=1}^p \left(\frac{x_{i\ell}}{2} - \mu_\ell\right)\left(\frac{x_{j\ell}}{2} - \mu_\ell\right)
-\E[(q_{i\ell}-\mu_\ell)(q_{j\ell}-\mu_\ell)] \bigg|<t
\end{align}
We note that conditional on $q_{i\ell}$ and $q_{j\ell}$, the genotypes $x_{i\ell}$ and $x_{j\ell}$ are independent.
Therefore, 
$$\E\Big[\left(\frac{x_{i\ell}}{2}-\mu_\ell\right)\left(\frac{x_{j\ell}}{2}-\mu_\ell\right)\Big] =
\E\Big[\E\Big[\left(\frac{x_{i\ell}}{2}-\mu_\ell\right)\left(\frac{x_{j\ell}}{2}-\mu_\ell\right)\Big|q_{i\ell}, q_{j\ell}\Big]\Big]
= \E[(q_{i\ell}-\mu_\ell)(q_{j\ell}-\mu_\ell)]\,.$$
Further, the genotypes $x_{i\ell}$ are independent for different SNPs $\ell$.
By applying Hoeffding's inequality, we obtain $\prob(\cA_{i,j}^c)\le 2e^{-2pt^2}$ for any fixed pair $i$ and $j$. 

To bound the second term in \eqref{etaij-dec}, we write,
\begin{align}\label{eq:term2-1}
\frac{1}{2p}\sum_{\ell=1}^p (x_{i\ell} + x_{j\ell}) |\hmu_\ell-\mu_\ell| \le \frac{2}{p} \sum_{\ell=1}^p |\hmu_\ell-\mu_\ell|
\end{align}
Note that the summands $|\mu_\ell-\hmu_\ell|$ are independent. We define the event $\cE$ as,
\begin{align}\label{eq:term2-2}
\cE:\quad \frac{1}{p} \sum_{\ell=1}^p |\mu_\ell-\hmu_\ell| - \frac{1}{p} \sum_{\ell=1}^p \E(|\mu_\ell-\hmu_\ell|) < t\,. 
\end{align}
By applying Hoeffding's inequality, we obtain $\prob(\cE^c)\le e^{-2pt^2}$.

We next bound $\E(|\mu_\ell-\hmu_\ell|)$. For each $\ell$, $1\le \ell \le p$, define the events $\cB_{\ell}$ and $\cC_{\ell}$ as follows,
\begin{align*}
&\cB_{\ell}: \quad \bigg| \frac{1}{n} \sum_{i=1}^n \left(\frac{x_{i\ell}}{2} - q_{i\ell}\right) \bigg| <t\\
&\cC_\ell:\quad \bigg|\frac{1}{n} \sum_{i=1}^n q_{i\ell} - \mu_\ell \bigg| < t\,.
\end{align*}
Note that conditional on the allele frequencies $q_{i\ell}$, the genotypes $x_{i\ell}$ are independent across index $i$. We can apply Hoeffding's inequality again to get $\prob(\cB_\ell^c~|~\{q_{i\ell}\}_{1\le i\le n}) \le 2e^{-2nt^2}$. Hence, $\prob(\cB_\ell^c) \le 2e^{-2nt^2}$ as well.
Bounding the probability of $\cC_\ell$ requires more work since the summands $q_{i\ell}$ are dependent. We construct $K\in \reals^{n \times n}$ with $K_{ij} = \eta(\bfz_i-\bfz_j)$. Let ${\bf 1} = (1/\sqrt{n},\dotsc, 1/\sqrt{n})^\sT$, $\bq = (q_{1\ell}-\mu_\ell, q_{2\ell}-\mu_\ell, \dotsc, q_{n\ell}-\mu_\ell)^\sT$ and set $\tbq = K^{-1/2} \bq$.  We write
\begin{align}
\frac{1}{n} \sum_{i=1}^n(q_{i\ell} -\mu_\ell) = \frac{1}{\sqrt{n}}{\bf 1}^\sT K^{1/2} \tbq\,. \label{eq:q_deviation_centered}
\end{align}
Since the coordinates of $\tbq$ are mean zero and are uncorrelated, using Chebyshev's inequality and \eqref{eq:q_deviation_centered}, we get
\begin{align*}
\prob(\cC_\ell^c) \le \prob(|{\bf 1}^\sT K^{1/2} \tbq|> t\sqrt{n})\le \frac{1}{nt^2} ({\bf 1}^\sT K {\bf 1})\,.
\end{align*}
We are now ready to bound $\E(|\mu_\ell-\hmu_\ell|)$. Using the notation $\kappa:= {\bf 1}^\sT K {\bf 1}$, we have
\begin{align}
\prob\Big(|\hmu_\ell-\mu_\ell|>2t \Big) \le \prob(\cB_\ell^c) + \prob(\cC_\ell^c) \le 2e^{-2nt^2} + \frac{1}{nt^2} \kappa\,,
\end{align}
where the first inequality follows from triangle inequality.
Hence,
\begin{align*}
\E(|\mu_\ell-\hmu_\ell |) &= \int_0^{\infty} \prob(|\mu_\ell-\hmu_\ell |>t) \\
&\le s+ \int_{s}^{\infty} \left(2e^{-nt^2/2} + \frac{4\kappa}{nt^2} \right) \de t\\
&\le s+\frac{2}{n}+ \frac{4\kappa}{sn}\,,
\end{align*}
for any $s > 0$. Choosing $s = 2\sqrt{\kappa/n}$, we arrive at
\begin{align}\label{eq:Emu}
\E(|\mu_\ell-\hmu_\ell|) \le \frac{2}{n} + 4\sqrt{\frac{\kappa}{n}}\,.
\end{align}
Using equations~\eqref{eq:term2-1}, \eqref{eq:Emu} and recalling definition~\eqref{eq:term2-2} we conclude that on event $\cE$ the following is true,
\begin{align}\label{eq:T2}
\frac{1}{2p}\sum_{\ell=1}^p (x_{i\ell} + x_{j\ell}) |\hmu_\ell-\mu_\ell|  < 2t+\frac{4}{n} + 8\sqrt{\frac{\kappa}{n}} \,.
\end{align}
By a similar argument, on event $\cE$ we have the following bound on the third term in equation~\eqref{etaij-dec}:
\begin{align}\label{eq:T3}
\frac{1}{p}\sum_{\ell=1}^p (\hmu_{\ell} + \mu_{j\ell}) |\hmu_\ell-\mu_\ell|  < 2t+\frac{4}{n} + 8\sqrt{\frac{\kappa}{n}} \,.
\end{align}
Combining \eqref{eq:Aij}, \eqref{eq:T2}, and \eqref{eq:T3}, we obtain that on event $\cA_{ij}\cap \cE$, the following is true,
\begin{align}\label{eq:B1}
|\hat{\eta}_{i,j} - \eta(\bfz_i - \bfz_j)| \le 5t+ \frac{8}{n}+ 16\sqrt{\frac{\kappa}{n}}\,.
\end{align}

We next proceed to bound $|\heta_0-\eta(\bfzero)|$. Note that $\E[x_{i\ell}^2-x_{i\ell}] = 2\E[q_{i\ell}^2]$. For $1\le i\le n$, define the event $\cD_i$ as,
\begin{align}
\cD_i :\quad \frac{1}{p} \sum_{\ell=1}^p \left(\frac{x_{i\ell}^2-x_{i\ell}}{2} - \E[q_{i\ell}^2]\right) < t\,.
\end{align}
Recalling that $x_{i\ell}$ are independent for different $\ell$, using Hoeffding's inequality, we obtain $\prob(\cD_i^c)\le 2e^{-2pt^2}$. 
On the event $\cap_{i=1}^n \cD_i \cap \cE$, we have,
\begin{align}
|\heta_0-\eta(0)| &\le \frac{1}{n} \sum_{i=1}^n \bigg|\frac{1}{p}\sum_{\ell=1}^p \Big(\frac{x_{i\ell}^2-x_{i\ell}}{2}
- \E[q_{i\ell}^2]\Big)\bigg| + \bigg|\frac{1}{p} \sum_{\ell=1}^p (\hmu_\ell^2 - \mu_\ell^2) \bigg|\nonumber\\
&\le \frac{1}{n} \sum_{i=1}^n \bigg|\frac{1}{p}\sum_{\ell=1}^p \Big(\frac{x_{i\ell}^2-x_{i\ell}}{2}
- \E[q_{i\ell}^2]\Big)\bigg|+
\frac{1}{p}\sum_{\ell=1}^p (\hmu_\ell+\mu_\ell)\, |\hmu_\ell-\mu_\ell|\nonumber\\
|\heta_0-\eta(0)| &\le 3t+\frac{4}{n} + 8\sqrt{\frac{\kappa}{n}}\,,\label{eq:B2}
\end{align}
where we used \eqref{eq:T3} in the last inequality.
Finally, by utilizing the union bound over $\cA_{i,j}$, $\cD_i$, and $\cE$ for all $1\le i<j\le n$,
equations \eqref{eq:B1} and \eqref{eq:B2} are true with probability at least
\begin{align}
1- (n^2-n) e^{-2pt^2} -2n e^{-2pt^2} - e^{-2pt^2} \ge 1-(n+1)^2e^{-2pt^2}\,.
\end{align}
Choosing $t = \sqrt{2\log(n+1)/p}$ gives the desired result.

\end{document}